\documentclass[12pt]{article}
\usepackage{jheppubMOD}
\usepackage{amsfonts,ulem,bbold}
\usepackage{amsmath,amssymb, braket}
\usepackage{ifthen}
\usepackage{amstext}
\usepackage{graphicx,epsfig}
\usepackage{verbatim} 
\usepackage{fancybox}
\usepackage{color,xcolor}
\usepackage{enumitem}
\usepackage{subfigure}
\usepackage{bbm}
\usepackage{parskip}

\graphicspath{{Figures/}}

\newcommand{\Comment}[1]{{}}
\definecolor{mycol1}{RGB}{85,26,139}
\definecolor{mycol2}{RGB}{255,0,0}
\definecolor{mycol3}{RGB}{85,26,139}
\definecolor{mycol4}{RGB}{0,0,255}

\newcommand\newc{\newcommand}

\newc{\Tr}{\mathrm{Tr}}
\newc{\md}{\mathrm{d}}
\newc{\beqn}{\begin{eqnarray}}
\newc{\eeqn}{\end{eqnarray}}
\newc{\bea}{\begin{array}}
\newc{\eea}{\end{array}}
\newc{\be}{\begin{equation}}
\newc{\ee}{\end{equation}}
\newc{\nn}{\nonumber}
\newc{\td}{\mathrm{d}}
\newc{\dd}{\mathrm{d}}
\newc{\p}{\partial}
\newc{\wt}{\widetilde}

\newc{\gmn}{g_{\mu\nu}}
\newc{\bgmn}{\bar g_{\mu\nu}}
\newc{\fmn}{f_{\mu\nu}}
\newc{\emn}{\eta_{\mu\nu}}
\newc{\bfmn}{\bar f_{\mu\nu}}
\newc{\sgf}{\sqrt{g^{-1}f}}
\newc{\Gmn}{G_{\mu\nu}}
\newc{\Mmn}{M_{\mu\nu}}
\newc{\hmn}{h_{\mu\nu}}
\newc{\humn}{h^{\mu\nu}}
\newc{\meff}{M_\mathrm{eff}}
\newc{\dx}{\mathrm{d}^4x}
\newc{\dg}{\delta{g}}
\newc{\df}{\delta{f}}
\newc{\dm}{\delta{M}}
\newc{\dS}{\delta{S}}
\newc{\dMG}{\delta M^G}
\newc{\dMg}{\delta M}
\newc{\bphi}{\bar{\Phi}}
\newc{\bpsi}{\bar{\Psi}}
\newc{\dphi}{\delta{\Phi}}
\newc{\dpsi}{\delta{\Psi}}
\newc{\dgg}{\delta{G}}
\newc{\bgg}{\bar{G}}
\newc{\bg}{\bar{g}}
\newc{\mA}{\mathbb{A}}
\newc{\mB}{\mathbb{B}}
\newc{\tr}{\mathrm{Tr}}
\newc{\Lag}{\mathcal{L}}
\newc{\E}{\bar{\mathcal{E}}}
\newc{\EE}{\tilde{\mathcal{E}}}
\newc{\mfp}{m_{\mathrm{FP}}}
\newc{\Mc}{M^{\mathrm{c}}}
\newc{\te}{\tilde\epsilon}
\newc{\cN}{{\cal N}}
\newc{\bbeta}{\bar{\beta}}
\newc{\mpl}{M_{\mathrm{Pl}}}

\def\ph{\phantom}
\newc{\la}{\langle}
\newc{\ra}{\rangle}
\newc{\prof}{\mathcal{P}}
\newc\eg{{\rm {e.g.~}}}
\newc\etal{{\rm {et al.}}}
\newc\ie{{\rm i.e.~}}
\newc\etc{{\rm {etc}}}

\newcommand{\gae}{%
  \ensuremath{\lower 2pt \hbox{%
    $\, \buildrel {\scriptstyle >}\over {\scriptstyle \sim}\,$}%
    }%
  }
\newcommand{\lae}{%
  \ensuremath{\lower 2pt \hbox{%
    $\, \buildrel {\scriptstyle <}\over {\scriptstyle \sim}\,$}%
    }%
  }

\title{Recent developments in bimetric theory}
\author[1]{Angnis~Schmidt-May,}
\author[2]{Mikael von Strauss}
\affiliation[1]{Institut f\"ur Theoretische Physik, Eidgen\"ossische Technische Hochschule Z\"urich,\\
Wolfgang-Pauli-Strasse 27, 8093 Z\"urich, Switzerland}
\affiliation[2]{UPMC-CNRS, UMR7095,
 Institut d'Astrophysique de Paris, GReCO,\\
 98bis boulevard Arago, F-75014 Paris, France.}
\emailAdd{angniss@itp.phys.ethz.ch}
\emailAdd{strauss@iap.fr}

\abstract{ This review is dedicated to recent progress in the field of classical, interacting, massive spin-2 theories, with a focus on ghost-free bimetric theory. We will outline its history and its development as a nontrivial extension and generalisation of nonlinear massive gravity. We present a detailed discussion of the consistency proofs of both theories, before we review Einstein solutions to the bimetric equations of motion in vacuum as well as the resulting mass spectrum. We introduce couplings to matter and then discuss the general relativity and massive gravity limits of bimetric theory, which correspond to decoupling the massive or the massless spin-2 field from the matter sector, respectively. More general classical solutions are reviewed and the present status of bimetric cosmology is summarised.
An interesting corner in the bimetric parameter space which could potentially give rise to a nonlinear theory for partially massless spin-2 fields is also discussed. Relations to higher-curvature theories of gravity are explained and finally we give an overview of possible extensions of the theory and review its formulation in terms of vielbeins.}
\date{\today}

\begin{document} 
\maketitle
\newpage

\section{Introduction}

\subsection{Motivation}

The Standard Model of particle physics contains massive and massless fields with spin 0, 1/2 and 1. Gravitational interactions are attributed to a spin-2 field which, in the standard framework of General Relativity (GR), is massless and possesses nonlinear self-interactions. Even though the Standard Model and GR are experimentally and observationally well-tested, several phenomena still remain unexplained and motivate the study of theories beyond the standard picture. In particular, two of the biggest unresolved problems concern the presently unknown nature of dark matter and dark energy and, in order to account for these constituents in a satisfactory way, introducing new physics becomes unavoidable.
Additional degrees of freedom could be of the same type as the fields already present in the standard scenarios or, more interestingly, they could arise from heretofore unknown field theories. While the field theories for spin 0, 1/2 and 1 are well-understood, the treatment of higher spins turns out to be much more difficult. 
One of the simplest, or at least most natural, new ingredients that could be added to the known models is a massive spin-2 field whose presence is expected to mostly affect the gravitational sector. This may be desirable since modifying gravity is motivated by the fact that the Standard Model of particle physics is based on the very solid framework of a renormalisable quantum field theory, while a quantum theory of gravity does not yet exist and hence GR is not expected to be complete. Moreover, both the dark energy and the dark matter problems are intimately related to gravity but cannot be solved in the context of GR without raising additional questions. 

Interactions for massive spin-2 fields have long been thought to inevitably give rise to ghost instabilities and only recently a ghost-free theory for nonlinear interactions between massive and massless spin-2 fields has been found. Since the construction of nonlinear massive gravity and its extension to bimetric theory there has been significant progress in the field, both on the theoretical and the phenomenological side. Two long review articles on the subject have already been written by Hinterbichler~\cite{Hinterbichler:2011tt} and de Rham~\cite{deRham:2014zqa}. While these references focus on massive gravity, its gauge invariant formulation in terms of St\"uckelberg fields and its cosmology, this review is mostly dedicated to the manifestly covariant and dynamical bimetric theory. We focus on theoretical aspects and the structure of the theory (which is enforced by consistency) and mention its application to cosmology only as an aside. Moreover, all our considerations in this article will be at the classical level, although quantum corrections have been under preliminary investigations and are discussed in more detail in de Rham's review~\cite{deRham:2014zqa}, mostly in the context of massive gravity.

The following subsection summarises the interesting historical developments that led to the construction of consistent spin-2 interactions. It is no prerequisite for the subsequent discussions and the reader more interested in the details of massive spin-2 theories may directly proceed to section~\ref{sec:spin2fix}.

\subsection{Historical background}

Since massive gravity and bimetric theory have a rather long history starting with the work by Fierz and Pauli in 1939, it is impossible to give sufficient credit to all the groups that have contributed to the field over the last 75 years. We try here to collect the most important inputs towards the modern construction and pay attention to historical accuracy to the best of our knowledge.

\subsubsection{Early attempts}
The program of investigating massive spin-2 fields was initiated by Fierz and Pauli, who derived the unique classically consistent linearised theory of a free massive spin-2 field $h_{\mu\nu}$ propagating in Minkowski space-time \cite{Fierz:1939ix,Pauli:1939xp}. They demonstrated that the corresponding Lagrangian is of the form,
\begin{align}\label{FPorig}
	\Lag_{\mathrm{FP}} = \tfrac{1}{4}\big(-\p_\mu h_{\alpha\beta}\p^\mu h^{\alpha\beta}
	-2\p_\mu h\,\p_\nu h^{\mu\nu}
	+2\p_\mu h^{\mu\nu}\p^\alpha h_{\alpha\nu}
	&+\p_\mu h\p^\mu\, h\big)\nn\\
	&-\tfrac{\mfp^2}{4}\left(h_{\mu\nu}h^{\mu\nu}-h^2\right)\,,
\end{align}
where $h=\eta^{\mu\nu}h_{\mu\nu}$ and $\mfp$ is the mass parameter. The first line here corresponds exactly to the kinetic operator obtained by linearising the Einstein-Hilbert action of GR around flat space, while the second line encodes the quadratic non-derivative self-interactions of $h_{\mu\nu}$ that render it massive. The equations of motion derived from the above Lagrangian are equivalent to the system of equations,
\be
	\left(\square-\mfp^2\right)h_{\mu\nu}=0\,,\quad \p^\mu h_{\mu\nu}=0\,,\quad h=0\,.
\ee
The first of these is a massive wave equation, while the latter two are constraints on components of $h_{\mu\nu}$. In particular, Fierz and Pauli showed that if the relative coefficient ($-1$) of the two parts in the mass term of (\ref{FPorig}) is changed in any way, the on-shell condition of tracelessness is lost and a ghost-like scalar mode inside $h_{\mu\nu}$ becomes propagating. At the classical level a ghost is a field with negative kinetic energy which gives rise to an unbounded Hamiltonian and thus causes fatal instabilities; at the quantum level ghosts must be avoided in order to ensure unitarity. It is therefore crucial to work with the above Lagrangian with correct relative coefficient in the mass term.
In four space-time dimensions, it describes the on-shell propagation of a traceless, transverse and symmetric tensor field $h_{\mu\nu}$ with five massive degrees of freedom. This allows us to identify $h_{\mu\nu}$ with a massive spin-2 field with helicities $\pm2, \pm1, 0$. In all that follows, when using the terms mass and spin, we refer to their correspondence to degrees of freedom of relativistic field equations; we will rarely speak about the quantum nature of these concepts. In a similar fashion, we will frequently employ the particle perspective when dealing conceptually with nonlinear generalisations of the Fierz-Pauli Lagrangian, such as bimetric theory.

During the following years much of the efforts in field theory was directed elsewhere, developing a firm understanding of GR and exploring the booming realm of particle physics. Not much consideration was paid to Fierz and Pauli's theory of massive spin-2 until the early 1970's. We note however that related questions concerning bimetric theories were raised early on by Rosen \cite{Rosen:1940zz,Rosen:1975kk} and later by Aragone \& Deser in \cite{Aragone:1971kh} and by Isham, Salaam \& Strathdee in \cite{Isham:1971gm}. A major development came about when van Dam \& Veltman \cite{vanDam:1970vg} and Zakharov \cite{Zakharov:1970cc} (see also \cite{Iwasaki:1971uz}) independently investigated particular consequences of the Fierz-Pauli Lagrangian interpreted as a theory of a massive graviton. They realised that, in the presence of matter sources, the zero-mass limit of the theory is discontinuous, a property which is now referred to as the \textit{vDVZ discontinuity}. More precisely, this limit does not result in a theory for a single massless spin-2 field like linearised GR, but also contains a propagating scalar field which couples to the trace of the stress-energy source. The observational consequence would be an inferred difference in the bending of light around massive sources which was so severe that the theory would have been ruled out already at this time. Shortly thereafter, however, Vainshtein recognised a loophole in this reasoning~\cite{Vainshtein:1972sx}. He argued that, due to the presence of more scales in the theory when coupled to a source, a scalar mode becomes strongly coupled below some distance $r_V$ (the \textit{Vainshtein radius}). Hence the linearised analysis breaks down in this regime and a nonlinear completion of the theory is necessary in order to address any questions at distances within $r_V$ in a consistent manner. In particular these findings made it possible again to recover GR at short distances in the zero-mass limit. That this recovery can indeed be realised at the nonlinear level was demonstrated in~\cite{Babichev:2013pfa} for the case of ghost-free bimetric theory (for a recent review on the Vainshtein mechanism see~\cite{Babichev:2013usa}).

As an immediate response to Vainshtein's idea, Boulware and Deser studied the consistency of a wide class of possible nonlinear extensions of the massive Fierz-Pauli theory. They concluded that it was inevitable to introduce an extra propagating ghost-like scalar mode in any nonlinear extension of the theory \cite{Boulware:1973my}. In fact, this scalar mode, the Boulware-Deser ghost, is the same mode that at the linear level was removed by the trace constraint on~$h_{\mu\nu}$. The implication of this analysis was that no consistent nonlinear theory of massive spin-2 fields could exist. As we now know this strong conclusion was incorrect for two reasons: (1) Boulware and Deser did not consider the most general nonlinear extensions possible in their analysis. They considered only non-derivative self-interactions of $h_{\mu\nu}$ that were given through a general analytic function $f(h^2-h_{\mu\nu}^2)$, which naturally Taylor-expands to the Fierz-Pauli mass term at lowest order. As we will see later, in the consistent theory the self-interactions are contained in very specific scalar functions (the elementary symmetric polynomials) constructed out of the matrix argument $[\sqrt{\mathbb{1}+\eta^{-1}h}]^\mu_{\ph\lambda\nu}$. The correct field dependence of the interactions is therefore not of the form assumed in the proof by Boulware and Deser. (2) In their Hamiltonian analysis, they expected one specific equation (the equation of motion for the lapse variable of the metric) to provide the constraint that removes the ghost. The analysis of the consistent theory reveals that, in fact, it is a different equation (a rather contrived combination of lapse and shift equations) that gives the constraint. 

The conclusions of Boulware and Deser's analysis were so widely accepted that no further progress was made in the field for another 30 years and it would be almost 40 years until a consistent theory was fully developed. In hindsight this is unfortunate since, in the meantime, some interesting ideas did not receive the deserved attention due to the strong no-go theorem. For example, the correct structure of interactions for massive spin-2 fields was in fact partly suggested very early on by Zumino and also Chamseddine~\cite{Zumino:1970tu, Chamseddine:1978yu}, but without addressing the ghost problem (in fact it can be noted that~\cite{Zumino:1970tu} actually predated the no-go theorem). Similarly, the correct structure in the vielbein formulation was partly written down in \cite{Nibbelink:2006sz}, again failing to address the ghost issue. It has also recently been pointed out in~\cite{Pitts:2015aqa} that some attempts to construct a theory of nonlinear massive spin-2 fields were made by Maheshwari~\cite{Maheshwari:1972mb} using ideas from the works of Ogievetsky and Polubarinov \cite{OP,OPmassive2}. These constructions however went largely unnoticed and did not contribute towards the modern ideas and understanding of nonlinear massive gravity.

Another approach that has become popular in recent years and deserves mentioning is the attempt to construct consistent spin-2 mass terms that break Lorentz invariance~\cite{Rubakov:2004eb, Dubovsky:2004sg, Blas:2009my}. Based on the experience gained from the recent progress made in constructing the Lorentz invariant theories, similar progress has also been made in constructing Lorentz-breaking theories of massive gravity~\cite{Comelli:2013paa,Comelli:2013txa,Comelli:2014xga}. In this review we will however restrict ourselves to the class of theories that respects Lorentz invariance.

\subsubsection{A renewed interest}
After the very precise confirmation of the accelerated expansion of the universe in 1998 \cite{Riess:1998cb,Perlmutter:1998np}, a huge effort was devoted towards understanding better the underlying physics of this discovery. Within the context of standard GR this observed accelerated expansion of spacetime itself requires the addition of an additional source term with rather strange behaviour as compared to standard matter sources. The main-stream philosophy at this time was to hold firm in GR and not worry too much about the origin of such a source but simply collectively call this mysterious source ``dark energy". The simplest explanation for the acceleration is the presence of a constant source term in Einstein's equations. Adopting this view led to the celebrated $\Lambda$CDM (``$\Lambda$ Cold Dark Matter") concordance model of cosmology, which is in excellent agreement with observational data thus far~\cite{Adam:2015rua}. On the other hand, a growing community of theoretical physicists with a particle physics oriented mind was now realising the pressing nature of the cosmological constant problem~\cite{Weinberg:1988cp}: the small value of the observed acceleration does not fit in with expectations from a particle physics perspective, where a cosmological constant is naturally associated with vacuum energy. Unless additional symmetries (such as supersymmetry) are at work, a natural value for the vacuum energy scale is the mass of the heaviest field in the theory, which in any scenario is many orders of magnitude higher than the observed value for the cosmological constant.
\\
In addition to the poorly understood nature of dark energy, cosmologists seek to explain the presence of an unidentified matter component, commonly referred to as ``dark matter", which, in the context of GR, is required to account for the observational data at distances ranging from galactic to cosmological scales~\cite{Zwicky:1933gu, Rubin:1970zza, Adam:2015rua}. 

Since quantum field theory is such a rigorous framework and both the nature of dark matter and in particular dark energy seems so deeply connected with the large scale behaviour of gravity, which was really only tested within the solar system, many theorists started to look for an answer by modifying the gravitational sector of field theory.
\\
In the beginning much attention was given to extra dimensional setups which were mainly inspired by the additional dimensions arising in string theory and by the braneworld scenarios geared towards addressing the Higgs hierarchy problem of the Standard Model as well as supersymmetry breaking through anomaly mediation \cite{ArkaniHamed:1998rs,Antoniadis:1998ig,Randall:1998uk}. With respect to the cosmological constant problem, a particular interest was paid to models of brane induced gravity \cite{Dvali:2000hr,Dvali:2001gm,Rubakov:2008nh} and similar constructions (\eg cascading gravity \cite{deRham:2007rw,deRham:2007xp}). Brane induced gravity models, and in particular the codimension-one DGP model \cite{Dvali:2000rv,Dvali:2000hr} (for related investigations see~\cite{Gregory:2000jc}), were historically very important for a renewed interest in studying massive gravitons, since a generic feature of these models was the appearance of massive spin-2 resonance states on the brane that makes up our Universe in these models (see e.g.~\cite{Hassan:2010ys}). Due to the generic Yukawa suppression of the gravitational potential mediated by a massive field, a massive graviton also became interesting in itself for addressing the cosmological constant problem. The hope was that this exponential suppression could sufficiently weaken gravity at large distance scales to screen out a large vacuum energy coming from the matter sector and thus lead to a small effective cosmological constant. Even though this picture is correct in linear Fierz-Pauli theory, it eventually turned out that it would not work in the nonlinear theory without fine-tuning (as discussed in e.g.~\cite{Hassan:2011vm}).

In 2002 Arkani-Hamed, Georgi and Schwartz proposed a new perspective on studying effective theories with explicitly broken general covariance, in close analogy to symmetry breaking in spin-1 theories and the associated emergence of Goldstone modes \cite{ArkaniHamed:2002sp}. These ideas were quite general but in particular provided a new language for analysing the internal consistency of massive gravity. More explicitly, the setup was based on intuition from the Goldstone boson equivalence theorem \cite{Lee:1977eg,Chanowitz:1985hj}, which relates the physics of longitudinal modes of spin-1 gauge bosons to the physics of Goldstone modes at high energies. The authors of~\cite{ArkaniHamed:2002sp} suggested a similar correspondence in gravitational theories and that, in certain energy regimes, the complex problem of self-interacting spin-2 fields could be simplified to studying only their scalar longitudinal components.

In 2005 Creminelli, Nicolis, Papucci and Trincherini followed up on the ideas of Arkani-Hamed, Georgi and Schwartz (further inspired by the results of \cite{ArkaniHamed:2003uy,Rubakov:2004eb,Dubovsky:2004sg}) and attempted to explicitly construct a consistent theory of massive gravity using a bottom-up approach \cite{Creminelli:2005qk}.\footnote{At the same time Deffayet and Rombouts independently used the same formalism to study the ghost and its relation to the Vainshtein mechanism~\cite{Deffayet:2005ys}.} As a bitter irony of history (and humbling lesson in importance of rigor), despite a very beautiful analysis they reached the erroneous conclusion that the ghost problem of nonlinear massive gravity could not be resolved and found the Boulware-Deser ghost reappearing again. This result was based on an unfortunate sign mistake which arose from copying a basic equation of \cite{ArkaniHamed:2002sp} (for which the sign was not important). It is worth mentioning that without this sign mistake the discovery of a consistent theory of massive gravity could have been made already in 2005.

\subsubsection{Massive gravity rediscovered: dRGT theory}
In 2009 the issue of massive gravitons was further pursued by Gabadadze who modified GR by introducing an auxiliary extra dimension~\cite{Gabadadze:2009ja}. This work was shortly followed up by (and with) de Rham in \cite{deRham:2009rm,deRham:2010gu}. Their approach was based on an interesting field theoretical tool to generate a mass term for a vector (or scalar) field by imposing boundary conditions in an auxiliary extra dimension. Even though introducing an extra unphysical dimension and fixing boundary conditions by hand seems rather {\it ad hoc}, the procedure itself straightforwardly extends to the spin-2 case. The massive spin-2 model obtained in this way was demonstrated to be ghost-free to cubic order in a ``decoupling limit" analysis \cite{deRham:2010gu}. Although the same setup was subsequently shown to be inconsistent in a fully nonlinear analysis by Hassan and Rosen \cite{Hassan:2011zr}, it provided important inspiration that pushed developments further. The belief that the auxiliary-dimension model was consistent motivated de Rham and Gabadadze to revisit the ghost analysis of nonlinear massive gravity by Creminelli {\it et~al.}

In 2010 de Rham and Gabadadze studied generic extensions of the Fierz-Pauli Lagrangian~(\ref{FPorig}) by higher-order interactions of the massive spin-2 fluctuation $\hmn$~\cite{deRham:2010ik}. Their analysis went to quintic order in the longitudinal component of the massive spin-2 field and demonstrated that its interactions could in fact be made ghost-free in a decoupling limit, correcting the conclusions of \cite{Creminelli:2005qk}. The decoupling limit analysis relies heavily on the aforementioned Goldstone boson analogy suggested by Arkani-Hamed, Georgi and Schwartz and requires taking a double scaling limit in order to study the dynamics of the longitudinal mode separately. 
As a follow up to \cite{deRham:2010ik}, de Rham, Gabadadze and Tolley (henceforth dRGT) presented a nonlinear theory of massive gravity in whose decoupling limit they proved the absence of ghost for all nonlinear self-interactions of the longitudinal component~\cite{deRham:2010kj}. The dRGT action is of the form,
\be\label{SdRGT1}
	S_{\mathrm{dRGT}} = m_g^{2}\int\td^4x\sqrt{g}\left(R(g)
	+\frac{m^2}{2}\sum_{n=2}^4n!(4-n)!\alpha_ne_n(\mathcal{K})\right)\,,
\ee
where the first term is the ordinary Einstein-Hilbert term of GR with Planck mass $m_g$ and the second term is the interaction potential for the graviton whose mass is set by the scale $m$. Furthermore, $\alpha_2=1$ while the two remaining $\alpha_n$ are arbitrary interaction parameters. The $e_n(\mathcal{K})$ are the elementary symmetric polynomials (see appendix~\ref{app:technical}) constructed out of the matrix $\mathcal{K}^\mu_{\ph\mu\nu}=\delta^\mu_\nu-[\sqrt{g^{-1}\eta}]^\mu_{\ph\mu\nu}$.\footnote{It should be noted that Ref.~\cite{deRham:2010kj} did actually not present the action precisely in the above form. The interactions were resummed to give the square-root matrix structure, but it was not obvious from~\cite{deRham:2010kj} that the sum in~(\ref{SdRGT1}) would terminate at $n=4$. This fact was pointed out by Hassan \& Rosen in~\cite{Hassan:2011vm}.} Expanding the action in terms of $\hmn=m_g(\gmn-\eta_{\mu\nu})$ indeed results in a nonlinear extension of the Fierz-Pauli theory~(\ref{FPorig}).

The authors of \cite{deRham:2010ik, deRham:2010kj} also made the important observation of a loophole in the Hamiltonian analysis by Boulware and Deser~\cite{Boulware:1973my} which had long been believed to forbid any consistent theory of nonlinear massive spin-2 interactions.\footnote{In fact, to our knowledge, this caveat was first encountered in~\cite{Gabadadze:2009ja}.} 
By avoiding the ghost in the interactions of the longitudinal component, the dRGT model satisfied one of the necessary requirements on the complete theory to be consistent and consequently became the most promising candidate for a consistent theory of massive gravity. 
However, as noted, the proof of absence of the Boulware-Deser ghost in \cite{deRham:2010kj} was not made for the full theory since the analysis focused on self-interactions of one scalar component only. One reason for the difficulty of conclusively proving the consistency was the complexity of the nonlinear spin-2 interactions whose structure required better understanding before a full Hamiltonian analysis could be performed.

\subsubsection{Important generalisations}
The dRGT formulation of massive gravity was constructed with a perturbative expansion around a fixed flat background in mind. In 2011 Hassan and Rosen presented a reformulation of the dRGT action, which clarified the non-perturbative structure of the theory and identified consistent generalisations to more general backgrounds that were not apparent in the original formulation. 
Using basic properties of the elementary symmetric polynomials, they showed that the dRGT massive gravity action (\ref{SdRGT}) can be reformulated and then generalised to \cite{Hassan:2011vm},
\be\label{Shr}
	S_{\mathrm{HR}} = m_g^{2}\int\td^4x\sqrt{g}\left(R(g)
	-2m^2\sum_{n=0}^4\beta_ne_n\left(\sqrt{g^{-1}f}\right)\right)\,,
\ee
where $\beta_n$ are five arbitrary parameters whose role will be explained in detail later.
Most notably, the new formulation involves an arbitrary (but fixed) reference metric $\fmn$, which in dRGT is strictly taken to be $\eta_{\mu\nu}$.
It is worth to stress that even though the structures of \eqref{Shr} and \eqref{SdRGT1} are very similar, there is no obvious way of getting to \eqref{Shr} from the original dRGT formulation in \cite{deRham:2010kj} (whereas obtaining \eqref{SdRGT1} from \eqref{Shr} is straightforward) and thus the new formulation truly represents an important generalisation of the original massive gravity theory. In particular, the new formulation is indispensable for  addressing any question about massive spin-2 interactions on a curved background. Even more importantly, as we will see, it suggests how to arrive at a fully dynamical theory of interacting spin-2 fields.

Going beyond the decoupling limit analysis of \cite{deRham:2010kj}, Hassan and Rosen quickly utilised the formulation (\ref{Shr}) to give a fully nonlinear consistency proof for dRGT massive gravity with flat reference metric~\cite{Hassan:2011hr}: In a Hamiltonian analysis based on the ADM formalism \cite{Arnowitt:1962hi}, they showed that the complete nonlinear theory gives rise to a constraint that removes the Boulware-Deser ghost. 
Shortly after this, the proof was generalised to the case of an arbitrary reference metric $\fmn$ in \cite{Hassan:2011tf}. These consistency proofs (see also~\cite{Hassan:2011ea, Hassan:2012qv}~and~\cite{Golovnev:2011aa, Kluson:2012wf, Kluson:2012zz, Alexandrov:2012yv, Alexandrov:2013rxa, Kugo:2014hja,Comelli:2013txa} for subsequent confirmations) for the massive gravity action (\ref{Shr}) essentially completed the program of finding a Lorentz-invariant theory for a massive spin-2 field initiated by Fierz and Pauli in 1939.

\subsubsection{Hassan-Rosen bimetric theory}
A very important outcome of the generalised investigations of Hassan and Rosen laid in obtaining an extension of the massive gravity theory in which the reference metric $\fmn$ receives its own dynamics. As a consequence, the two metrics $\gmn$ and $\fmn$ are treated on the same footing in this bimetric theory and all fields in the action are determined dynamically. Bimetric theories of gravity have been subject of earlier investigations in, for instance,~\cite{Rosen:1975kk,Isham:1971gm,Aragone:1971kh,Chamseddine:1978yu,Damour:2002ws,Damour:2002wu,Blas:2005yk,Berezhiani:2009kv,Berezhiani:2009kx,Deffayet:2011uk} but, just as nonlinear massive gravity, they generically suffer from the Boulware-Deser ghost instability.

Shortly after the nonlinear theory for massive gravity had been developed, the unique ghost-free bimetric theory was presented by Hassan and Rosen in~\cite{Hassan:2011zd}. Its form is reminiscent of the massive gravity action,
\begin{align}\label{Sbm}
	S_\mathrm{HR} =\, m_g^2\int\td^4x\sqrt{g}\,R(g)&+m_f^2\int\td^4x\sqrt{f}\,R(f)\nn\\
	&-2m^4\int\td^4x\sqrt{g}\sum_{n=0}^4\beta_ne_n\left(\sqrt{g^{-1}f}\right)\,,
\end{align}
but it includes an Einstein-Hilbert term of $\fmn$ in which $m_f$ is the ``Planck mass" for the second metric. Most importantly, bimetric theory provides a covariant formulation for massive spin-2 fields, in which now both metrics are dynamical and the structure of the action is in fact symmetric with respect to the interchange of $\gmn$ and $\fmn$.
The consistency of bimetric theory was first demonstrated in~\cite{Hassan:2011zd, Hassan:2011ea} and, as we shall see later, this established the first theory describing consistent nonlinear interactions of massive spin-2 fields with massless ones.

\subsection{Outline of the review}

We have chosen the following structure for this review: Our starting point will be the linear theories for massless and massive spin-2 fields in flat background in section~\ref{sec:spin2fix}. The possibility to generalise the massless theory to arbitrary backgrounds by constructing the nonlinear theory of general relativity motivates us to look for a nonlinear completion of the mass term. In section~\ref{sec:nonlin}, after reviewing the ghost problem, we pursue this goal by introducing a suitable set of ADM variables and provide a constructive proof for the consistent spin-2 interaction potential. Section~\ref{sec:gfnlmg} summarises the main features of ghost-free nonlinear massive gravity with general and flat reference metric, in both gauge fixed and in gauge invariant form. We then move on to the fully dynamical bimetric theory in section~\ref{sec:HRbim} where we write down its mass spectrum and also introduce couplings to matter. Classical solutions to the bimetric equations are discussed in section~\ref{sec:classsol}, first in general and then with a focus on black hole and cosmological solutions. In section~\ref{sec:PM} we review the phenomenon of partially massless spin-2 fields on de Sitter space and present the idea of realising partial masslessness at the nonlinear level. In this context, we also reveal a connection between bimetric theory and certain higher-derivative theories of gravity. Possible extensions of bimetric theory, including new kinetic terms in higher dimensions as well as multiple spin-2 interactions, are discussed in section~\ref{sec:extbim}. Finally, we conclude this review article with a list of several open questions in section~\ref{sec:discuss}.

\newpage
\section{Spin-2 Fields on Fixed Backgrounds}\label{sec:spin2fix}

Having outlined the history of massive gravity in the introduction, we now turn to the detailed description of spin-2 field theories. 
In this section we first discuss the linear theory for a massless spin-2 field in a flat background and see how the correct number of physical degrees of freedom emerges due to gauge invariance. The nonlinear completion of the corresponding Lagrangian is the Einstein-Hilbert action of GR from which one can derive the linear theory around general curved backgrounds. Thereafter we review the linear theory of massive spin-2 in flat space, where the presence of constraints lead to the correct number of propagating modes. The quest for a generalisation to curved backgrounds again requires a nonlinear completion of the spin-2 mass term and paves the way towards nonlinear massive gravity.

\subsection{Massless spin-2 field}

\subsubsection{Flat space}\label{sec:FPflat}

The Lagrangian for a massless spin-2 field $\hmn$ to quadratic order in the field in flat space with Minkowski metric $\emn$ reads,
\beqn\label{masslesslin1}
\mathcal{L}_\mathrm{lin}=\tfrac1{2}h^{\mu\nu}\mathcal{E}_{\mu\nu}^{\phantom{\mu\nu}\rho\sigma}h_{\rho\sigma}\,,
\eeqn
where the structure of the kinetic terms is captured by the two-derivative operator,
\begin{align}\label{kinstruc1}
	\mathcal{E}_{\mu\nu}^{\phantom\mu\phantom\nu\rho\sigma} = \tfrac{1}{2}\Big(
	\eta_\mu^{\phantom\mu\rho}\eta_\nu^{\phantom\nu\sigma}\partial^2
	-\eta_\nu^{\phantom\nu\sigma}\partial_\mu\partial^\rho
	-\eta_\mu^{\phantom\mu\sigma}\partial_\nu\partial^\rho
	+\eta_{\mu\nu} \partial^\sigma\partial^\rho
	+\eta^{\rho\sigma}\partial_\mu\partial_\nu
	-\eta_{\mu\nu}\eta^{\rho\sigma}\partial^2\Big)
	\,. 
\end{align}
The corresponding equations of motion are, 
\beqn\label{massllineq}
\mathcal{E}_{\mu\nu}^{\phantom{\mu\nu}\rho\sigma}h_{\rho\sigma}=0\,.
\eeqn
As can easily be verified, the Lagrangian and equations for the massless spin-2 particle are invariant under the following gauge transformations,
\beqn\label{lindiff}
\hmn(x)\longmapsto \hmn(x) + \partial_\mu\xi_\nu(x)+\partial_\nu\xi_\mu(x)\,,
\eeqn
with vector gauge parameter $\xi_\mu(x)$. An equivalent manifestation of this gauge invariance is the linearised Bianchi identity: the divergence of the left-hand side of the equations of motion~(\ref{massllineq}) is identically zero, i.e.~$\partial^\mu\mathcal{E}_{\mu\nu}^{\phantom{\mu\nu}\rho\sigma}h_{\rho\sigma}=0$.

As should be familiar from the spin-1 case (Maxwell's theory), we can use the symmetry transformation \eqref{lindiff} to pick a convenient gauge. The de Donder gauge condition for the spin-2 field is $\partial^\mu\hmn-\frac{1}{2}\partial_\nu h=0$. This is the analogue of the Lorenz gauge for massless vectors and constrains four of the ten components in $h_{\mu\nu}$. In this gauge the equations of motion assume the simple form,
\beqn
\Box\hmn=0\,.
\eeqn
After fixing the de Donder gauge there still exist residual transformations, namely those with $\Box\xi_\mu=0$, that leave the de Donder gauge intact. Since the four residual gauge parameters satisfy the same equation as the field, they can directly be invoked to remove further redundant degrees of freedom.
In total, gauge invariance therefore removes eight of the ten components in the symmetric field $\hmn$, and the only propagating modes are the two degrees of freedom of a massless spin-2 particle.

It is also possible to couple the massless spin-2 field to other fields by introducing an external source $T_{\mu\nu}$ into the Lagrangian,
\beqn\label{masslesslin}
\mathcal{L}_\mathrm{lin}=\tfrac{1}{2}\Big(h^{\mu\nu}\mathcal{E}_{\mu\nu}^{\phantom{\mu\nu}\rho\sigma}h_{\rho\sigma}-\kappa h^{\mu\nu} T_{\mu\nu}\Big)\,,
\eeqn
where $\kappa$ is a coupling constant of mass dimension minus one (here $h_{\mu\nu}$ is taken to have mass dimension one). The linearised Bianchi identity now implies $\partial^\mu T_{\mu\nu}=0$, i.e.~conservation of the source.

\subsubsection{Curved space}

The graviton in GR is a massless spin-2 particle and should hence be described by the Lagrangian~(\ref{masslesslin}). On the other hand, the equivalence principle tells us that the field should couple to all kinds of energy in the same manner, including its own stress-energy tensor. Implementing this requirement in~(\ref{masslesslin}) iteratively introduces nonlinearities in $\hmn$ and eventually leads to the Einstein-Hilbert action for GR (see~e.g.~\cite{Butcher:2009ta,Deser:2009fq}),
\beqn\label{EHGR}
S_{\mathrm{GR}}=M_{\mathrm{Pl}}^{2}\int \mathrm{d}^4x\,\sqrt{-g}~R ~+ \int \mathrm{d}^4x\,\sqrt{-g}~\mathcal{L}_\mathrm{m} \,,
\eeqn
The corresponding equations of motion are Einstein's equations,
\beqn\label{EE}
R_{\mu\nu}-\frac{1}{2}\gmn R =\frac{1}{M_{\mathrm{Pl}}^{2}}T_{\mu\nu}\,,
\eeqn
where $R_{\mu\nu}$ is the Ricci tensor with trace $R$ and the stress-energy tensor is derived from the matter Lagrangian~$\mathcal{L}_\mathrm{m}$ as,
\beqn
T_{\mu\nu}\equiv-\frac{1}{\sqrt{-g}}\frac{\partial(\sqrt{-g}\,\mathcal{L}_\mathrm{m})}{\partial g^{\mu\nu}}\,.
\eeqn
The Einstein-Hilbert action is the gauge invariant nonlinear extension of~(\ref{masslesslin}). It can be defined uniquely, as the field theory that describes nonlinear self-interactions of a massless spin-2 particle. The gauge transformations~(\ref{lindiff}) are linearised general coordinate transformations (GCTs), under which the metric at the nonlinear level transforms as,
\beqn
\gmn\longmapsto\gmn+\nabla_\mu\xi_\nu+\nabla_\nu\xi_\mu\,.
\eeqn
The presence of a gauge symmetry at the nonlinear level ensures that the full theory propagates the same number of degrees of freedom as its linear version.

The Einstein equations~(\ref{EE}) allow for flat space solutions, $\gmn=\emn$, in the case of vanishing matter source. Linearising the action around this background in the perturbation $\hmn=\kappa^{-1}(\gmn-\emn)$ results in the linear theory for a massless spin-2 field~(\ref{masslesslin}). 
\\
On the other hand, a non-vanishing background value of the source will give rise to a curved background metric $\bar{g}_{\mu\nu}$ and around this solution the action can be linearised as,
\beqn\label{lincurv}
\mathcal{L}'_\mathrm{lin}=\tfrac{1}{2}\Big(h^{\mu\nu}\bar{\mathcal{E}}_{\mu\nu}^{\phantom{\mu\nu}\rho\sigma}h_{\rho\sigma}
-\kappa h^{\mu\nu} \delta T_{\mu\nu}\Big)\,,
\eeqn
where now the linearised Einstein operator takes the covariant form,
\begin{align}\label{kincurv}
	\bar{\mathcal{E}}_{\mu\nu}^{\phantom\mu\phantom\nu\rho\sigma} = \tfrac{1}{2}\Big(
	\delta^\rho_{~\mu}\delta^\sigma_{~\nu}\bar\nabla^2
	&-\delta_{~\nu}^{\sigma}\bar\nabla_\mu\bar\nabla^\rho
	-\delta_{~\mu}^{\sigma}\bar\nabla_\nu\bar\nabla^\rho
	+\bar{g}_{\mu\nu} \bar\nabla^\sigma\bar\nabla^\rho\nn\\
	&
	+\bar{g}^{\rho\sigma}\bar\nabla_\mu\bar\nabla_\nu
	-\bar{g}_{\mu\nu}\bar{g}^{\rho\sigma}\bar\nabla^2
	-\bar{g}_{\mu\nu}\bar{R}^{\rho\sigma}
	+\delta^\rho_{~\mu}\delta^\sigma_{~\nu}\bar{R}
	\Big)
	\,, 
\end{align}
where $\bar\nabla_\mu$ and $\bar{R}$ are defined with respect to $\bar g_{\mu\nu}$. At this stage let us make an important remark on covariantisation of a theory whose form is known in flat space. Suppose we were given only the flat space Lagrangian~(\ref{masslesslin}) and asked to derive its covariant generalisation. Na\"ively, one would replace all partial derivatives $\partial_\mu$ by covariant ones $\nabla_\mu$ and all occurrences of the Minkowski metric $\emn$ by the more general background $\bar{g}_{\mu\nu}$. This procedure works for lower spin-particles coupled to gravity and results in well-known formulations of spin-0, spin-1/2 and spin-1 theories in curved space. For the spin-2 field, however, the procedure fails because na\"ive covariantisation of~(\ref{masslesslin}) does not result in the consistent form~(\ref{kincurv}) obtained from linearising GR. In particular, ambiguities arise because the covariant derivatives do not commute and since there is no obvious guideline telling us which curvature terms to include in the linearised Einstein operator. On the other hand, knowledge of the nonlinear theory~(\ref{EHGR}) allows us to straightforwardly arrive at the correct Lagrangian for a massless spin-2 field in curved background.

\subsection{Massive spin-2 field in flat space}\label{sec:linmflatsp}

As has been known since the work of Fierz and Pauli from 1939~\cite{Fierz:1939ix}, the quadratic Lagrangian for a massive spin-2 excitation in flat space has the form,
\beqn\label{FPlag}
\mathcal{L}_\mathrm{FP}=\frac{1}{2}\Big(h^{\mu\nu}\mathcal{E}_{\mu\nu}^{\phantom{\mu\nu}\rho\sigma}h_{\rho\sigma}-\frac{m_\mathrm{FP}^2}{2}(h^{\mu\nu}\hmn-h^2)-\kappa h^{\mu\nu} T_{\mu\nu}\Big)\,,
\eeqn
where the kinetic operator is the same as in the massless case, given by (\ref{kinstruc1}), and we have included a source term~$T_{\mu\nu}$.
The equations of motion obtained from the Lagrangian for the massive particle are
\beqn\label{FPeqs}
\mathcal{E}_{\mu\nu}^{\phantom{\mu\nu}\rho\sigma}h_{\rho\sigma}-\frac{m_\mathrm{FP}^2}{2}(\hmn-\emn h)=\frac{\kappa}{2} T_{\mu\nu}\,.
\eeqn
The mass term breaks the gauge invariance of the massless theory but, as a consequence of the linearised Bianchi identity, the divergence and trace of these equations give rise to five constraints,
\begin{subequations}\label{twoconstr}
\beqn
\partial^\mu \hmn-\partial_\nu h&=&-\frac{\kappa}{m_\mathrm{FP}^2}\partial^\mu T_{\mu\nu}\,,\\
h&=&\frac{\kappa}{3m_\mathrm{FP}^2}T+\frac{2\kappa}{3m_\mathrm{FP}^4}\partial^\nu\partial^\mu T_{\mu\nu}\,.\label{trconstr}
\eeqn
\end{subequations}
The source is not necessarily conserved but, for simplicity, we shall anyway assume $\partial_\mu T^{\mu\nu}=0$. This assumption certainly holds for any source that is derived from a diffeomorphism invariant matter coupling.
For vanishing sources, the constraint equations imply that the massive spin-2 field is transverse and traceless. 

Using (\ref{twoconstr}) in the equations of motion, we can rewrite them as,
\beqn
(\Box-m_\mathrm{FP}^2)\hmn=\kappa\left(T_{\mu\nu}-\frac{1}{3}\left[\eta_{\mu\nu}-\frac{1}{m_\mathrm{FP}^2}\partial_\mu\partial_\nu T\right]\right) \,,
\eeqn
and see that the massive spin-2 field satisfies a sourced Klein-Gordon equation.
The constraints (\ref{twoconstr}) remove five of the ten components in $\hmn$, leaving us with the five propagating degrees of freedom of a massive spin-2 particle. 

An important observation made by Fierz and Pauli is that modifying the structure of the mass term, i.e.~changing the numerical factor in front of $h^2$ in (\ref{FPlag}), introduces an additional degree of freedom into the theory. This happens because the trace constraint (\ref{trconstr}) is lost and $h$ satisfies a dynamical equation of motion instead. It can furthermore be shown that the propagator corresponding to the extra degree of freedom comes with a residue of the wrong sign and therefore gives rise to a ghost instability~\cite{Boulware:1973my}.\footnote{See \cite{VanNieuwenhuizen:1973fi} for a detailed analysis of the propagator in Fierz-Pauli theory and its deformed version containing the ghost.} This unwanted dynamical field is referred to as the \textit{Boulware-Deser ghost}. It is this ghost mode that complicates the construction of a nonlinear interaction potential for a spin-2 field because, even when banned from the linear theory, the ghost notoriously reappears through the higher-order interactions.

On the other hand, tuning the linear mass potential to the Fierz-Pauli structure comes with its own problems that threaten the phenomenological viability of the theory as a modification of GR. Namely, as we already mentioned in the introduction, it was shown by van Dam, Veltman and Zakharov (vDVZ) \cite{vanDam:1970vg, Zakharov:1970cc} that the $m_\mathrm{FP}\rightarrow 0$ limit of linear massive gravity does not continuously approach linearised GR.
Nonlinear self-interactions for the spin-2 field may be able to cure this problem if they exhibit the Vainshtein mechanism~\cite{Vainshtein:1972sx}. Historically, this was one of the main motivations for the construction of a nonlinear theory of massive gravity.
Another reason to search for a completion of the Fierz-Pauli mass term is the existence of the nonlinear closed form~(\ref{EHGR}) in the massless theory.

Since consistency of linearised massive gravity in flat space requires tuning a coefficient in the mass potential, one can expect that a consistent (i.e.~ghost-free) nonlinear potential cannot contain arbitrary interaction terms, but that the coefficients of certain terms will be related to each other by demanding the absence of the extra degree of freedom.
If the fully nonlinear theory for massive spin-2 was known, it could be linearised around general backgrounds to give the covariantised version of~(\ref{FPlag}). However, as in the massless case, it would be a very difficult task to derive the linear theory on arbitrary backgrounds by covariantising the flat-space Lagrangian. As we will see later in section~\ref{sec:linarb}, the most general linear theory for massive spin-2 can also be derived from a nonlinear action and has a rather complicated form.

\newpage
\section{Towards Nonlinear Spin-2 Interactions}\label{sec:nonlin}

In the previous chapter we considered massless and massive theories for a linear fluctuation $\hmn$ in a fixed Minkowski background $\emn$. In the nonlinear massless theory given by the Einstein-Hilbert action of GR, it was possible to combine the background and the fluctuation into a single nonlinear field $\gmn$. We would now like to do something similar in the massive case and construct a nonlinear self-interaction term without derivatives for the metric.

\subsection{General structure}
Historically it was often assumed that any nonlinear theory for massive gravity must give rise to flat space solutions and a Fierz-Pauli mass term in the linear theory. On the other hand, it is well known that a consistent mass term may also be written down on maximally symmetric backgrounds\footnote{Interestingly, the vDVZ discontinuity turns out to be absent on de-Sitter backgrounds~\cite{Porrati:2000cp}.} \cite{Higuchi:1986py,Higuchi:1989gz,Porrati:2000cp,Kogan:2000uy,Deser:2001pe,Deser:2001us,Deser:2001wx}, or even more generally on homogeneous and isotropic backgrounds \cite{Grisa:2009yy,Berkhahn:2010hc,Berkhahn:2011hb}. This possibility motivates us to consider a general nonlinear theory without reference to flat Minkowski solutions and only demand that the correct number of degrees of freedom propagate at the nonlinear level (or, equivalently, around any background). Of course, when restricting to certain backgrounds the linearised version of the nonlinear theory must also reduce to the correct known structure.

What we call a mass term for a rank-2 tensor $\gmn$ must be a scalar density, i.e.~it has to be a nontrivial scalar function $V(g)$ multiplied by the scalar density $\sqrt{g}$. Obviously, the scalar function $V(g)$ cannot have any loose covariant indices and, by definition, it should not contain any derivatives. But then, the only object at hand to contract the indices of the metric tensor $\gmn$ is the metric itself which, since $g^{\rho\mu}\gmn=\delta^\rho_{~\nu}$, leads to a trivial cosmological constant contribution in the action. We conclude that there is no possibility to construct a covariant nonlinear interaction term for a spin-2 field using only one tensor field.\footnote{Note the difference to e.g.~the vector example, where the metric $\gmn$ can be used to contract the indices in nonlinear interaction terms for $A_\mu$ in a nontrivial way. Furthermore, in the massless spin-2 theory this problem did not occur because in the kinetic terms the indices of the metric could be contracted with derivative operators.} 
Hence, we are forced to introduce another field in order to build nonlinear contractions with $\gmn$. In principle, this could be any object with sufficient amount of indices, but the minimal choice is to work with a second rank-2 tensor which we shall call $\fmn$. The interaction potential will then be given by $\sqrt{g}$ multiplying a scalar function of $g^{\rho\mu}\fmn$. Note that, due to the existence of an additional tensor $\fmn$, we could in principle consider densitising by using for example $\sqrt{f}$ or $g^{1/4}f^{1/4}$. Since it is always possible to factor out $\sqrt{g}$ (for instance, by writing $\sqrt{f}=\sqrt{g}~\det\sqrt{g^{-1}f}$ and absorbing the second factor into the potential) we may use that as the scalar density without any loss of generality.

In summary, we expect the nonlinear massive gravity action to be of the form,
\beqn\label{mggenact}
S_\mathrm{MG}=m_g^{2}\int \mathrm{d}^4x~\sqrt{g}~\left[R(g) - m^2\, V(g^{-1}f)\right]\,,
\eeqn
where $m$ is an arbitrary energy scale that sets the mass of the spin-2 field. In this setup, the second ``metric" $\fmn$ is a fixed background field that needs to be put into the theory by hand. From the viewpoint of field theory this is somewhat unusual because $\fmn$ is not determined by an equation of motion. In fact, there is no need to worry about this, since we will resolve this slightly confusing point later, when we introduce the fully dynamical bimetric theory that treats $\gmn$ and $\fmn$ on the same footing. For now, let us accept the possibility to work with a fixed reference metric $\fmn$ and investigate the consistency of this class of theories. 
\\
A simple example for a possible interaction term would be $V(g^{-1}f)=\mathrm{Tr}(g^{-1}f)$. Interestingly, the corresponding action is closely related to another modification of gravity that goes under the name Eddington-inspired Born-Infeld theory~\cite{Banados:2008fi}. Unfortunately, having only this term in the action gives rise to the Boulware-Deser ghost and is thus not a viable choice.

A necessary requirement on the interaction potential in (\ref{mggenact}) is that it reduces to the Fierz-Pauli mass term for parameter choices that permit a linearisation around flat space (i.e.~when the reference metric $\fmn$ is flat and when flat background solutions for $\gmn$ exist), otherwise it will certainly propagate the Boulware-Deser ghost. But this requirement alone is not sufficient for consistency: A generic nonlinear interaction potential, even if it incorporates the Fierz-Pauli structure around flat space, will reintroduce the extra degree of freedom which leads to instabilities at the nonlinear level. This is anticipated from a simple degree-of-freedom counting in the full theory:
The introduction of the interaction potential breaks the diffeomorphism invariance of GR. Therefore the four gauge symmetries are lost and the theory generically will propagate four additional degrees of freedom. Since $2+4=6$, this does not give the correct number for a massive spin-2 particle, but there is an extra degree of freedom in the theory. This is the nonlinear Boulware-Deser ghost and a constraint is needed to remove it from the spectrum of propagating modes. Tuning the linear mass term to the Fierz-Pauli combination ensures the presence of this constraint only in the linear theory.
\\
We will see that, in order to obtain an additional constraint and thereby ensure the absence of the ghost beyond the linear level, it is necessary to impose strong restrictions on the structure of interactions. In fact, demanding the presence of a constraint will fix all but three coefficients of all possible interaction terms.

\subsection{The Boulware-Deser ghost}

In the construction of the consistent theory, we will follow the approach of Boulware and Deser who studied massive gravity in the Hamiltonian formulation~\cite{Boulware:1972zf, Boulware:1973my}. They claimed that, even though it is possible to remove the ghost from the spectrum of propagating modes at the linear level, it will return for any nonlinear interaction terms that are added to the Lagrangian (\ref{FPlag}). In order to show that this result is in fact incorrect, we first need to familiarise ourselves with variables suitable for the Hamiltonian formulation of GR.

\subsubsection{ADM variables for general relativity}

The Hamiltonian formulation of GR traces back to the work by Arnowitt, Deser, and Misner (ADM) from 1962~\cite{Arnowitt:1962hi}. The authors decomposed the metric $\gmn$ into a scalar $N$ (lapse), a three-dimensional metric $\gamma_{ij}$, and a three-component vector $N_i$ (shift) as follows: 
\beqn\label{admmetr}
\gmn=\begin{pmatrix}
-N^2+N_i \gamma^{ij}N_j& N_j\\
N_i & \gamma_{ij}
\end{pmatrix}\,,
\eeqn
where $\gamma^{ij}$ denotes the inverse of $\gamma_{ij}$. This parametrisation essentially splits the metric into its time ($0\mu$) and spatial ($ij$) components. From (\ref{admmetr}) we can also compute the inverse of the metric,
\beqn\label{invadm}
g^{\mu\nu}=\frac{1}{N^2}\begin{pmatrix}
-1& N^j\\
N^i & N^2\gamma^{ij}-N^iN^j
\end{pmatrix}\,.
\eeqn
Here and in the following we raise the indices on the shift vector $N_i$ using the inverse spatial metric~$\gamma^{ij}$.
\\
It turns out that, in any theory with kinetic structure given by the Einstein-Hilbert term, the lapse $N$ and shift $N_i$ are non-dynamical gauge degrees of freedom because the Ricci scalar of $\gmn$ contains no derivatives on those fields. Hence, all of the propagating modes are contained in the spatial metric $\gamma_{ij}$ which has six independent components. The gauge invariance of GR further reduces the number of propagating degrees of freedom to two, along with the same number of corresponding canonical momenta.

More explicitly, in terms of the ADM variables the Einstein-Hilbert action (\ref{EHGR}) in vacuum becomes (up to a boundary term which, although crucial for certain applications, is not important for our considerations) \cite{Arnowitt:1962hi}
\beqn\label{gradm}
S_{\mathrm{GR}}
=M_\mathrm{P}^{2}\int \mathrm{d}^4x~\left(\pi^{ij}\partial_t \gamma_{ij}-NR^0-N_iR^i\right)\,,
\eeqn
where, in terms of the curvature scalar $R_{(3)}$ of the metric $\gamma_{ij}$,
\beqn
R^0&=& -\sqrt{\gamma}~ \left[R_{(3)} +\gamma^{-1}\left(\frac{1}{2}({\pi^i}_i)^2-\pi^{ij}\pi_{ij}\right)\right] \,,\nonumber\\
R^i&=& -2\partial_j\pi^{ij}.
\eeqn
The conjugate momenta $\pi^{ij}$ of the six metric components $\gamma_{ij}$ are computed from the GR Lagrangian in the standard way which leads to expressions in terms of derivatives of the four-dimensional metric.
\\
It is now evident that the action (\ref{gradm}) does not contain dynamical terms for the scalar $N$ nor for the vector $N^i$. On top of that, these variables appear only linearly and therefore act as Lagrange multipliers whose equations of motion do not contain $N$ and $N^i$ themselves. This implies that these equations in fact correspond to four constraints~$R^\mu=0$, with~$R^\mu=(R^0, R^i)$, on the remaining twelve variables $\gamma_{ij}$ and $\pi^{ij}$.

According to the theory of constrained Hamiltonian systems,\footnote{For a review of this subject see~e.g.~\cite{Date:2010xr}.} the presence of a gauge symmetry can now be seen in the Poisson algebra of the constraints, $\{R^\mu, R^\nu\}$. Here, the Poisson bracket for functions $A,B$ is defined as
\beqn\label{defpoisb}
\{A(x), B(y)\}\equiv\int \mathrm{d}^3z~\left(  \frac{\delta A(x)}{\delta\gamma_{ij}(z)}\frac{\delta B(y)}{\delta \pi^{ij}(z)} -\frac{\delta A(y)}{\delta \pi^{ij}(z)}\frac{\delta B(x)}{\delta\gamma_{ij}(z)}\right)\,.
\eeqn
The result for the constraint algebra of GR reads~\cite{DeWitt:1967yk, Hojman:1976vp},
\beqn
\{R^0(x), R^0(y)\}&=& R^i(y)\frac{\partial}{\partial y^i}\delta^3(x-y)-R^i(x)\frac{\partial}{\partial x^i}\delta^3(x-y)\, , \nonumber\\
\{R_i(x), R_j(y)\}&=& R_i(y)\frac{\partial}{\partial y^j}\delta^3(x-y)-R_j(x)\frac{\partial}{\partial x^i}\delta^3(x-y)\,,\nonumber\\
\{R^0(x), R_i(y)\}&=& -R^0(y)\frac{\partial}{\partial x^i}\delta^3(x-y)\,.
\eeqn
Since these brackets are proportional to the constraints themselves, they vanish on the constraint surface, as they should in the presence of a gauge symmetry. In the language of Dirac, non-trivial gauge symmetries generate first class constraints. All four primary constraints will serve to put conditions on the remaining variables $\gamma_{ij}$ and $\pi^{ij}$, while the Lagrange multipliers $N$ and $N^i$ remain undetermined at this stage. Since the Hamiltonian $H$ is itself a linear combination of $R$ and $R^i$, all constraints are automatically preserved in time since the time-evolution of any quantity is determined by $\tfrac{\td}{\td t} A(x)=\{A(x),H\}$.
\\
After imposing the four conditions on $\gamma_{ij}$ and $\pi^{ij}$, we can still use gauge transformations to remove another four degrees of freedom whose equations of motion will eventually determine $N$ and $N^i$.
In total we therefore end up with four dynamical variables, corresponding to the two helicity states ($\pm2$) of the massless graviton and their canonical momenta. This shows that, even nonlinearly, GR propagates the correct number of degrees of freedom for describing a massless spin-2 particle.

Note that throughout the above discussion we never actually wrote down the Hamiltonian, but remained in the Lagrangian formulation. In order to investigate the positivity of the energy\footnote{The positivity of the Hamiltonian has not been proven in the bimetric theory or in massive gravity in general, but has been studied for spherically symmetric configurations in~\cite{Volkov:2014ida}.} this is naturally insufficient, but since here we were only interested in counting degrees of freedom there was no need to work directly with the Hamiltonian, which can trivially be obtained from (\ref{gradm}).

\subsubsection{The no-go theorem}

The ADM variables turn out to be very useful for investigating the consistency of massive gravity. We saw that at the linearised level the only consistent mass term is the one proposed by Fierz and Pauli given in (\ref{FPlag}).
In order to demonstrate the presence of the ghost instability in the ADM formalism, in their paper \cite{Boulware:1973my} from 1972, Boulware and Deser studied the more general ``mass term" with arbitrary coefficient~$a$,
\beqn
\mathcal{L}_{\mathrm{BD}}=\frac{1}{2}\Big(h^{\mu\nu}\mathcal{E}_{\mu\nu}^{\phantom{\mu\nu}\rho\sigma}h_{\rho\sigma}-\frac{m^2}{2}(h^{\mu\nu}\hmn-ah^2)\Big)\,.
\eeqn
The trace constraint $h=0$ which removes the ghost mode does not exist for $a\neq 1$.
Another way of seeing the additional mode is by noticing that, precisely for the Fierz-Pauli choice $a=1$, the mass term is linear in the component $h_{00}$, which furthermore appears without time derivatives in the linearised Einstein operator. Thus, for $a=1$, the equation of motion for $h_{00}$ is a constraint which removes one dynamical variable. For other values of $a$, the equation depends on $h_{00}$ itself in which case it does not constrain other components. As a direct consequence, a sixth degree of freedom is propagating.

We can make this more explicit using the variables introduced for the Hamiltonian analysis of GR in the previous subsection. From the ADM decomposition \eqref{admmetr} we read off the decomposition of the fluctuation,
\beqn
\hmn=\gmn-\emn=\begin{pmatrix}
1-N^2+N_i \gamma^{ij}N_j~~&~~ N_j\\
N_i ~~&~~ h_{ij}
\end{pmatrix}\,,
\eeqn
where $h_{ij}=\gamma_{ij}-\delta_{ij}$. Inserting this into the mass terms one finds,
\beqn\label{admfpfp}
\frac{m^2}{4}\big(h^{\mu\nu}\hmn-ah^2\big)=-\frac{m^2}{4}\Big(h^{ij}h_{ij}&-&a({h^i}_i)^2-2N_iN_i+2a(1-N^2+N_i\gamma^{ij}N_j){h^i}_i \nonumber \\
&+&(1-a)(1-N^2+N_i\gamma^{ij}N_j)^2\Big)\,. 
\eeqn
The last term gives rise to nonlinearities in $h_{00}=1-N^2+N_i\gamma^{ij}N_j$ and vanishes only for $a=1$. 
Moreover, when $\hmn = \gmn-\emn$ is viewed as a small perturbation of $\emn$, its ADM variables can be written as small fluctuations as well,
\beqn
\delta N=N-1\,, \quad \delta N_i= N_i\,,\quad h_{ij}=\gamma_{ij}-\delta_{ij}\,,
\eeqn
and we can study their appearance to quadratic order in the mass term.
For $a=1$, the expression in (\ref{admfpfp}) turns out to be linear in $\delta N$ at the quadratic level, such that the equation of motion of $\delta N$ gives rise to a constraint. On the other hand, we also see that the shift vector $\delta N_i$ does no longer appear only linearly which is of course consistent with the breaking of diffeomorphism symmetry by the mass term.
Together with its associated secondary constraint, the $\delta N$ equation removes two degrees of freedom, one field plus its canonical momentum.\footnote{The Fierz-Pauli theory is known to give rise to a secondary constraint coming from requiring the primary constraint to be preserved in time.} We end up with $12-2=10$ degrees of freedom, describing the five polarisation states and corresponding conjugate momenta of the massive graviton.
\\
Contrarily, for $a\neq 1$, there is a term involving $\delta N^2$ at the quadratic level. The constraint arising from the equation of motion of $\delta N$ is lost in that case because the equation now determines $\delta N$ itself instead of constraining other variables. There are thus 12 degrees of freedom, describing six propagating modes, one of which is the Boulware-Deser ghost. 
\\
The same situation occurs if we do not consider $\hmn$ as a small perturbation, i.e.~look at more general backgrounds than $\emn$. In that case, the ADM variables of $\hmn$ are no longer small fluctuations and we have to consider the theory beyond quadratic order. Boulware and Deser studied a class of corrections to the Fierz-Pauli mass term and found that those higher-order terms in $\hmn$ could never result in an expression that is linear in the lapse $N$. From this they concluded that the constraint that removes the ghost in the linear theory is destroyed and hence a theory of nonlinear spin-2 interactions can never be consistent~\cite{Boulware:1973my, Boulware:1972zf}.

As pointed out in~\cite{deRham:2010ik, deRham:2010kj} and as will be discussed in the next subsection, there exists a loophole in this no-go theorem for nonlinear massive gravity and it turns out that a consistent theory can be constructed.\footnote{Note also that the arguments given by Boulware and Deser are of perturbative nature. An idea that is rather different from everything we discuss here is that the negative norm states which plague the general higher-order interactions could be avoided non-perturbatively~\cite{Iglesias:2011it, Kakushadze:2013dba}.} 
In the following, we will deviate from the historical path and construct the consistent interaction potential directly in the redefined ADM variables that were used in the consistency proof of~\cite{Hassan:2011hr, Hassan:2011tf} instead of presenting the derivation in the decoupling limit performed in~\cite{deRham:2010ik, deRham:2010kj}. From our point of view this construction is the most efficient way to arrive at the action for nonlinear massive gravity and it has the further advantages that it (a) automatically ensures the absence of ghost in the full theory (i.e.~away from the decoupling limit) and (b) immediately results in the generalised form of the action with arbitrary reference metric $\fmn$.

\subsection{ADM variables for massive gravity}
Our aim is to arrive at a nonlinear theory for massive spin-2 fields of the form (\ref{mggenact}) and, to this end, we shall discuss interactions in terms of ADM variables. For definiteness we will work in four dimensions but all our considerations and conclusions generalise straightforwardly to any dimension. Before we start, let us briefly recapitulate the situation.

Since the kinetic term for the metric $\gmn$ in (\ref{mggenact}) is the same as in GR, the lapse and shift functions $N$ and $N^i$ will still appear without derivatives. However, the interaction potential will in general no longer be linear in these functions. Therefore, their equations of motion, instead of imposing constraints on the remaining variables, will now determine $N$ and $N^i$ themselves. The four gauge constraints are lost and, as explained in the previous subsection, the number of propagating modes will now generically be six, plus their corresponding canonical momenta. These are two phase-space degrees of freedom too many for the theory of a massive spin-2 field. Moreover, the Hamiltonian of a generic theory will not be positive definite~\cite{Boulware:1972zf, Boulware:1973my}, signalling that the extra propagating mode is a ghost.
A necessary requirement on any consistent interaction potential is therefore the presence of an additional constraint that removes the nonlinear Boulware-Deser ghost.
 
Before we set out to construct the potential featuring the constraint it should be noted that there are two other conditions which need to be fulfilled by a fully consistent theory. Firstly, the preservation of this constraint in time must itself provide a constraint in order to remove also the conjugate momentum and hence the full phase-space pair associated with the pathological degree of freedom. Secondly, the resulting Hamiltonian must be positive definite so that none of the surviving five spin-2 modes gives rise to an instability. It should be emphasised that positivity of the Hamiltonian of the nonlinear theories that we are about to discuss has never been proven in general and in fact seems not to be true without additional physical assumptions, see e.g.~\cite{Volkov:2014ida}. For instance, ghosts different from the Boulware-Deser mode may still propagate around certain backgrounds, an example being the Higuchi instability of the helicity-zero mode in de Sitter space~\cite{Higuchi:1986py}. 
In the literature, whenever massive gravity and bimetric theory are labelled as ``consistent"  or ``ghost-free", one is usually referring only to the complete removal of the Boulware-Deser mode and its conjugate momentum. Throughout this review we frequently adhere to this conventional abuse of terminology.

\subsubsection{The loophole in Boulware \& Deser's argument}\label{sec:loophole}
In order to investigate whether a particular structure in the potential $V(g^{-1}f)$ can give rise to a constraint and thus satisfy the first necessary condition on any consistent theory, we first decompose the second rank-2 tensor $f_{\mu\nu}$ into its own ADM variables,\footnote{In fact, it is not automatically guaranteed that a simultaneous ADM split for $\gmn$ and $\fmn$ exists or, equivalently, that $N^2$ and $L^2$ are both positive definite. The assumption of simultaneous ADM decompositions, which shall be made here and in the following, is related to the existence of intersecting light cones for the two metrics. The details of this will be discussed in~\cite{Fawad+Mikica}.}
\beqn\label{admf}
f_{\mu\nu}=\begin{pmatrix}
-L^2+L_l \phi^{lk}L_k& L_j\\
L_i & \phi_{ij}
\end{pmatrix}\,.
\eeqn
Here, $\phi^{ij}$ denotes the inverse of the three-dimensional metric $\phi_{ij}$, $L_i$ is the shift-vector, and $L$ is the lapse of~$f_{\mu\nu}$. 
We furthermore express the measure factor $\sqrt{g}$ in terms of the lapse $N$ and the determinant $\gamma$ of $\gamma_{ij}$,
\beqn\label{measur}
\sqrt{g}=N \sqrt{\gamma}\,.
\eeqn
With these expressions at hand, we can write a generic interaction potential in terms of ADM variables as,\footnote{We will often make use of the notations $N^i\equiv \gamma^{ij}N_j$ and $L^i\equiv \phi^{ij}L_j$.}
\beqn
\sqrt{g}~V(g^{-1}f)=N \sqrt{\gamma}~V(\gamma_{ij}, N, N^i; \phi_{ij}, L, L^i)\,.
\eeqn
According to the argument by Boulware and Deser, the right-hand side of this equation needs to be linear in the lapse $N$ in order to provide a constraint that removes the ghost mode. However, this is not entirely correct and this is exactly where the loophole lies:
In fact, the necessary condition on the potential is slightly weaker because the constraint may be obtained after combining several of the equations for $N$ and $N^i$. In other words, if there exists a particular combination of these equations that is independent of $N^i$ and $N$, then this equation will constrain the remaining variables. At the level of the action, this means that we should allow for a field-dependent redefinition of the shift components, $N^i\rightarrow n^i$, that renders the Lagrangian linear in the lapse $N$. Let us explain this in a bit more detail. Suppose that a linear combination of the equations for $N$ and $N^i$ is independent of the lapse and thus corresponds to the constraint,
\beqn\label{constrcomb}
\mathcal{C}=\frac{\delta S(N, N^j,\hdots)}{\delta N}+C^i\frac{\delta S(N, N^j,\hdots)}{\delta N^i}=0\,,
\eeqn
where the $C^i$ are some functions of the ADM variables. We can now make a redefinition of variables, $N^i(N, n^j,\hdots)\equiv {C^i}_k(N,n^j,\hdots) n^k$, such that the variation of the action with respect to the lapse becomes,
\beqn
\frac{\delta S(N, n^j,\hdots)}{\delta N}&=&\left.\frac{\delta S(N, N^j,\hdots)}{\delta N}\right|_{N^i}+\left.\frac{\delta S(N, N^j,\hdots)}{\delta N^i}\right|_N\frac{\delta N^i(N, n^k,\hdots)}{\delta N}\nn\\
&=&\left.\frac{\delta S(N, N^j,\hdots)}{\delta N}\right|_{N^i}+\left.\frac{\delta S(N, N^j,\hdots)}{\delta N^i}\right|_N\frac{\delta{C^i}_k(N,\hdots) }{\delta N}n^k\,.
\eeqn
Here $|_{N^i}$ means that the function $N^i$ is kept fixed when the functional derivative is taken.
From this we see that if we choose the redefined shift components $n^k$ such that they satisfy $\frac{\delta{C^i}_k(N,\hdots) }{\delta N}n^k=C^i$ then the variation of the action with respect to the lapse gives precisely the constraint~(\ref{constrcomb}) which by assumption does not involve $N$. We shall thus look for a redefinition of the shift vector that renders the action linear in the lapse~$N$.

Certainly, the redefinition, i.e.~the matrix ${C^i}_k(N,\hdots)$, must be linear in $N$ itself since the $N^i$ appear linearly in the kinetic term. Furthermore, since the redefined shift components $n^i$ are expected to appear in the constraint, they must be fully determined by their own equation of motion which therefore must not depend on $N$.  To summarise, in order to fulfil the first necessary condition of obtaining a constraint we make two requirements on the potential in terms of the redefined shift $n^i$:
\begin{itemize}
\item[(i)] linearity of the Lagrangian in $N$,
\item[(ii)]  absence of $N$ in the $n^i$ equations of motion.
\end{itemize}
For now, we focus on the first requirement and we will see later that the potential that we construct by demanding only (i) automatically satisfies (ii).

\subsubsection{Redefinition of ADM variables}

We now observe that the full interaction potential density, $\sqrt{g}~V(g^{-1}f)$ in (\ref{mggenact}), already has a factor of $N$ in front coming from the measure factor (\ref{measur}). Hence, in order to satisfy requirement (i) listed above, the potential $V(g^{-1}f)$ written in redefined ADM variables must be of the form,
\beqn
V(g^{-1}f)=\frac{1}{N}V_1+V_2\,,
\eeqn
where $V_1$ and $V_2$ are functions only of $(\gamma_{ij}, n^i)$ (apart from the non-dynamical components of $\fmn$ which we choose to omit here), i.e.~they are independent of~$N$. 

Recalling the ADM decomposition (\ref{admmetr}) of $\gmn$, we notice that $\gmn$ is quadratic in $N$ and in order to obtain inverse powers of $N$ we need to consider the inverse metric~(\ref{invadm}). The latter is quadratic in $1/N$ and the best we can achieve by a field redefinition which is linear in $N$ is to complete the dependence on $N$ into a perfect square such that taking a square-root can result in an expression linear in $1/N$. 
In other words, the only quantity that has a chance of giving something linear in $1/N$ after a linear redefinition of the $N^i$ is an object whose square is proportional to the inverse metric~(\ref{invadm}). 
 We are thus led to consider a potential~$V$ that is a function of the matrix $S\equiv\sqrt{g^{-1}f}$, defined via $S^2=g^{-1}f$. 
This square-root matrix has a very nontrivial ADM decomposition and is certainly highly nonlinear in $1/N$ before any redefinition. However, we will now make use of the allowed redefinition of $N^i$ and demand that in terms of the new shift-vectors $n^i$ the square-root matrix $S$ is of the form~\cite{Hassan:2011hr, Hassan:2011tf},
\beqn\label{sadm}
S=\frac{1}{N}\mA+\mB\,,
\eeqn
where $\mA$ and $\mB$ are matrix-valued functions of $(\gamma_{ij}, n^i)$. The redefinition that leads to (\ref{sadm}) as well as the explicit expressions for $\mA$ and $\mB$ can be obtained straightforwardly by the following method: Square the right-hand side of the ansatz (\ref{sadm}) and equate it with the ADM expression for $g^{-1}f$ obtained from (\ref{admmetr}) and (\ref{admf}), using the most general ansatz for the redefinition, $N^i=c^i+Nd^i$. Then compare the expressions on both sides order by order in $1/N$ to determine the vectors $c^i$ and $d^i$ in the shift redefinition as well as the matrices $\mA$ and $\mB$.
This derivation was given in~\cite{Hassan:2011tf} and we discuss it in more detail in appendix~\ref{App:redefinition}; here we simply state the result. The redefinition that renders the square-root matrix $S$ linear in $1/N$ takes a rather simple form~\cite{Hassan:2011hr, Hassan:2011tf},
\beqn\label{shiftred}
N^i=Ln^i+L^i+N{D^i}_kn^k\,.
\eeqn
The $3\times3$ matrix $D$ on the right-hand side is a function of the variables $(\gamma_{ij}, n^i)$ and the non-dynamical spatial metric $\phi_{ij}$. Explicitly, it can be written in matrix notation as,
\beqn\label{defD}
D=\sqrt{\gamma^{-1}\phi\, Q}\,Q^{-1}\,,
\eeqn
where we have defined another matrix $Q$ through,
\beqn
{Q^i}_j=x{\delta^i}_j+n^in^k\phi_{kj}\,,\qquad x=1-n^i\phi_{ij}n^j\,.
\eeqn
Note that the definition (\ref{defD}) of the matrix $D$ that enters the redefinition involves a $3\times3$ square-root matrix. By introducing the shift vectors $n^i$ we have therefore reduced the dimension of the square-root matrix that appears in $S$ by one and, most importantly, simplified the dependence on the lapse $N$ which no longer appears under any square-root in (\ref{sadm}). One may worry that a real solution for the $3\times3$ square-root in (\ref{defD}) does not always exist. However, we will show now that the variables can be further redefined to demonstrate the existence of real solutions for $D$ and, in fact, to remove the square-root matrix entirely.

\subsubsection{On the existence of the redefinition}\label{sec:exofred}

The form of the redefinition~(\ref{shiftred}) is not entirely unique. In fact, the original papers~\cite{Hassan:2011hr, Hassan:2011tf} on massive gravity mainly worked with a set of variables that slightly differs from the one presented here, whereas the choice of variables we made above is more suitable for application to bimetric theory~\cite{Hassan:2011zd}. Moreover, as was shown later in~\cite{Hassan:2014gta}, it is possible to arrive at simpler expressions which are also more symmetric between the two metrics $\gmn$ and $\fmn$.
In order to see this, let us decompose the two spatial metrics into ``spatial dreibeins",
\beqn
\gamma_{ij}=e^a_{~i}\delta_{ab}e^b_{~j}\,,\qquad
\phi_{ij}=\varphi^a_{~i}\delta_{ab}\varphi^b_{~j}\,.
\eeqn
These expressions are invariant under rotations of the dreibeins, which means that $\tilde{\varphi}=R\varphi$ with $R^\mathrm{T}=R^{-1}$ is an equivalent dreibein of $\phi$ and the rotations are a local symmetry of the theory. This freedom can be used to resolve the square-root appearing in $D$ (c.f.~\eqref{defD}).
Next, we redefine the shift vectors $n^i$ according to,
\beqn
n^i\equiv (\tilde{\varphi}^{-1})^i_{~a}\delta^{ab}v_b\,,
\eeqn
where the new shifts $v^a$ carry a spatial Lorentz index and $\tilde{\varphi}=R\varphi$ is a gauge-fixed dreibein. The rotation matrix is chosen such that it satisfies,
\beqn\label{defR}
\left[\left(\hat{I}+\tfrac{1}{x+\sqrt{x}}vv^\mathrm{T}\right)R\varphi e^{-1}\right]^\mathrm{T}=\left(\hat{I}+\tfrac{1}{x+\sqrt{x}}vv^\mathrm{T}\right)R\varphi e^{-1}\,,
\eeqn
where $\hat{I}_{ab}=\delta_{ab}$.
It is straightforward to obtain a solution for $R$ from this equation and this solution always exists, which follows directly from the polar decomposition theorem stating that any matrix can be symmetrised by a rotation. Inserting the new variables into (\ref{defD}) and using (\ref{defR}) to evaluate the square-root matrix, we find after some algebraic manipulations that $D$ takes the much simpler form,
\beqn
D=\tfrac{1}{\sqrt{x}}e^{-1}\left(\hat{\mathbb{1}}-\tfrac{1}{1+\sqrt{x}}\hat{I}^{-1}vv^\mathrm{T}\right)\tilde{\varphi}\,,
\eeqn
where we are using matrix notation and $\hat{\mathbb{1}}^a_{~b}=\delta^a_{~b}$. Introducing the dreibeins and the new shift vectors has thus enabled us to get rid of any matrix square-root in the equations. The only square-root left is the scalar $\sqrt{x}=\sqrt{1-v^\mathrm{T}\hat{I}^{-1}v}$. This square-root has real solutions provided that the metric components satisfy the bound $v_a \delta^{ab}v_b<1$. Interestingly, this bound has an interpretation connected with Lorentz transformations and it turns out that the redefined shift vector $v_a$ can be interpreted as a Lorentz velocity. We will come back to this point at the end of section~\ref{sec:vbform}. For now, let us assume that the bound is not violated and therefore the redefinition of shift vectors exists. In terms of the new variables, it takes the following simple and symmetric form~\cite{Hassan:2014gta},
\beqn
N^i=L^i+\Big(L(\tilde{\varphi}^{-1})^i_{~a}+N(e^{-1})^i_{~a}\Big)\delta^{ab}v_b\,.
\eeqn
Even though the structure of this expression is a little less complicated than~(\ref{shiftred}), in order to stay closer to the conventions in the literature, we choose to continue using the original redefinition.

\subsection{Construction of the ghost-free potential}\label{sec:construction}

As we show in appendix~\ref{App:redefinition}, the matrices $\mA$ and $\mB$ in $\sqrt{g^{-1}f}=\frac{1}{N}\mA+\mB$ depend on the redefined ADM variables in the following way,
\beqn\label{exprab}
\mA&=&\frac{1}{\sqrt{x}}
\begin{pmatrix}
L+n^kL_k~&~n^k\phi_{kj}\\
-(L+n^kL_k)(Ln^i+L^i)~&~-(Ln^i+L^i)n^k\phi_{kj}
\end{pmatrix}\,,\nn\\
\mB&=&\sqrt{x}
\begin{pmatrix}
0~&~0\\
{D^i}_jL^j~&~{D^i}_j
\end{pmatrix}\,,
\eeqn
in which $x=1-n^i\phi_{ij}n^j$ as before and the index on the shift vector $L_i$ is raised using the inverse spatial metric~$\phi^{ij}$.
Although the above expressions seem rather complicated, what is essential for the construction of the ghost-free potential is the structure of the matrices $\mA$ and $\mB$. Most importantly, $\mA$ is a matrix of rank one and it can be written as the outer product of two vectors,
\beqn\label{vwa}
\mA=u w^\mathrm{T}\,,~~\quad~~\text{where}~~u=\begin{pmatrix}1\\-c^i \end{pmatrix}\,~~\text{and}~~~w=\frac{1}{\sqrt{x}}\begin{pmatrix}a_0\\a_i\end{pmatrix}\,.
\eeqn
This will be the key property entering our constructive proof in the following.  
Having obtained the ADM expression with the desired dependence on $N$ for the square-root matrix $S=\sqrt{g^{-1}f}=\frac{1}{N}\mA+\mB$, we are ready to construct the interactions which give rise to an additional constraint that removes the Boulware-Deser ghost.

As explained above, the mass potential has to be a function of the square-root matrix $S=\sqrt{g^{-1}f}=\frac{1}{N}\mA+\mB$, multiplied by $\sqrt{g}=N\sqrt{\gamma}$\,. We assume this function to be, at least formally, expandable as a Taylor series in $S$. This commonly defines what one means by a matrix valued function anyway so is not really a serious restriction. Generically, this will give an expression that is nonlinear in the lapse $N$. 
The only way to ensure linearity in $N$ is to demand the absence of higher powers of~$\frac{1}{N}\mA$ in the expansion. Obviously, the simplest possible term is $\sqrt{g}~\tr(S)=\sqrt{\gamma}~\tr(\mA+N\mB)$.
At first sight it seems that all higher powers of $S=\frac{1}{N}\mA+\mB$ will involve higher powers of~$1/N$. However, due to the special structure of the matrix $\mA$ in (\ref{vwa}) this is not quite correct and specific terms of higher order in $S$ can still be linear in $1/N$.
Since $\mA$ has rank one, it is a projection operator on a one-dimensional subspace. Owing to this property, there is a unique way of building polynomials of $\frac{1}{N}\mA+\mB$ that are linear in $\frac{1}{N}\mA$. Namely, only an antisymmetric product of $\mA$'s will automatically contain only one power of~$\mA$. Let us see how this works in detail by considering
\beqn\label{potantisym}
V(S)=\sum_{n=0}^4 ~b_n~\epsilon^{\mu_1\mu_2\hdots\mu_n\lambda_{n+1}\hdots\lambda_4}\epsilon_{\nu_1\nu_2\hdots\nu_n\lambda_{n+1}\hdots\lambda_4}~{S^{\nu_1}}_{\mu_1}\hdots{S^{\nu_n}}_{\mu_n}\,,
\eeqn
with arbitrary coefficients $b_n$ and totally antisymmetric tensors $\epsilon_{\mu\nu\rho\sigma}$. 
Since the simultaneous exchange of $\mu_i$, $\mu_j$ and $\nu_i$, $\nu_j$ only changes the sign of both $\epsilon$-tensors, it leaves the whole term invariant.
Therefore, at $n$th order, the product of $S$'s under the sum can be written in the form,
\beqn
\sum_{l=0}^n {n \choose l}\left(\frac{1}{N}\right)^l   {\mA^{\nu_1}}_{\mu_1}\hdots {\mA^{\nu_l}}_{\mu_l}{\mB^{\nu_{l+1}}}_{\mu_{l+1}}\hdots {\mB^{\nu_n}}_{\mu_n}\,.      
\eeqn
We insert this expression into (\ref{potantisym}) and obtain
\beqn\label{vsdef}
V(S)=\sum_{n=0}^4 b_n\sum_{l=0}^n {n \choose l}\left(\frac{1}{N}\right)^l V_n(\mA,\mB)\,,
\eeqn
with
\beqn
V_n(\mA,\mB)&=&\epsilon^{\mu_1\mu_2\hdots\mu_n\lambda_{n+1}\hdots\lambda_4}\epsilon_{\nu_1\nu_2\hdots\nu_n\lambda_{n+1}\hdots\lambda_4} \nn\\
&~&\hspace{50pt}\cdot~{\mA ^{\nu_1}}_{\mu_1}\hdots {\mA ^{\nu_l}}_{\mu_l}{\mB^{\nu_{l+1}}}_{\mu_{l+1}}\hdots {\mB^{\nu_n}}_{\mu_n}  \nonumber\\
&=&\epsilon^{\mu_1\mu_2\hdots\mu_n\lambda_{n+1}\hdots\lambda_4}\epsilon_{\nu_1\nu_2\hdots\nu_n\lambda_{n+1}\hdots\lambda_4}\nn\\
&~&\hspace{50pt} \cdot~{u ^{\nu_1}}w_{\mu_1}\hdots {u ^{\nu_l}}w_{\mu_l}{\mB^{\nu_{l+1}}}_{\mu_{l+1}}\hdots {\mB^{\nu_n}}_{\mu_n}\,.
\eeqn
Here we have used (\ref{vwa}) in the second equality.
Since all indices of the symmetric products of either $u^{\nu_i}$ or $w_{\mu_i}$ are contracted with the totally antisymmetric indices of the corresponding $\epsilon$-tensors, we find that in $V_n(\mA,\mB)$ only terms with at most one $\mA$ contribute. This implies that the sum over $l$ in (\ref{vsdef}) actually terminates at $l=1$ and hence $V(S)$ is linear in $1/N$, which is precisely the property that we were looking for.   
We therefore conclude that the most general form of the complete potential density which is linear in $N$ after the redefinition (\ref{shiftred}) is
\beqn\label{completep}
\sqrt{g}\,\sum_{n=0}^4 ~b_n~\epsilon^{\mu_1\mu_2\hdots\mu_n\lambda_{n+1}\hdots\lambda_4}\epsilon_{\nu_1\nu_2\hdots\nu_n\lambda_{n+1}\hdots\lambda_4}~{S^{\nu_1}}_{\mu_1}\hdots{S^{\nu_n}}_{\mu_n}\,,
\eeqn
and thus satisfies criterion~(i) that we wrote down in section~\ref{sec:loophole}. In order to give rise to a constraint it needs to also meet criterion~(ii), that the $n^i$ equations following from the action with the above potential need to be independent of the lapse $N$. As one can verify in a lengthy but straightforward computation, this second requirement does not impose further restrictions on the form of the potential but is automatically satisfied by~(\ref{completep}). More explicitly, the $n^i$ equations are of the form~\cite{Hassan:2011hr, Hassan:2011tf}, 
\beqn
\frac{\delta S}{\delta n^i}=\left(L\delta^j_{~i}+N\frac{\delta \big(D^j_{~k}n^k\big)}{\delta n^i}\right)E_j=0\,,
\eeqn
where $E_j$ does not involve $N$. Since the matrix multiplying $E_j$ is exactly equal to the Jacobian $\frac{\delta N^j}{\delta n^i}$ of the redefinition, it is invertible by assumption (because otherwise the redefinition would not be well-defined). We can thus multiply the equations by its inverse to arrive at the equivalent equations $E_j=0$ which do not depend on $N$. 

We have thus derived the unique form of the interaction potential which gives rise to an additional constraint. Note in particular that there is only a finite number of terms giving a potential which is linear in $N$. Due to the antisymmetric structure of the interactions, the possible terms are limited by the dimension of spacetime. Before discussing their structure and properties in more detail in the next section, let us make some final remarks on the existence of the associated secondary constraint which arises from demanding the primary constraint to be preserved in time.

\paragraph{The secondary constraint:}

The above requirements that we used to construct the consistent interaction potential were necessary but not sufficient: A secondary constraint is crucial for the absence of the Boulware-Deser ghost because two constraints are needed to remove both the ghost mode and its canonical momentum from the set of dynamical variables.

It was already motivated in~\cite{Hassan:2011hr, Hassan:2011tf} that there is in fact a secondary constraint arising from demanding the primary constraint to be constant in time.  In order to compute the time evolution of the constraint one uses the Poisson bracket (\ref{defpoisb}) and the Hamiltonian, which for the massive gravity action~(\ref{mggenact}) with potential (\ref{completep}) is of the form
 \beqn\label{hamsecc}
H=\int \mathrm{d}^{3} y~(\mathcal{H}_0-N\mathcal{C})\,.
\eeqn
Here $\mathcal{H}_0$ is independent of $N$ and $\mathcal{C}$ is the Hamiltonian constraint obtained from varying the action with respect to $N$. The time evolution of $\mathcal{C}$ then reads
\beqn\label{seccons}
\frac{\mathrm{d}\mathcal{C}(x)}{\mathrm{d}t}=\{\mathcal{C}(x),H\}=0\,.
\eeqn  
As in the case of the primary constraint, this equation should be independent of $N$, because otherwise it would determine $N$ instead of constraining $\gamma_{ij}$ and $\pi^{ij}$. Inserting~(\ref{seccons}) into~(\ref{hamsecc}) gives
\beqn
\int \mathrm{d}^{3} x~\Big(\{\mathcal{C}(x),\mathcal{H}_0(y)\}-N\{\mathcal{C}(x),\mathcal{C}(y)\}\Big)=0\,.
\eeqn  
Since $\mathcal{C}$ and $\mathcal{H}_0$ are independent of $N$, we need the Poisson bracket among the constraints $\{\mathcal{C}(x),\mathcal{C}(y)\}$ to vanish. Before this was actually demonstrated to be the case, there had been objections against the theory claiming that $\{\mathcal{C}(x),\mathcal{C}(y)\}$ could be nonzero \cite{Kluson:2011qe}. The issue was resolved when the secondary constraint was eventually shown to exist in a detailed calculation in~\cite{Hassan:2011ea}. We will not repeat this analysis here but instead refer the interested reader to the original reference as well as other subsequent independent confirmations of these results~\cite{Hassan:2012qv, Alexandrov:2012yv, Alexandrov:2013rxa, Comelli:2013txa, Kugo:2014hja}.

\newpage
\section{Ghost-free Nonlinear Massive Gravity}\label{sec:gfnlmg}

In the previous section we derived the nonlinear interaction potential for massive gravity equipped with an additional constraint that removes the Boulware-Deser ghost. Here we will study its properties in more detail and discuss the subclass of dRGT models as well as the gauge invariant St\"uckelberg formulation.

\subsection{Most general action}

For notational purposes, a more convenient way of writing the consistent potential in (\ref{completep}) is to introduce the elementary symmetric polynomials $e_n(S)$ of the matrix $S$. These can be defined in terms of totally antisymmetric tensors (with unit weight), 
\beqn
e_n(S)=\frac{1}{n!(4-n)!}\,\epsilon^{\mu_1\mu_2\hdots\mu_n\lambda_{n+1}\hdots\lambda_4}\epsilon_{\nu_1\nu_2\hdots\nu_n\lambda_{n+1}\hdots\lambda_4}~{S^{\nu_1}}_{\mu_1}\hdots{S^{\nu_n}}_{\mu_n}\,,
\eeqn 
with $e_0(S)\equiv1$. More properties of the elementary symmetric polynomials as well as precise definitions of the $\epsilon$-tensors as anti-symmetrisation operators are summarised in appendix~\ref{app:technical}.
In terms of these the complete action for ghost-free massive gravity with general reference metric takes the form,
\beqn\label{HRMG}
S_\mathrm{MG}=m_g^{2}\int \mathrm{d}^4x\,\sqrt{g}\,\left[R(g) -2 m^2\sum_{n=0}^4 ~\beta_ne_n\left(S\right)\right]\,,
\eeqn
with $S=\sqrt{g^{-1}f}$, Planck mass $m_g$ and spin-2 mass scale $m$. Furthermore, we have introduced the rescaled coefficients $\beta_n=b_n n!(4-n)!/2 $ for $n=0,\hdots 4$. 
Out of these five parameters, only three are truly measuring interaction strengths: Since $e_0(S)=1$, the $\beta_0$-term is simply a cosmological constant for the dynamical metric $\gmn$. Moreover, the last term in the sum that is proportional to $\beta_4$ is just a cosmological constant term for $\fmn$ because $e_4(S)=\det S$ and hence $\sqrt{g}\,e_4(S)=\sqrt{f}$. This term is therefore independent of $\gmn$ and does not contribute to the equations of motion. Nevertheless, we choose to include it in the action because it will become relevant when we give dynamics to $\fmn$ later on.

The above action is a nontrivial generalisation of the de-Rham-Gabadadze-Tolley (dRGT) model~\cite{deRham:2010ik, deRham:2010kj} which we shall discuss in the next subsection. Its above form (with general reference metric $\fmn$, finite sum over $n$ and in terms of the elementary symmetric polynomials) was first presented in~\cite{Hassan:2011vm}.

The equations of motion for $\gmn$ obtained from~(\ref{HRMG}) are,
\be\label{geommg}
	R_{\mu\nu}(g)-\frac1{2}\gmn R(g)+m^2V^g_{\mu\nu}(g,f;\beta_n)
	=0\,,
\ee
where the first two terms are the usual Einstein tensor while the contribution from the interaction potential is,
\beqn
V^g_{\mu\nu}=g_{\mu\rho}\sum_{n=0}^3(-1)^n\beta_n(Y_{(n)})^\rho_{~\nu}(S)\,,
\eeqn
where we have defined the matrix functions,
\beqn\label{yndef}
(Y_{(n)})^\rho_{~\nu}(S)\equiv\sum_{k=0}^n(-1)^k e_k(S)\,(S^{n-k})^\rho_{~\nu}\,.
\eeqn
There is no equation of motion for $\fmn$ whose form therefore needs to be put in by hand.\footnote{Imposing an equation of motion for the reference metric, one can integrate out $\fmn$ from the action but in this case the theory becomes dynamically equivalent to GR.} 
Taking the divergence of the above equations and using the Bianchi identity $\nabla^\mu\mathcal{G}_{\mu\nu}=0$ satisfied by the Einstein tensor $\mathcal{G}_{\mu\nu}$, we arrive at a set of Bianchi constraints,
\beqn
\nabla^\mu V^g_{\mu\nu}=0\,.
\eeqn
These remove four degrees of freedom while the remaining extra scalar (the Boulware-Deser ghost) is eliminated by the additional constraint present in the special structure of~(\ref{HRMG}). A covariant expression for this scalar constraint is difficult, if not impossible, to obtain in general. Its explicit form for certain regions in the parameter space is provided in~\cite{Deffayet:2012nr, Deser:2014hga} which make use of the vierbein formulation that we shall discuss in section~\ref{sec:vierbein}. For the same restricted parameter choices, it is also possible to identify a covariant constraint in the linear theory around arbitrary backgrounds, see section~\ref{sec:linarb}.

\subsection{dRGT theory}

The theory first derived by de Rham, Gabadadze and Tolley (dRGT) in \cite{deRham:2010ik, deRham:2010kj} was defined for flat reference metric $\fmn=\emn$. The original construction uses St\"uckelberg fields (c.f.~section~\ref{sec:stuckmg}) and its ``ghost proof" is valid only in the scalar sector of a decoupling limit that strongly relies on the flat reference metric.\footnote{For the sake of historical accuracy, let us emphasise again, that \cite{deRham:2010ik, deRham:2010kj} also pointed out the loophole in Boulware and Deser's argument. Moreover, the lowest-order terms of the shift redefinition in a perturbative expansion around flat space were computed, but the results do not agree with the nonlinear result~(\ref{shiftred}) derived in \cite{Hassan:2011hr, Hassan:2011tf}.} In contrast, the construction that we have presented here is based on the results of~\cite{Hassan:2011hr, Hassan:2011tf} and is valid for all reference metrics. It demonstrates the consistency of the full nonlinear theory away from any limiting approximation.

It is worth discussing the dRGT model and its relation to the Hassan-Rosen formulation in a bit more detail. Using the observation that the sum in the interaction potential terminates and can be given in terms of elementary symmetric polynomials~\cite{Hassan:2011vm}, the dRGT action can be written in the form~\cite{deRham:2010ik, deRham:2010kj} (using the conventions of~\cite{deRham:2014zqa}),
\be\label{SdRGT}
	S_{\mathrm{dRGT}} = m_g^{2}\int\td^4x\sqrt{g}\left(R(g)
	+\frac{m^2}{2}\sum_{n=0}^4n!(4-n)!\,\alpha_ne_n(\mathcal{K})\right)\,,
\ee
where
\beqn
\mathcal{K}^\mu_{\ph\mu\nu}=\delta^\mu_{~\nu}-\big(\sqrt{g^{-1}\eta}\,\big)^\mu_{\ph\mu\nu}\,,\qquad
\alpha_0=\alpha_1=0\,,\qquad
\alpha_2=1\,.
\eeqn
We can arrive at this action starting from~(\ref{HRMG}) by setting $\fmn=\emn$ and taking,
\beqn
\beta_n=-\frac{(-1)^n}{4}\sum_{k=n}^4 {4-n \choose k-n}k!(4-k)!\,\alpha_k\,.
\eeqn
This follows from the identity \eqref{app:enrel} satisfied by the elementary symmetric polynomials. Hence the consistency of massive gravity with general $\fmn$ implies the absence of ghost in the dRGT action. In contrast, there is no obvious way of getting to~(\ref{HRMG}) from~(\ref{SdRGT}) and results obtained in the latter do not generalise automatically to the former. 

Let us briefly explain why the lowest-order $\alpha_n$ parameters in the dRGT action are fixed while they remain arbitrary in the Hassan-Rosen formulation, since the exact reason for this is sometimes obscured. For $\gmn=\emn$ to be a solution to the equations of motion following from~(\ref{SdRGT}), the cosmological constant for $\gmn$ must vanish. It is straightforward to verify that this requires\footnote{It should be noted that this first requirement can actually be avoided by demanding only that $\gmn=c^2\emn$ is a solution. This is still flat of course and fixes $c$ instead of a parameter of the action.} $4\alpha_0+\alpha_1=0$, since this combination is proportional to the effective cosmological constant for $\gmn$. Next, in order to remove terms linear in the perturbation $h_{\mu\nu}=\gmn-\eta_{\mu\nu}$ in the quadratic action (so-called ``tadpoles"), one must enforce $\alpha_1=0$.\footnote{From the bimetric perspective, this second requirement would mean that the effective cosmological constant for $\fmn$ vanishes since the latter is proportional to $\alpha_1\sim \beta_1+3\beta_2+3\beta_3+\beta_4$, c.f.~equation~\eqref{lambdas}. The first requirement of vanishing cosmological constant for $\gmn$ can also be written as $4\alpha_0+\alpha_1\sim\beta_0+3\beta_1+3\beta_2+\beta_3=0$.}
Finally, the interaction parameters contain a redundant overall scale that can be absorbed into~$m^2$. One way to get rid of this redundancy is to demand that $m^2$ corresponds to the squared Fierz-Pauli mass in the quadratic action around flat space. This requirement finally fixes $\alpha_2=1$.

The above action~(\ref{SdRGT}) thus gives rise to Minkowski solutions for $\gmn$ and linearising the theory around these backgrounds gives precisely the Fierz-Pauli Lagrangian~(\ref{FPlag}) with $m^2_\mathrm{FP}=m^2$. In this sense the dRGT theory can be viewed as the consistent nonlinear completion of the Lagrangian for a massive spin-2 field in flat space. In the following, we will turn to the linear theory around arbitrary background solutions of the more general theory~(\ref{HRMG}).

\subsection{Linear theory on arbitrary background}\label{sec:linarb}

As we pointed out in section~\ref{sec:linmflatsp}, na\"ively covariantising the linear theory for a massive spin-2 field in flat background does not result in a consistent action. Attempts to find the correct equations describing five helicity states around any background were made already before the ghost-free nonlinear theory was known. For instance, the authors of~\cite{Buchbinder:1999ar} were able to write down the consistent linearised theory to first order in a small-curvature expansion.
\\
On the other hand, the knowledge of the full nonlinear action~(\ref{HRMG}) that avoids the ghost mode makes it possible to derive the linear theory around any background solution.\footnote{Note that in order to work with the most general backgrounds, we have to start from nonlinear massive gravity in the generalised formulation~(\ref{HRMG}) where $\fmn$ is entirely arbitrary. The set of solutions for the metric $\gmn$ in the dRGT model~(\ref{SdRGT}) with flat reference metric is much smaller and hence does not allow us to derive the most general form for the linearised theory. For instance, it is not possible to obtain the quadratic action for massive spin-2 in de Sitter space from~(\ref{SdRGT}).} This computation was first carried out in~\cite{Bernard:2014bfa, Bernard:2015mkk} (see also~\cite{Guarato:2013gba, Cusin:2015tmf, ourpaper}). The analysis is complicated by the presence of the square-root matrix $\sqrt{g^{-1}f}$ whose perturbation around general backgrounds is rather nontrivial due to the matrices $\gmn$ and $\fmn$ in general being non-commuting. The full expression can however be obtained using the Cayley-Hamilton theorem for matrices. An alternative way of arriving at the linearised theory is to redefine the dynamical fluctuation variables in order to simplify the square-root variation~\cite{ourpaper}.

The general structure of the quadratic Lagrangian for perturbations $\delta \gmn$ around arbitrary backgrounds $\bgmn$ is,
\beqn\label{quadlaggb}
\mathcal{L}=\delta\gmn\mathcal{E}^{\mu\nu\rho\sigma}\delta g_{\rho\sigma} 
+\frac{m^4}{m_g^2}\, \delta\gmn \mathcal{V}^{\mu\nu\rho\sigma}\delta g_{\rho\sigma} \,,
\eeqn
where $\mathcal{E}^{\mu\nu\rho\sigma}$ is a function of the background metric $\bgmn$ and quadratic in its associated covariant derivative $\bar{\nabla}^\mu$, whereas $\mathcal{V}^{\mu\nu\rho\sigma}$ depends on the background metric, its curvatures and the reference metric $\fmn$. 
The corresponding equations of motion for the fluctuations read,
\beqn\label{lineqgb}
\delta E_{\mu\nu}\equiv \mathcal{E}_{\mu\nu}^{~~\rho\sigma} \delta g_{\rho\sigma}
+\frac{m^4}{m_g^2}\, \mathcal{V}_{\mu\nu}^{~~\rho\sigma} \delta g_{\rho\sigma}=0\,.
\eeqn
Using the background equations~(\ref{geommg}), it is always possible to convert curvatures of $\bgmn$ into functions of the background and reference metric. Eliminating the reference metric through the background equations is more difficult since the equations are nonlinear in $\fmn$. In general it is not possible to obtain a closed form for the solution. However, quite remarkably, it turns out that if $\beta_2=\beta_3=0$ a closed form solution exists and the equations can be used to get rid of all appearances of the reference metric. Then~(\ref{lineqgb}) describes five massive spin-2 degrees of freedom propagating in an arbitrary background given solely by $\bgmn$ and its curvature $\bar{R}_{\mu\nu}$.
In the more general case it is still possible to solve the background equations perturbatively in curvatures and reproduce the results of~\cite{Buchbinder:1999ar}.

The absence of the Boulware-Deser ghost in the linearised equations follows from the existence of the additional constraint in the nonlinear theory from which~(\ref{quadlaggb}) was obtained.
In order to bypass the ADM analysis and show explicitly that the above equations contain the same number of propagating degrees of freedom as in the Fierz-Pauli case~(\ref{FPeqs}) in flat space, one needs to identify the analogues of the constraints~(\ref{twoconstr}). The vector constraints, i.e.~the generalisation of $\partial^\mu h_{\mu\nu}-\partial_\nu h=0$, are easily obtained by taking the (covariant) divergence of the equations and using the linearised version of the Bianchi identity. The additional scalar constraint which removes the ghost is more difficult to identify. For models with $\beta_3=0$, it has been found in~\cite{Bernard:2014bfa, Bernard:2015mkk, ourpaper}. It corresponds to the following combination of equations,
\beqn
\frac{1}{m^2}(S^{-1})^\nu_{~\rho}\nabla^\rho \nabla^\mu \delta E_{\mu\nu}+\frac{\beta_1}{2}\bar{g}^{\mu\nu}\delta E_{\mu\nu}+\beta_2S^\nu_{~\rho}\bar{g}^{\rho\mu}\delta E_{\mu\nu}=0\,,
\eeqn
in which all terms containing two derivatives on $\delta\gmn$ cancel out. This equation is the analogue of $h=0$, which it reduces to when the background is an Einstein spacetime.
When $\beta_3\neq 0$, it seems to be impossible to obtain the constraint in a covariant way~\cite{Deser:2014hga, ourpaper} and the reason for that is not yet fully understood.

\subsection{Gauge invariant massive gravity}\label{sec:stuckmg}

A common perspective is to take the fixed reference metric $\fmn$ to be fully specified in a given coordinate frame which implies that the invariance under general coordinate transformations (GCTs) of GR is broken by the mass term for the graviton. This is because the metric $\gmn$ itself transforms as a covariant rank-two tensor under GCTs while the fixed background metric $\fmn$ does not if it is taken to be fixed in a given coordinate system. Therefore, in this view, objects like $g^{\rho\mu}\fmn$ do not transform as tensors under diffeomorphisms. Of course covariance is not a true symmetry and it is always possible to simply view $\fmn$ as a fixed reference geometry but still allow it to transform as a tensor under GCTs, e.g.~to be flat but not necessarily on the cartesian Minkowski form. A constructive way of restoring any gauge symmetry in the action is the so called St\"uckelberg trick: By introducing new gauge degrees of freedom, so-called St\"uckelberg fields, that transform in a certain way under the gauge group, one can rewrite the theory in a manifestly gauge invariant way whilst keeping track of the thereby introduced redundant degrees of freedom. In turn, fixing the so-called physical (or unitary) gauge will give back the original action.

\subsubsection{The St\"uckelberg trick}
        
One possibility to reintroduce diffeomorphism invariance in massive gravity is to perform a gauge transformation on the dynamical metric~$\gmn$ and treat the gauge parameters as new fields in the action. Under a GCT, $x^\alpha\mapsto x'^\alpha(x)$, the metric transforms as
\beqn
\gmn(x)\longmapsto\frac{\partial x'^\alpha}{\partial x^\mu}\frac{\partial x'^\beta}{\partial x^\nu}g_{\alpha\beta}(x')\,.
\eeqn
After performing this GCT, one interprets the $x'^\alpha$ as dynamical fields. Another equivalent way, that turns out to be more convenient for studying massive gravity, is to mimic this kind of transformation on the reference metric $\fmn$, which is then taken to not transform under GCTs.  But since a simultaneous coordinate transformation of $\gmn$ and $\fmn$ would be a symmetry of the massive gravity action, instead of transforming $\gmn$ we can choose to perform the transformation on $\fmn$. 
That is to say, we replace $\fmn$ by its covariantised form,
\beqn\label{etcov}
\fmn\longrightarrow\frac{\partial \varphi^\alpha}{\partial x^\mu}\frac{\partial \varphi^\beta}{\partial x^\nu}\bar{f}_{\alpha\beta}\,.
\eeqn
Here, $\bar{f}_{\alpha\beta}$ is a new fixed background metric that does not transform under the gauge group. The ``coordinates'' $\varphi^\alpha$ are promoted to dynamical fields that transform as scalars under GCTs. Introducing these St\"uckelberg fields ensures that objects such as
\beqn
 \Tr\, (g^{-1}f) =g^{\mu\nu}\frac{\partial \varphi^\alpha}{\partial x^\mu}\frac{\partial \varphi^\beta}{\partial x^\nu}\bar{f}_{\alpha\beta}
 \eeqn
now are manifestly diffeomorphism invariant. In addition to the equations of motion for the metric $\gmn$, we now also consider the $\varphi^\alpha$ equations and, as we will see below, this system of equations is equivalent to the original $\gmn$ equation without St\"uckelberg fields. This method of restoring gauge invariance, in contrast to treating $\fmn$ as a fixed geometry tensor, is particularly useful for considering so called decoupling limits of a theory: Limits/regimes where certain operators become dominant and physical modes decouple.

\subsubsection{The decoupling limit}
Let us take $\bar{f}_{\alpha\beta}=\eta_{\alpha\beta}$ and expand the St\"uckelberg fields around the identity transformation, which corresponds to considering infinitesimal GCTs,
\beqn\label{pimu}
\varphi^\alpha=x^\alpha-\pi^\alpha\,.
\eeqn 
Furthermore, let us decompose the perturbations into a transverse and a longitudinal mode,
\beqn\label{decpi}
\pi^\alpha=\frac{\hat{\pi}^\alpha}{mm_g}+\frac{\partial^\alpha\pi}{m^2 m_g}\,,
\eeqn
with $\p_\alpha\hat{\pi}^\alpha=0$.
The ghost-free dRGT potential~(\ref{SdRGT}) was originally constructed using the gauge invariant formulation and performing a scaling limit to separate the interactions for the longitudinal component~$\pi$.

The power of the St\"uckelberg formalism lies in a conjecture made in~\cite{ArkaniHamed:2002sp}, stating that the ghost instability of nonlinear massive gravity can be traced back to higher-derivative interactions of the $\pi$ fields in flat space, where $\bar{f}_{\alpha\beta}=\eta_{\alpha\beta}$. This assumption relies on an analogy to the Goldstone equivalence theorem for spin-1 fields \cite{Lee:1977yc, Chanowitz:1985hj}. In that case, the longitudinal modes of the gauge fields have been shown to carry all information needed for computing scattering processes at high energies. Although the theorem has not been proven for the spin-2 case, it is reasonable to start with the assumption of its validity in order to investigate the stability of massive gravity in a certain parameter limit defined through,
\beqn
m_g\longrightarrow \infty\,,\qquad
m\longrightarrow 0\,,\qquad
\Lambda_3\equiv (m^2 m_g)^{\frac{1}{3}}=\mathrm{const}.
\eeqn
The metric is decomposed into flat background $\emn$ and fluctuations $\hmn$ according to $\gmn=\emn+\frac{\hmn}{m_g}$.
The fields in the St\"uckelberg decomposition~(\ref{decpi}) already have the correct normalisations and, in a first approach, the vector modes are set to zero. Then the decomposition of the tensor $g^{-1}f$ that appears in the interaction potential into flat background and perturbation reads, in matrix notation,
\beqn\label{stuckS}
g^{-1}f=\big(\mathbb{1}+\tfrac{\eta^{-1} h}{m_g}\big)^{-1}\big(\mathbb{1}+\tfrac{\Pi}{\Lambda_3^3}\big)^2\,,\qquad
\Pi^\mu_{~\nu}\equiv \eta^{\mu\rho}\partial_\rho\partial_\nu \pi\,.
\eeqn

The conjecture of \cite{ArkaniHamed:2002sp} suggests now to construct the interaction potential in such a way that there appear no higher-order interactions of the longitudinal modes. Since each of these modes always comes with two derivatives, their interactions generically give rise to higher-derivative terms. These can be shown to lead to ghost instabilities due to the famous Ostrogradsky theorem~\cite{ostrogradsky} and must therefore be avoided in a consistent theory (see~\cite{Woodard:2015zca} for a recent discussion of this theorem).
De Rham, Gabadadze and Tolley managed to tune the coefficients of the $\partial_\mu\partial_\nu\pi$ interaction terms in such a way that the higher-derivative terms combined into total derivatives that could be dropped from the action~\cite{deRham:2010ik, deRham:2010kj}. Interestingly, the resulting action for the longitudinal modes (after the kinetic terms have been diagonalised) resembles precisely the Galileon interactions~\cite{Nicolis:2008in}. In fact, from \eqref{stuckS} it is a trivial observation to see that by taking the square-root the expression becomes linear in $\Pi_{\mu\nu}$ and then any antisymmetric products of this square-root will at most carry two derivatives on any single $\pi$.

A few more comments are in order.
It is sometimes argued (see \eg \cite{Mirbabayi:2011aa,deRham:2011rn, deRham:2014zqa}) that any massive gravity theory can be described as a theory of GR plus four scalar fields. In other words, this would imply that it is always possible to reduce the number of degrees of freedom of the massive fluctuation $h_{\mu\nu}$ to two (corresponding to a massless spin-2 mode) and let the remaining three reside in the St\"uckelberg fields. Let us briefly clarify why this picture is not quite correct. 
Firstly, an equivalent statement would be that it is always possible to render the background solution for $\gmn$ flat by performing a coordinate transformation. As already pointed out in~\cite{Hassan:2012wr}, this is obviously not correct since the flat space action and equations are not equivalent to the ones in curved space.
Secondly, the counting of degrees of freedom in GR is an on-shell statement and only holds for the massless equations of motion, e.g.~$\Box\hmn=0$ on a flat background, which do not arise in a massive gravity theory. Therefore, the number of degrees of freedom in $\hmn$ cannot be reduced to two by simply performing a gauge transformation. 
Thirdly, from the transformation properties of the $\pi^\alpha$ fields in (\ref{pimu}) under linearised coordinate transformations ($\delta_\xi\pi^\alpha=-\xi^\alpha$), it is clear that these fields can only mix with the four gauge modes of $\gmn$. In other words, they are gauge trivial by construction.
Physical degrees of freedom can therefore not be fully transferred to the gauge modes because otherwise they would not contribute to interactions between conserved sources (for which $\nabla_\mu T^{\mu\nu}=0$). 

Although these caveats cast some serious doubts on the conclusiveness of the consistency analysis in the decoupling limit (other objections were raised in \eg \cite{Alberte:2010qb}), its historical importance for the discovery of ghost-free massive gravity should of course not be underestimated. The complete expression for the dRGT theory in the decoupling limit, including the transverse vector modes in~(\ref{decpi}), was later worked out in~\cite{Ondo:2013wka}. The limit is furthermore useful for understanding the validity regime of the effective field theory and the identification of the strong-coupling scale of massive gravity. 
We refer the interested reader to the review~\cite{deRham:2014zqa} where these subjects are discussed in more detail.

\subsubsection{Absence of ghost in the St\"uckelberg formulation}\label{secgenargst}

While the proofs of absence of the Boulware-Deser ghost given in \cite{Hassan:2011hr, Hassan:2011tf} were formulated in the gauge fixed version of Massive Gravity, i.e.~without introducing the St\"uckelberg fields, it is expected that restoring the gauge invariance does not alter the dynamics of the theory. In fact, since gauge symmetry is merely a redundancy in the description of the underlying physics, the two formulations are completely equivalent. In particular, absence of ghost in the gauge fixed version of massive gravity implies that the St\"uckelberg formulation cannot exhibit the instability either.
Nevertheless, in \cite{Chamseddine:2011mu} it was claimed that the ghost generically reappears in the St\"uckelberg sector and the author of \cite{Kluson:2011rt, Kluson:2012gz} could not find a constraint to remove the ghost. 
First attempts to disprove these statements were made in~\cite{deRham:2011rn, deRham:2011qq}, before the absence of ghost was conclusively shown in~\cite{Hassan:2012qv}.

The fact that the dynamics remain unaltered in the gauge invariant theory can be demonstrated by considering the equations of motions for the St\"uckelberg fields. After making the replacement (\ref{etcov}) for $f_{\mu\nu}$, the mass potential $V(g,f)$  becomes a function of the scalar fields and their equations of motion read
\beqn\label{varphieoms}
\frac{\delta}{\delta\varphi^\alpha} V(g,\varphi)=0\,.
\eeqn
However, this equation will not give rise to any new dynamics because it is, in fact, already implied by the Bianchi constraint,
\beqn
\nabla^\mu V_{\mu\nu}(g,\varphi)&=&0\,,
\eeqn
where,
\beqn
V_{\mu\nu}(g,\varphi)&\equiv&\frac{-2}{\sqrt{g}}\frac{\delta}{\delta g^{\mu\nu}}\Big(\sqrt{g}~ V(g,\varphi)\Big)\,.
\eeqn
In order to see the equivalence, consider a gauge transformation $\delta x^\mu=\xi^\mu$ of the action \eqref{HRMG} involving St\"uckelberg fields,
\begin{align}
&\delta S=m_g^{2}\int\mathrm{d}^4x\sqrt{g}\left[\left( \mathcal{G}_{\mu\nu}
+m^2V_{\mu\nu}\right)\delta g^{\mu\nu}
-2m^2\frac{\delta V}{\delta\varphi^\alpha}\delta\varphi^\alpha\right]\,,
\end{align}
where $\mathcal{G}_{\mu\nu}$ is the Einstein tensor and the variations of the fields under the gauge transformation are $\delta g^{\mu\nu}=2\nabla^{(\mu} \xi^{\nu)}$ and $\delta\varphi^\alpha=-\xi^\mu\partial_\mu\varphi^\alpha$. The action is now invariant under GCTs, $\delta S=0$. Thus integrating by parts and using the Bianchi identity, $\nabla^\mu\mathcal{G}_{\mu\nu}=0$, we can derive the following identity,
\beqn\label{coordconst}
\nabla^\mu V_{\mu\nu} = \frac{\delta V}{\delta\varphi^\alpha}\partial_\nu\varphi^\alpha\,.
\eeqn
Since the St\"uckelberg fields were introduced to mimic nonsingular coordinate transformations, the matrix $\partial_\nu\varphi^A$ is invertible and hence the Bianchi constraint $\nabla^\mu V_{\mu\nu}=0$ is equivalent to the $\varphi^\alpha$ equations of motion~(\ref{varphieoms}). As expected, the redundant gauge degrees of freedom $\varphi^\alpha$ therefore do not introduce new dynamics into the theory.

From these general arguments it is obvious that the consistency proof of~\cite{Hassan:2011hr, Hassan:2011tf} has to be valid for both the gauge fixed and the gauge invariant version of massive gravity. This was eventually confirmed in a Hamiltonian analysis where the constraint was shown to exist in the massive gravity action (with $\beta_2=\beta_3=0$) including St\"uckelberg fields~\cite{Hassan:2012qv}.

\subsection{Potential shortcomings of massive gravity}\label{sec:shortc}

In order to conclude the presentation of nonlinear massive gravity, we list a few drawbacks of the theory in its present form. These can be viewed as motivations to look for possibilities of going beyond the setup with fixed reference metric.

\begin{itemize}

\item The reference metric $\fmn$ in the general massive gravity action~(\ref{HRMG}), or $\emn$ in the dRGT theory~(\ref{SdRGT}), is not dynamically determined. It is put into the theory by hand and there is no obvious fundamental principle determining its form. From a field theoretical point of view, this and the related fact that the theory (without additional fields) breaks diffeomorphism invariance are rather undesirable features and it is not entirely clear how to interpret the presence of the fixed reference metric. The notion of a pre-geometric structure clearly goes against the main spirit of GR.

\item Possibly related to the above point is the occurrence of superluminal and in particular acausal propagation in massive gravity~\cite{Deser:2012qx, Deser:2013eua, Deser:2013qza, Deser:2014hga, Deser:2014fta, Deser:2015wta}.\footnote{The analyses of these references do not extend to bimetric theory but in prinicple it is possible that similar problems arise there as well.} This raises serious questions for the physical viability of the theory, but see~\cite{deRham:2014zqa} for a discussion of some counter-arguments.

\item The equations of motion of massive gravity (with flat or general reference metric) cannot give rise to homogeneous and isotropic solutions that lead to a viable cosmology~\cite{D'Amico:2011jj, Gumrukcuoglu:2011ew, Gumrukcuoglu:2011zh, Vakili:2012tm, DeFelice:2012mx, Fasiello:2012rw, DeFelice:2013awa}. Lacking this feature, the model cannot serve as a serious alternative to GR.

\item More generally, in the parameter space of massive gravity there is no good limit which brings the equations and their solutions close to those of GR, which are well-tested. In principle, this issue could be resolved by the Vainshtein mechanism, but this causes serious tension with the cosmological Higuchi bounds~\cite{Fasiello:2012rw, Fasiello:2013woa}. The underlying reason is that the metric that couples to matter contains additional degrees of freedom with respect to GR. This generically changes the physics significantly, as becomes apparent through the vDVZ discontinuity already at the linearised level.
\end{itemize}

From our point of view, the above shortcomings strongly motivate the extension of the ghost-free massive gravity action to a fully dynamical bimetric theory, which we shall focus on in the remainder of this article.

\newpage
\section{Ghost-Free Bimetric Theory}\label{sec:HRbim}

A natural question that arises within the formulation of massive gravity with general reference metric~$\fmn$ is whether this second metric could be dynamical on its own without spoiling the consistency of the theory. The fact that one can extend the theory by a kinetic term and an equation of motion for $\fmn$ without reintroducing the Boulware-Deser ghost is not immediately evident from the consistency proofs of~\cite{Hassan:2011hr, Hassan:2011tf, Hassan:2012qv}. Nevertheless, Hassan and Rosen were able to show that one can indeed augment the action by an Einstein-Hilbert term for the second metric and let it be determined dynamically~\cite{Hassan:2011zd}. It has subsequently been confirmed that this also seems to be the unique kinetic term which can be added to the nonlinear massive gravity action in order to give dynamics for $\fmn$~\cite{deRham:2013tfa, deRham:2015rxa, Matas:2015qxa}.\footnote{Note that uniqueness of course only holds up to field redefinitions. The precise statement should therefore be that, assuming an Einstein-Hilbert term for $\gmn$ and a potential of the form given in~(\ref{HRMG}), the only possible kinetic term for $\fmn$ is also of the Einstein-Hilbert form. Here we do not discuss the possibility of including additional non spin-2 degrees of freedom. For example, by adding extra scalar degrees of freedom one can consider $f(R)$ extensions of the theory (see e.g.~\cite{Nojiri:2012zu, Nojiri:2012re, Kluson:2013yaa, Bamba:2013fha, Nojiri:2015qyc}).} In this section we shall present the resulting ghost-free bimetric theory, review its consistency proof and discuss some of its most important features.

\subsection{Action and equations of motion}
\label{sec: Bimetricdetails}

The ghost-free action for Hassan-Rosen bimetric theory is given through~\cite{Hassan:2011zd}
\begin{align}\label{SbmHR}
	S_\mathrm{HR} =\, m_g^2\int\td^4x\sqrt{g}\,R(g)&+m_f^2\int\td^4x\sqrt{f}\,R(f)\nn\\
	&-2m^4\int\td^4x\sqrt{g}\sum_{n=0}^4\beta_ne_n\left(\sqrt{g^{-1}f}\right)\,,
\end{align}
where $m_f$ is the ``Planck mass" for $\fmn$. It corresponds to the massive gravity action~(\ref{HRMG}), extended by an Einstein-Hilbert term for the reference metric $\fmn$ which is promoted to a dynamical field on the same footing as $\gmn$. Consistency of the action now of course requires the inverse $f^{\mu\nu}$ to exist which, in analogy with GR, we shall always assume except perhaps at isolated physical singularities. The corresponding equations of motion for the two metrics read,
\begin{subequations}\label{bimeom}
\beqn
	R_{\mu\nu}(g)-\frac1{2}\gmn R(g)+\frac{m^4}{m_g^{2}}V^g_{\mu\nu}(g,f;\beta_n)
	&=&0\,,\label{geomp}
\\
	R_{\mu\nu}(f)-\frac1{2}\fmn R(f)+\frac{m^4}{m_f^{2}}V^f_{\mu\nu}(g,f;\beta_n)
	&=&0\,,\label{feomp}
\eeqn
\end{subequations}
in which the contributions from the interaction potential in terms of $S=\sqrt{g^{-1}f}$ are of the form,
\beqn\label{potconbim}
	V^g_{\mu\nu}(g,f;\beta_n)&=&g_{\mu\rho}\sum_{n=0}^3(-1)^n\beta_n(Y_{(n)})^\rho_{~\nu}(S)\,,\\
	V^f_{\mu\nu}(g,f;\beta_n)&=&f_{\mu\rho}\sum_{n=0}^3(-1)^n\beta_{4-n}(Y_{(n)})^\rho_{~\nu}(S^{-1})\,.
\eeqn
The matrix functions $Y_{(n)}(S)$ have already been defined in (\ref{yndef}). Due to the overall covariance of the interaction potential there is an identity between the divergences of these interaction contributions~\cite{Damour:2002ws} (note that this identity follows from covariance and is otherwise independent of the form of the bimetric interactions),
\be\label{bianchiId}
\sqrt{g}\,g^{\mu\rho}\nabla_\rho V^g_{\mu\nu}=-\sqrt{f}\,f^{\mu\rho}\tilde\nabla_\rho V^f_{\mu\nu}\,,
\ee
where $\tilde{\nabla}$ is the covariant derivative compatible with~$\fmn$.
Due to this identity there is only one set of independent Bianchi constraints just as in massive gravity. Two important remarks are in order: Firstly, note that the $\beta_4$-term $\sqrt{g}\,e_4(S)=\sqrt{f}$ is no longer non-dynamical but now contributes to the $\fmn$ equations of motion. Secondly, due to the more general symmetry property of the elementary symmetric polynomials (which is just a rewriting of~\eqref{enSSI}), 
\beqn\label{esymprop}
\sqrt{g}\,e_n\big(\sqrt{g^{-1}f}\,\big)=\sqrt{f}\,e_{4-n}\big(\sqrt{f^{-1}g}\,\big) \,,
\eeqn
the structure of the above action is symmetric in the two metrics. At the level of equations, this symmetry manifests itself through the identity,
\beqn\label{sympotcontr}
V^f_{\mu\nu}(g,f;\beta_n)=V^g_{\mu\nu}(f,g;\beta_{4-n})\,.
\eeqn
The metrics are therefore treated on the same footing and in section~\ref{sec:matter} we will see how a ``physical metric" is selected only by the choice of matter couplings.
Various further aspects of the Hassan-Rosen action will be discussed in more detail in the remaining parts of this review.

\subsection{Absence of ghost}\label{sec:aogbt}

The interaction potential of bimetric theory breaks the two diffeomorphism invariances of $\gmn$ and $\fmn$ down to their diagonal subgroup. In other words, the bimetric action is not invariant under independent coordinate transformations of the two metrics but only under those $\delta x^\mu=\xi^\mu$ that transform both metrics simultaneously in the same way: $\delta_\xi\gmn=-2g_{\rho(\mu}\nabla_{\nu)}\xi^{\rho}$ and $\delta_\xi\fmn=-2f_{\rho(\mu}\tilde{\nabla}_{\nu)}\xi^{\rho}$, where $\tilde{\nabla}$ is again the covariant derivative compatible with~$\fmn$. For a general interaction potential, the degree of freedom counting therefore goes as follows: There are $2\times 10=20$ components to start with; $2\times 4=8$ of these get removed by gauge constraints and gauge fixing. 
Just as in the massive gravity case there is one set of Bianchi constraints, which can be taken to be either $\nabla^\mu V^g_{\mu\nu}=0$ or $\tilde{\nabla}^\mu V^f_{\mu\nu}=0$, since these are related by the identity \eqref{bianchiId}. These vector constraints thus removes four additional degrees of freedom, leaving us with a total number of eight propagating modes. These correspond to the two degrees of freedom of a massless spin-2 field, five of a massive spin-2 and one additional scalar which gives rise to the Boulware-Deser ghost instability. In a consistent bimetric theory we therefore also need an additional constraint that eliminates the ghost mode.  

In the ADM language, this means that we need the action to be linear in both lapses $N$ and $L$ of the two metrics as well as one set of three-dimensional vectors. In total there will then be five non-dynamical variables whose equations of motion become constraints: four corresponding to the gauge constraints associated to the overall diffeomorphism invariance and one extra constraint that removes the Boulware-Deser ghost.

Using the same variables as for massive gravity (c.f.~section~\ref{sec:construction}) resulting from the redefinition~(\ref{shiftred}), Hassan and Rosen were able to show that the bimetric action~(\ref{SbmHR}) indeed assumes the form~\cite{Hassan:2011zd},
\beqn
S_\mathrm{HR} = \int\td^4x~\big( L_i\mathcal{C}^i+L\tilde{\mathcal{C}}+ N\mathcal{C} \big)\,,
\eeqn
in which the constraints $\mathcal{C}^i$, $\tilde{\mathcal{C}}$ and $\mathcal{C}$ are independent of the lapses $N$ and $L$ and the shifts $L^i$. As in the massive gravity case, the action is nonlinear in the redefined shift vectors $n^i$, which is a consequence of the breaking of one set of general coordinate transformations.  The scalar constraints $\tilde{\mathcal{C}}$ and $\mathcal{C}$ contain terms coming from the interaction potential, whereas the vector constraints $\mathcal{C}^i$ entirely originate from the Einstein-Hilbert terms. Note that due to the redefinition of the form $N^i=Ln^i+L^i+N{D^i}_kn^k$, all constraints receive contributions from the Einstein-Hilbert term for $\gmn$ which originally contains a term $N_iR^i$, c.f.~(\ref{gradm}).
\\
These results have subsequently been confirmed by explicit calculations and independently verified in various other approaches (see e.g.~\cite{Hassan:2012qv, Alexandrov:2012yv, Alexandrov:2013rxa, Kugo:2014hja}) which we shall not review here. Instead we note that the key observation which motivated Hassan and Rosen to study the fully dynamical extension of massive gravity was that due to the symmetry property~(\ref{esymprop}), linearity of the interaction potential in the lapse $N$ implies that it must also be linear in $L$.\footnote{There could of course be terms proportional to $NL$, but a closer inspection of the structure of the matrices $\mA$ and $\mB$ in~(\ref{exprab}) as well as the interactions in~(\ref{completep}) reveals that such terms do not arise.} The corresponding secondary constraint which removes the canonical momentum of the ghost mode was also shown to exist~\cite{Hassan:2011ea}. Bimetric theory therefore gives rise to the correct amount of constraints in order to propagate the $5+2=7$ degrees of freedom, corresponding to a massive and a massless spin-2 field. 

At the nonlinear level there is no unique split of degrees of freedom into ``massive" and ``massless" ones. As discussed in the following, the mass eigenstates can only be properly defined in the linearised theory around particular backgrounds.

\subsection{Mass spectrum}

The notion of mass is intimately related to the isometries of space-time. Its definition is most concise in Minkowski space where mass arises as a Casimir invariant of the Poincar\'e isometry group and it is possible to generalise that concept to space-times with the same amount of symmetries, i.e.~Anti-de Sitter and de Sitter isometries. For less symmetric spaces it becomes more difficult to obtain a clear identification of mass, but one option is to classify a field as massless or massive depending on its number of propagating degrees of freedom. More precisely, if there is a parameter which when taken to zero increases the amount of gauge redundancy and thus reduces the number of propagating degrees of freedom to that of a massless theory then that parameter can loosely be identified with mass. This notion is implicit when we use the term ``nonlinear massive gravity" or when we speak of ``massless" and ``massive" degrees of freedom in bimetric theory. Nevertheless, around their maximally symmetric background solutions, we expect the nonlinear theories to give rise to a well-defined mass spectrum. In bimetric theory, such backgrounds are most easily obtained by making a proportional ansatz for the metrics.\footnote{For general parameter values, there can exist additional maximally symmetric solutions, c.f.~section~\ref{sec:genprop}, but since they do not correspond to proportional metrics, we do not expect the perturbations to be diagonalisable into spin-2 mass eigenstates.}

\subsubsection{Proportional background solutions}\label{sec:propbg}

Probably the simplest and yet remarkably important class of solutions to the bimetric equations of motion in vacuum is obtained by making an ansatz that conformally relates the two metrics, $\bfmn=c(x)^2\bgmn$, where $c(x)$ is a space-time dependent function. Having plugged this ansatz into the equations of motion, we first note that the Bianchi constraint, $\nabla^\mu V^g_{\mu\nu}=0$, immediately enforces $c(x)=$\,const. This is simply because this equation reduces to a polynomial in $c$ with constant coefficients that multiplies $\p_\nu c$. This restricts our ansatz to proportional metrics,
\beqn
\bfmn=c^2\bgmn\qquad \text{with}\quad c=\mathrm{const.}
\eeqn
Using this in the bimetric equation (\ref{bimeom}) we find that they simply reduce to two copies of Einstein's equations,
\begin{align}\label{bgeq}
\mathcal{G}_{\mu\nu}(\bar{g})
+\Lambda_g\bgmn=0\,, \qquad
\mathcal{G}_{\mu\nu}(\bar{g})
+\Lambda_f\bgmn=0 \,.
\end{align}
In this we have defined the cosmological constant contributions arising from the interaction potential,
\beqn\label{lambdas}
\Lambda_g(c)&\equiv&\frac{m^4}{m_g^{2}}\left(\beta_0+3c\beta_1+3c^2\beta_2+c^3\beta_3\right)\,,\nn\\
\Lambda_f(c)&\equiv&\frac{m^4}{(m_f c)^{2}}\left(c\beta_1++3c^2\beta_2+3c^3\beta_3+c^4\beta_4\right)\,,
\eeqn
and the Einstein tensors are the same in both equations since the Einstein tensor is scale invariant, $\mathcal{G}_{\mu\nu}(c^2\bar{g})= \mathcal{G}_{\mu\nu}(\bar{g})$.
Proportional backgrounds thus simply correspond to solutions to Einstein's equations in GR. Importantly, it means that this class of solutions actually captures all the solutions of GR. The difference of the background equations~(\ref{bgeq}) implies that
\beqn\label{propbackeq}
\Lambda_g(c)=\Lambda_f(c)\,.
\eeqn
From (\ref{lambdas}) it is clear that this equation is a polynomial in $c$ with coefficients depending on the $\beta_n$ parameters and $\alpha=m_f/m_g$. In general, it serves to determine the proportionality constant $c$ of our ansatz in terms of the parameters of the theory and thereby specifies the solution completely. An important exception to this generic situation is the partially massless case, which we discuss in section~\ref{sec:PM}.

Apart from being able to capture all solutions of GR, the proportional background solutions are of particular interest because they allow for a definite mass spectrum of fluctuations around them. In general, the linear perturbation equations have a rather complicated structure because in order to derive them one needs to vary the square-root matrix $S=\sqrt{g^{-1}f}$. As we already discussed in section~\ref{sec:linarb} for the massive gravity setup, this is always possible but, for backgrounds giving rise to matrices $\bar{g}^{-1}\bar{f}$ that do not commute with the fluctuations, the resulting expressions are lengthy. In particular, the equations will not contain a mass term with Fierz-Pauli structure which makes it difficult, if not impossible, to uniquely identify the massive field. In contrast, for the proportional backgrounds we have $\bar{g}^{-1}\bar{f}=c^2\mathbb{1}$ which does commute with any other matrix and hence drastically simplifies the perturbation equations which now will exhibit the Fierz-Pauli structure.

\subsubsection{Spectrum of linear mass eigenstates}\label{sec:linspecBG}

We shall now consider small perturbations around the proportional backgrounds for both of the metrics,
\beqn
\gmn&=&\bgmn+\dg_{\mu\nu}\,,\qquad \,\,\fmn~=~c^2\bgmn+\df_{\mu\nu}\,.
\eeqn
The variation of the square-root matrix is easily obtained as,
\beqn
\bar{g}_{\mu\rho}\delta S^\rho_{~\nu}=\frac{1}{2c}\big(\df_{\mu\nu}-c^2 \dg_{\mu\nu}\big)\,.
\eeqn
Plugging these into the bimetric equations~(\ref{bimeom}), keeping only terms linear in the fluctuations and using the background equations, we obtain,
\begin{subequations}
\beqn
{\bar{\mathcal{E}}_{\mu\nu}}^{~~\rho\sigma}\delta{g}_{\rho\sigma}-
\bar\Lambda_g\left(\dg_{\mu\nu}-\tfrac1{2}\bar{G}_{\mu\nu}\bar{G}^{\rho\sigma}\dg_{\rho\sigma}\right)
-N \bar{G}_{\mu\rho}\left( \delta S^\rho_{~\nu}-\delta^\rho_{~\nu} \delta S^\alpha_{~\alpha}\right)&=&0\,,\\
{\bar{\mathcal{E}}_{\mu\nu}}^{~~\rho\sigma}\delta{f}_{\rho\sigma}-
\bar\Lambda_g\left(\df_{\mu\nu}-\tfrac1{2}\bar{G}_{\mu\nu}\bar{G}^{\rho\sigma}\df_{\rho\sigma}\right)
+\alpha^{-2}N\bar{G}_{\mu\rho}\left( \delta S^\rho_{~\nu}-\delta^\rho_{~\nu} \delta S^\alpha_{~\alpha}\right)&=&0\,.
\eeqn
\end{subequations}
Here $N$ depends on $\alpha$, $c$ and the $\beta_n$ parameters and the explicit dependence can be read off from the Fierz-Pauli mass below. Here, we have made use of the fact that the background metric can be conveniently rescaled by a constant without changing the structure of the linearised equations and expressed the equations with respect to a redefined background metric, 
\beqn
\bar{G}_{\mu\nu}\equiv \big(1+\alpha^2 c^2\big)\bgmn\,.
\eeqn
We have also redefined the cosmological constant with respect to this background,
\beqn\label{barlamb}
\bar{\Lambda}_g\equiv \big(1+\alpha^2 c^2\big)^{-1} \Lambda_g  \,, \qquad R_{\mu\nu}(\bar{G})=\bar{\Lambda}_g\bar{G}_{\mu\nu}\,,
\eeqn
and expressed the kinetic operator ${\bar{\mathcal{E}}_{\mu\nu}}^{~~\rho\sigma}$ with respect to the $\bar{G}_{\mu\nu}$ background,
\begin{align}\label{kinopds}
\bar{\mathcal{E}}^{~~\rho\sigma}_{\mu\nu}\delta G_{\rho\sigma} 
=-\tfrac{1}{2}\Big[\delta^\rho_\mu\delta^\sigma_\nu\bar{\nabla}^2
+\bar G^{\rho\sigma}\bar{\nabla}_\mu\bar{\nabla}_\nu 
&-\delta^\rho_\mu\bar{\nabla}^\sigma\bar{\nabla}_\nu
-\delta^\rho_\nu\bar{\nabla}^\sigma\bar{\nabla}_\mu \nn\\
&-\bar{G}_{\mu\nu}\bar G^{\rho\sigma}\bar{\nabla}^2 
+\bar{G}_{\mu\nu}\bar{\nabla}^\rho\bar{\nabla}^\sigma\Big]\delta G_{\rho\sigma}\,,
\end{align}
in which $\bar{\nabla}$ is the covariant derivative compatible with $\bar{G}_{\mu\nu}$.

The above equations can now easily be diagonalised into an equation for a massless and a massive perturbation. To this end, consider the following combinations of metric fluctuations, 
\beqn\label{dGanddM}
\delta G_{\mu\nu} \equiv\dg_{\mu\nu}+\alpha^{2}\df_{\mu\nu}\,,\qquad
\delta M_{\mu\nu} \equiv \frac{1}{2c}\big(\df_{\mu\nu}-c^2 \dg_{\mu\nu}\big)\,.
\eeqn
Now, taking the appropriate linear combinations of the original fluctuation equations for $\delta\gmn$ and $\delta\fmn$ decouples the massless from the massive mode. The resulting equations read~\cite{Hassan:2012wr, Hassan:2012rq},  
\begin{subequations}
\begin{align}
{\bar{\mathcal{E}}_{\mu\nu}}^{~~\rho\sigma}\delta G_{\rho\sigma}-\bar{\Lambda}_g\left(\delta G_{\mu\nu}-\tfrac{1}{2}\bar{G}_{\mu\nu}\delta G\right) &=0\,, \label{dGeq}\\
{\bar{\mathcal{E}}_{\mu\nu}}^{~~\rho\sigma}\delta M_{\rho\sigma}-\bar{\Lambda}_g\left(\delta M_{\mu\nu}-\tfrac{1}{2}\bar{G}_{\mu\nu}\delta M\right)
+\tfrac{\bar{m}^2_\mathrm{FP}}{2}\Big(\delta M_{\mu\nu}-\bgmn\delta M\Big)&=0\,,\label{dMeq}
\end{align}
\end{subequations}
where $\delta G=\bar{G}^{\mu\nu}\delta G_{\mu\nu}$ and $\delta M=\bar{G}^{\mu\nu}\delta M_{\mu\nu}$. The Fierz-Pauli mass in these equations is given by,
\beqn\label{FPmass}
\bar{m}^2_\mathrm{FP}=\frac{(1+(\alpha c)^2)^2}{\alpha^2 c}N=
\frac{m^4}{m_g^{2}} \frac{1}{\alpha^2 c^2}\left(c\beta_1 +2c^2\beta_2+c^3\beta_3\right)\,.
\eeqn
We remind that in all of the above expressions $c$ is to be regarded as a function of the Planck masses and the $\beta_n$ parameters, determined by the background equation $\Lambda_g=\Lambda_f$. 

As advertised, (\ref{dGeq}) and (\ref{dMeq}), respectively,  describe a massless and a massive spin-2 fluctuation in (Anti-) de Sitter background. At the linear level around proportional backgrounds, one can therefore assign two degrees of freedom to a massless fluctuation $\delta G_{\mu\nu}$ and the remaining  five to a massive fluctuation $\delta M_{\mu\nu}$. The linearised action in terms of the mass eigenstates is~\cite{Hassan:2012rq}, 
\beqn
S_\mathrm{PB}=\tfrac{1}{2}\int\td^4x\Big[ \delta G_{\mu\nu}\bar{\mathcal{E}}^{\mu\nu\rho\sigma}\delta G_{\rho\sigma}&-&\bar{\Lambda}_g\left(\delta G^{\mu\nu}\delta G_{\mu\nu}-\tfrac{1}{2}\delta G^2\right)\nn\\
+\delta M_{\mu\nu}\bar{\mathcal{E}}^{\mu\nu\rho\sigma}\delta M_{\rho\sigma}&-&\bar{\Lambda}_g\left(\delta M^{\mu\nu}\delta M_{\mu\nu}-\tfrac{1}{2}\delta M^2\right)\nn\\
&+&\tfrac{\bar{m}^2_\mathrm{FP}}{2}\Big(\delta M^{\mu\nu}\delta M_{\mu\nu}-\bgmn\delta M^2\Big) \Big]\,.
\eeqn
The main reason for choosing the new background $\bar{G}_{\mu\nu}$ was to render the final action and equations as simple as possible. Alternatively, we could have written all of the above expressions with respect to the original background $\bgmn$, in terms of $\Lambda_g$ in (\ref{lambdas}) and a properly rescaled Fierz-Pauli mass, $m^2_\mathrm{FP}\equiv \big(1+\alpha^2 c^{2}\big)\bar{m}^2_\mathrm{FP}$.


\subsection{Couplings to Matter}\label{sec:matter}

So far we have been dealing with theories involving only spin-2 degrees of freedom. In order to be accessible to any type of observations or experiments, the fields need to interact with ordinary matter. It may not come as a surprise that the set of allowed matter couplings which do not reintroduce the Boulware-Deser ghost is very restricted.

\subsubsection{Ghost-free matter couplings}

The only known couplings which can be added to the bimetric action~(\ref{SbmHR}) without exciting the ghost are,
\beqn\label{consistmc}
S_\mathrm{m}=\int\td^4 x\,\sqrt{g}~\mathcal{L}_\mathrm{m}(g,\Phi_g)+\int\td^4 x\,\sqrt{f}~\tilde{\mathcal{L}}_\mathrm{m}(f,\Phi_f)\,,
\eeqn
where $\mathcal{L}_\mathrm{m}$ and $\tilde{\mathcal{L}}_\mathrm{m}$ are standard minimally coupled matter Lagrangians of the same form as in GR. $\Phi_g$ and $\Phi_f$ schematically stand for sets of matter fields of any kind. Importantly, it was shown in~\cite{Yamashita:2014fga, deRham:2014naa} that it is not possible to couple the same (dynamical) matter field to both metrics using minimal couplings, and hence $\Phi_g$ and $\Phi_f$ must be entirely independent. That coupling a field to both $\gmn$ and $\fmn$ reintroduces the Boulware-Deser ghost can be understood as follows: In GR with only one metric, the matter action becomes linear in the lapse $N$ when written in the canonical variables for the metric and the matter fields. A simple calculation in an example with a free scalar field $\varphi$ shows that this happens because the variation of the action with respect to $\dot\varphi$ and hence also the canonical momentum of the scalar depend on $N$. This $N$-dependence is such that the action in terms of canonical variables is linear in $N$. In the bimetric case, however, when the free scalar is coupled to both $\gmn$ and $\fmn$, its canonical momentum will depend on $N$ and $L$ in a more complicated way and the action will not become linear in both of the lapses. As a consequence, the constraint which removes the Boulware-Deser ghost is lost. Note that this argument does not exclude coupling pure interaction terms (without appearance of time derivatives) to both of the metrics. This possibility seems quite contrived, however, and we do not discuss it any further but instead focus on couplings to two entirely independent matter sectors. 

In the equations of motion the matter couplings enter in the form of stress-energy tensors,
\beqn
T^g_{\mu\nu}\equiv-\frac{1}{\sqrt{g}}\frac{\delta \big(\sqrt{g}~\mathcal{L}_\mathrm{m}(g,\Phi_g)\big)}{\delta g^{\mu\nu}}\,,\qquad
T^f_{\mu\nu}\equiv-\frac{1}{\sqrt{f}}\frac{\delta \big(\sqrt{f}~\mathcal{L}_\mathrm{m}(f,\Phi_f)\big)}{\delta f^{\mu\nu}}\,.
\eeqn
The bimetric equations~(\ref{bimeom}) in the presence of matter sources thus become
\begin{subequations}\label{mateom}
\beqn
	\mathcal{G}_{\mu\nu}(g)+\frac{m^4}{m_g^{2}}V^g_{\mu\nu}(g,f;\beta_n)
	&=&\frac{1}{m_g^2}T^g_{\mu\nu}\,,\label{geomm}
\\
	\mathcal{G}_{\mu\nu}(f)+\frac{m^4}{m_f^{2}}V^f_{\mu\nu}(g,f;\beta_n)
	&=&\frac{1}{m_f^2}T^f_{\mu\nu}\,.\label{feomm}
\eeqn
\end{subequations}
We will comment on other matter couplings that have been studied in the literature below but first we shall discuss two important limits in the parameter space of bimetric theory which take it close to either general relativity or nonlinear massive gravity.

\subsubsection{General relativity limit}\label{sec:GRlim}

As we pointed out in section~\ref{sec:shortc}, the parameter space of massive gravity with a fixed reference metric does not include any region which is obviously close to GR. In the following we will see that, in contrast, a well-defined GR limit does exist for the bimetric theory. Since the structure of the bimetric action is completely symmetric in $\gmn$ and $\fmn$, either of the two metrics can play the role of the physical metric whose solutions will become similar to those of GR. We choose this metric to be $\gmn$ here but note that everything can analogously be derived with the roles of the metrics interchanged.

It is straightforward to verify that the contributions from the interaction potential $V$ to the bimetric equations of motion~(\ref{mateom}) satisfy the following identity~\cite{Baccetti:2012bk, Hassan:2014vja},\footnote{In fact, this identity can be derived for any form of covariant potential $V$ and does therefore not rely on the specific structure of the ghost-free bimetric action.}
\beqn\label{Vid}
\sqrt{g}~g^{\mu\rho}V^g_{\rho\nu}+\sqrt{f}~f^{\mu\rho}V^f_{\rho\nu}-\sqrt{g}~V\delta^\mu_{~\nu}=0\,.
\eeqn
Making use of this observation, the equations can be combined to give,
\beqn\label{combeq}
g^{\mu\rho}\mathcal{G}_{\rho\nu}(g)+\alpha^2\det\big(\sqrt{g^{-1}f}\,\big)\,f^{\mu\rho}\mathcal{G}_{\rho\nu}(f)+\frac{m^4}{m_g^2} V \delta^\mu_{~\nu}=\frac{1}{m_g^2}\left(g^{\mu\rho}T^g_{\rho\nu}+f^{\mu\rho}T^f_{\rho\nu}\right)\,.
\eeqn
This particular set of equations has interesting implications on classical solutions in bimetric theory which we shall come back to in section~\ref{sec:classsol}. For our purposes here it suffices to consider the dependence on the parameter $\alpha\equiv m_f/m_g$ in~(\ref{combeq}). In particular we note that in the limit $\alpha\rightarrow 0$, the equations reduce to~\cite{Baccetti:2012bk, Hassan:2014vja},
\beqn
g^{\mu\rho}\mathcal{G}_{\rho\nu}(g)+\frac{m^4}{m_g^2} V \delta^\mu_{~\nu}=\frac{1}{m_g^2}\left(g^{\mu\rho}T^g_{\rho\nu}+f^{\mu\rho}T^f_{\rho\nu}\right)\,.
\eeqn
When $T^f_{\mu\nu}=0$, i.e.~when there is no matter sector for the metric $\fmn$, we can use the covariant derivative compatible with $\gmn$ to take the divergence of the equations which for a covariantly conserved source then gives $V=$ constant on-shell. In this case, the equations thus reduce to Einstein's equations for the physical metric $\gmn$ with Planck mass $m_g$ and cosmological constant~$\frac{m^4}{m_g^2}V$,
\beqn\label{greql}
\mathcal{G}_{\mu\nu}(g)+\frac{m^4}{m_g^2} V \gmn=\frac{1}{m_g^2}T^g_{\mu\nu}\,.
\eeqn
Hence, the GR limit of bimetric theory is defined by,
\beqn 
\alpha\equiv\frac{m_f}{m_g}\longrightarrow 0\,, \qquad
m_g=\mathrm{const.}\,, \qquad 
T^f_{\mu\nu}=0\,,
\eeqn
in which case the solutions for the physical metric $\gmn$ coincide with those of GR.\footnote{One may be worried about occurrences of strong coupling for small values of $\alpha$. However, as has been discussed in detail in~\cite{Akrami:2015qga}, the strong coupling scale of the massive spin-2 mode in fact grows with decreasing $\alpha$.} Interestingly, a large value for the physical Planck mass $m_g$ automatically implies that bimetric theory is close to its GR limit, provided that $m_f$ is of reasonable size compared to other relevant scales (such as the electroweak scale, for instance).

The effect of taking the above limit on the $\fmn$ equation~(\ref{feomm}) is that it becomes purely algebraical, $V_{\mu\nu}^f=0$. The generic solutions to this equation are proportional backgrounds $\fmn=c^2\gmn$ with $c$ determined by the condition $\Lambda_f(c)=0$, where the function $\Lambda_f$ is the same as in~(\ref{lambdas}). Then the cosmological constant in~(\ref{greql}) is given by, 
\beqn
\frac{m^4}{m_g^2} V=\frac{m^4}{m_g^2} \left(\beta_0+4\beta_1c+6\beta_2c^2+4\beta_3c^3+\beta_4c^4\right)\,.
\eeqn
We further observe that, in the GR limit, the fluctuation of the physical metric $\gmn$ becomes massless as expected. This can be seen from (\ref{dGanddM}), which for $\alpha\rightarrow0$ gives $\delta G_{\mu\nu}\rightarrow\delta\gmn$. It is an important feature because, at least to our present knowledge, only the couplings of the original metrics to matter are ghost-free, whereas on the other hand the gravitational metric needs to behave like a massless spin-2 field for phenomenological reasons. In particular, since the massive mode decouples from the source, bimetric theory in the GR limit does not suffer from the vDVZ discontinuity and hence does not need to rely on the Vainshtein mechanism, a requirement which usually challenges the phenomenological viability of massive gravity models.

To summarise, in the limit of small $\alpha$, bimetric theory can be viewed as a smooth deformation of GR because the massive mode decouples in the limit. In particular, the gravitational metric satisfies Einstein's equations modified by a small correction and its fluctuations are dominated by the massless spin-2 mode. The dynamics of the massive field essentially decouple from the observable matter sector and its presence manifests itself only through the constant term in (\ref{greql}) which for a sufficiently small spin-2 mass could give rise to cosmic self-acceleration~\cite{Akrami:2015qga}. On the other hand, one can tune the $\beta_n$ parameters such that large contributions to the effective cosmological constant cancel (at the cost of giving up technical naturalness \`a la 't Hooft) and thus, even for a very large spin-2 mass, its observable effects could remain small.

Note also that, in this setup, the massive spin-2 field could potentially be a suitable dark matter candidate: It interacts with the Standard Model fields only very weakly, but couples non-minimally to the gravitational metric. A remarkable feature of this scenario would be that the closeness of the theory to GR goes hand in hand with the decoupling of the dark matter field, and both get related to the largeness of the Planck scale $m_g$. For related approaches in the context of bimetric theory, where not the spin-2 field itself is considered as the dark matter candidate but additional fields are invoked, see~\cite{Aoki:2014cla,Bernard:2014psa, Blanchet:2015sra, Blanchet:2015bia}.

\subsubsection{Massive gravity limit}\label{sec:mglimit}

Let us now consider the limit opposite to the one above, namely $\alpha\rightarrow \infty$. In this limit, bimetric theory reduces to massive gravity with a GR reference metric, i.e.~$\fmn$ solves the standard sourced Einstein equations~\cite{Volkov:2011an,vonStrauss:2011mq,Comelli:2011zm,Park:2012cq}. To see this, consider once more the bimetric equations of motion~(\ref{mateom}), this time along with the following scalings,
\beqn 
\alpha\equiv\frac{m_f}{m_g}\longrightarrow \infty\,, \qquad
m_g&=&\mathrm{const.}\,,\qquad 
\frac{1}{M^2}\tilde{T}_{\mu\nu}\equiv\frac{1}{m_f^2}T^f_{\mu\nu}=\mathrm{const.}\,,\nn\\
\beta_4'&\equiv&\frac{m_g^2}{m_f^2}\beta_4=\mathrm{const.}\,, \qquad
\beta_n=\mathrm{const.}~~\text{for}~n\leq3\,.
\eeqn
Note that we introduced a new mass scale $M$ and stress-energy $\tilde T$ here because now we also require a scaling of the matter fields in the $\fmn$ sector in order to be able to treat solutions for $\fmn$ that solve Einstein's equations in the presence of matter (for more on this see \eg discussions in~\cite{vonStrauss:2011mq,Baccetti:2012bk,Hassan:2012wr}). Similarly, the scaling of $\beta_4$ is required to keep a cosmological constant term for $\fmn$. The remaining $\beta_n$ are not allowed to scale since this would destroy interactions in the $\gmn$ equations (recall that $\beta_4$ only appear in the $\fmn$ equations).
The $\gmn$ equations~(\ref{geomm}) are then unaffected but in the $\fmn$ equations~(\ref{feomm}) all bimetric interaction terms drop out. Thus, $\fmn$ is now determined by,
\beqn\label{eefmn}
\mathcal{G}_{\mu\nu}(f)+\frac{m^4}{m_g^{2}}\beta'_4\fmn	&=&\frac{1}{M^2}\tilde{T}_{\mu\nu}\,,
\eeqn
which is an Einstein equation with cosmological constant $\frac{m^4}{m_g^{2}}\beta'_4$ and Planck mass $M$.
The exact limit thus decouples the dynamics of $\fmn$ which is now determined by equations that do not involve $\gmn$. We can obtain a solution to these equations and use it to replace $\fmn$ in the $\gmn$ equations. Effectively, the $\gmn$ equations are therefore the same as the ones obtained from varying the massive gravity action~(\ref{HRMG}), in which $\fmn$ is taken to be a fixed reference metric that solves Einstein's equations~(\ref{eefmn}). This picture explains the emergence of the fixed reference metric in massive gravity since the latter can be viewed as a particular point in the parameter space of the fully dynamical bimetric theory.

From the existence of the above limit we can infer that the solution space of bimetric theory is richer than that of massive gravity. Solutions $(g,f)$ to the bimetric equations of motion can be of the form $(g'+\mathcal{O}(\alpha^{-1}),f'+\mathcal{O}(\alpha^{-1}))$, with $(g',f')$ taken to be $\alpha$-independent. Only for such solutions is the limit $\alpha\rightarrow\infty$ well-defined and result in the massive gravity configurations $(g',f')$. 
Other bimetric solutions, however, do not possess a well-defined massive gravity limit, i.e.~they become singular for $\alpha\rightarrow\infty$. Such configurations for the metrics are known to exist and have no massive gravity counterpart, but rather constitute a distinct feature of bimetric theory. Simple examples of proportional solutions that become singular in the limit have been found in~\cite{Hassan:2014vja}. Moreover, see~\cite{Zhang:2014wia} for a study of Hawking-Moss instanton solutions in bimetric theory which do not seem to allow for a well-defined massive gravity limit. 

To summarise, all solutions of massive gravity can be viewed as arisen from bimetric theory, whereas it is quite easy to find solutions of the bimetric theory which does not have any massive gravity counterpart. Therefore, since the limit is singular, care has to be taken when arguing that results which hold in the massive gravity limit also hold in the full theory.

It has occasionally been argued that the above limiting procedure is somehow inferior and that the limit should instead be taken in the action. But, in fact, the two procedures of taking the limit in the equations or in the action are fully compatible when treated correctly. To see this let us briefly review the limiting procedure at the level of the action, as discussed, for instance, in~\cite{deRham:2014zqa}. In this case one starts by expanding the bimetric action \eqref{SbmHR} around, e.g.,~a constant curvature solution,\footnote{More generally one can couple appropriately scaled matter fields to the $\fmn$ metric and consider expanding around a solution of \eqref{eefmn}, but this does not change the main arguments.}
\be
\fmn\rightarrow \fmn^{\mathrm{E}}+\frac{\delta\fmn}{m_f}\,,
\ee 
and then considers the limit $m_f\rightarrow\infty$. Since $\fmn^{\mathrm{E}}$ is assumed to be a constant curvature metric the kinetic term for $\fmn$ simply reduces to the quadratic action for canonically normalised fluctuations on this background, i.e.
\be
m_f^2\sqrt{f}\,R(f)\rightarrow
\tfrac{1}{2}\delta\fmn\bar{\mathcal{E}}^{\mu\nu\rho\sigma}\delta f_{\rho\sigma}\,,
\ee
where $\bar{\mathcal{E}}^{\mu\nu\rho\sigma}$ is the operator defined in \eqref{kincurv}, written in terms of the metric $\fmn^{\mathrm{E}}$ and its curvatures. In the interaction potential on the other hand, all the $\delta\fmn$ fluctuations decouple in the limit and, assuming an appropriate scaling of $\beta_4$, only a cosmological term for the fluctuations remains. One then ends up with a non-covariant action in terms of a decoupled linear spin-2 field $\delta\fmn$ and a nonlinear spin-2 field $\gmn$ whose interaction potential contains the fixed reference metric $\fmn^{E}$. Now varying with respect to the dynamical fields $\gmn$ and $\delta\fmn$ results in the massive gravity equations for $\gmn$ (containing $\gmn$ and $\fmn^{E}$), supplemented with a completely decoupled linear equation for $\delta\fmn$. Of course, this procedure is only self-consistent provided that $\fmn^{\mathrm{E}}$ is really a finite constant curvature background solution in the limit. Ensuring this leads exactly to the limiting procedure of the equations we have discussed above. It is also straightforward to see that if we express our equations~\eqref{geomm} and~\eqref{eefmn} in the limit in terms of $\gmn$, $\delta\fmn$ and $\fmn^{E}$, then we end up with exactly the same result as that obtained from the action (the reason being that, if $\delta\fmn/(\alpha m_g)$ is assumed subdominant to $\delta\gmn/m_g$ in the action, the same will of course hold in the equations). Therefore taking the limit in the equations is equivalent to doing it in the action, but the former procedure deals more directly with solutions and the requirements for these to exist.

\subsubsection{Other matter couplings}

Matter couplings differing from the consistent ones in~(\ref{consistmc}) have also been studied in the literature. Particular attention has been paid to the ``doubly coupled" theory for which the matter sectors of $\gmn$ and $\fmn$ contain the same fields~\cite{Akrami:2013ffa, Akrami:2014lja} but, as explained above, these couplings reintroduce the Boulware-Deser ghost~\cite{Yamashita:2014fga, deRham:2014naa}. On the other hand, it turns out that a certain combination of the two metrics can be coupled to matter without exciting the ghost in the low-energy effective field theory~\cite{deRham:2014naa}. This ``effective metric" contains two arbitrary parameters $a$ and $b$ and is of the form,
\beqn\label{Geff}
G^\mathrm{eff}_{\mu\nu}=a^2\gmn +2ab\, g_{\mu\rho}\big(\sqrt{g^{-1}f}\,\big)^\rho_{~\nu}+b^2\fmn\,,
\eeqn
whose ``uniqueness"\footnote{More precisely, the metric \eqref{Geff} is unique up to multiplication of the right-hand side by an arbitrary matrix of unit determinant. When it is expressed in terms of vierbeins more ambiguities arise~\cite{Melville:2015dba}.} has been discussed in~\cite{Huang:2015yga, Heisenberg:2015iqa}. The reason this metric is special is because it can be written, in matrix notation, $G^{\mathrm{eff}}=a^2\,g\,(\mathbb{1}+(b/a)S)^2$, and hence $\sqrt{\det(G^{\mathrm{eff}})}=a^4\sqrt{g}\,\det(\mathbb{1}+(b/a)S)$. The identity \eqref{app:detrel} then reveals that any vacuum contribution generated by matter coupled to such a $G^{\mathrm{eff}}$ will not alter the ghost-free form of the bimetric interaction potential. Interestingly, this metric can be written as a Finsler metric \cite{Akrami:2014lja} and therefore provides a situation where the geometric interpretation is shifted from the standard (pseudo) Riemannian description to an effective Finsler geometry.\footnote{A Finsler geometry departs from the pseudo Riemannian geometry in that it characterises a manifold with a norm but it is not necessarily infinitesimally Minkowski~\cite{Cartan:1934, Bekenstein:1992pj}.} Since our main interest here lies in working with the full bimetric action and not in the effective field theory picture, we will not discuss phenomenological implications of such a matter coupling in this review. The interested reader is referred to the large variety of references~\cite{Hassan:2014gta, Schmidt-May:2014xla, deRham:2014fha, Solomon:2014iwa, Enander:2014xga, Gumrukcuoglu:2014xba, Heisenberg:2014rka, Gao:2014xaa, Gumrukcuoglu:2015nua, Comelli:2015pua, Blanchet:2015sra, Blanchet:2015bia, Lagos:2015sya} in which the effective coupling has been studied further, mostly in the context of cosmology.

\newpage
\section{Classical Solutions} \label{sec:classsol}

Despite the fact that the consistent theories have only been known for a few years, there already exists an extensive literature on classical solutions in ghost-free massive gravity and bimetric theory. Here we focus on bimetric theory whose solution spectrum is richer than that of massive gravity (c.f.~section~\ref{sec:mglimit}). Perhaps most interesting is the class of spherically symmetric solutions, which can potentially be used to study stars, galaxies, black holes and cosmology. An immediate problem that arises in this context is that the bimetric theory has no known analogue of Birkhoff's theorem\footnote{Which implies absence of monopole radiation in GR. See~e.g.~\cite{Deser:2004gi} for a nice GR oriented discussion which also gives appropriate credit to earlier independent findings of this important theorem.} and therefore many of the valuable uniqueness theorems of GR do not apply. This means that in many situations several solutions may exist for the same problem and one is left with the task of sorting out the most relevant one. A strong physically motivated guide here is to explore the stability of solutions under perturbations. Another complication is the analytical complexity of the nonlinear equations of motion, and in many cases only numerical solutions are known as of yet. In this review we have mainly kept our attention towards analytically tractable problems and will continue this practice. We will therefore restrict our discussion mainly to black hole and cosmological solutions, discuss features of these which can clearly be discerned analytically and only comment briefly on some phenomenological issues. For the spherically symmetric solutions and in particular their applications to black hole studies there already exist a few good reviews on the current status~\cite{Volkov:2013roa,Volkov:2014ooa,Babichev:2015xha}. We refer the interested reader to these for additional details. Before discussing particular features of the spherically symmetric solutions, we make some general remarks.
 
\subsection{General properties}\label{sec:genprop}

Recall the form of the bimetric equations of motion in the presence of matter sources,
\begin{subequations}\label{mateom2}
\beqn
	\mathcal{G}_{\mu\nu}(g)+\frac{m^4}{m_g^{2}}V^g_{\mu\nu}(g,f;\beta_n)
	&=&\frac{1}{m_g^2}T^g_{\mu\nu}\,,\label{geomm2}
\\
	\mathcal{G}_{\mu\nu}(f)+\frac{m^4}{m_f^{2}}V^f_{\mu\nu}(g,f;\beta_n)
	&=&\frac{1}{m_f^2}T^f_{\mu\nu}\,,\label{feomm2}
\eeqn
\end{subequations}
with the contributions from the interaction potential given in~(\ref{potconbim}). It is clear that due to the presence of these additional interaction terms, in general, the solutions to the bimetric equations will significantly differ from those obtained in GR. From a phenomenological point of view, such large new effects are not desirable since Einstein's theory is well-tested over a wide range of distance scales. In order not to be ruled out immediately, any modification of gravity needs to give rise to solutions that do not deviate too much from those of GR. As we already saw in section~\ref{sec:GRlim}, in the limit of small $\alpha\equiv m_f/m_g$ and a vanishing source for $\fmn$, all equations for the physical metric $\gmn$ will smoothly approach the form of the GR equations and so will their linear perturbations. This is already good news for the viability of bimetric theory. On the other hand, it may also render the theory less interesting because if all solutions are (almost) indistinguishable from those of GR in the weak-field limit then there is little room for predicting new signatures that could be observed. 
It is therefore interesting to see if the above equations, away from the GR limit, can still give rise to solutions that behave similarly to those of Einstein's theory.

In section~\ref{sec:propbg} we already encountered the proportional backgrounds as an example of GR solutions in bimetric theory. However, in the presence of matter, these backgrounds only exist if the stress-energy tensors satisfy the rather strict constraint $T^f_{\mu\nu}=\alpha^2 T^g_{\mu\nu}$. As we will see below, the bimetric equations can also reduce to Einstein equations after inserting ans\"atze for the metrics which possess particular symmetry properties (e.g.~spherical symmetry). In these cases, although the background solutions reproduce exactly the predictions of GR, differences will generically occur at the level of perturbations and the theory can in principle make testable predictions.

In this context, it is important to notice one more feature of bimetric theory: If either $\gmn$ or $\fmn$ is assumed to be an exact solution to the Einstein equations, then the bimetric equations will force the second metric to also solve (a different set of) Einstein's equations~\cite{Hassan:2014vja}.\footnote{In fact, this statement does not depend on the ghost-free structure but holds for any covariant bimetric interaction potential.}
In vacuum this can be seen as follows:
Recall that using the identity~(\ref{Vid}), we were able to combine the bimetric equations into,
\beqn\label{combeq1}
g^{\mu\rho}\mathcal{G}_{\rho\nu}(g)+\alpha^2\det\big(\sqrt{g^{-1}f}\,\big)\,f^{\mu\rho}\mathcal{G}_{\rho\nu}(f)+\frac{m^4}{m_g^2} V \delta^\mu_{~\nu}=0\,.
\eeqn
If in this equation we assume that, for instance, $\fmn$ is a GR solution with cosmological constant $\tilde{\Lambda}$,
\beqn
\mathcal{G}_{\mu\nu}(f)=-\tilde{\Lambda}\fmn\,,
\eeqn
then (\ref{combeq1}) takes the form,
\beqn\label{combeq2}
g^{\mu\rho}\mathcal{G}_{\rho\nu}(g)-\alpha^2\det\big(\sqrt{g^{-1}f}\,\big)\tilde{\Lambda}\,\delta^\mu_{~\nu}+\frac{m^4}{m_g^2} V \delta^\mu_{~\nu}=0\,.
\eeqn
Taking the covariant divergence of this equation implies that the terms proportional to $\delta^\mu_{~\nu}$ are constant and hence the equation reduces to an Einstein equation for $\gmn$,
\beqn
\mathcal{G}_{\mu\nu}(g)=-\Lambda\gmn\,,\qquad 
\Lambda\equiv-\alpha^2\det\big(\sqrt{g^{-1}f}\,\big)\tilde{\Lambda}+\frac{m^4}{m_g^2} V\,.
\eeqn
This proof can straightforwardly be generalised to the equations including matter sources. However, in this case GR solutions do not exist unless the sources for $\gmn$ and $\fmn$ are related in a particular way~\cite{Hassan:2014vja}. Solutions of this type (with stress-energy tensors assumed to resemble perfect fluids) have been found in~\cite{Gratia:2013uza}.

Another interesting general feature of the bimetric solutions was noticed in \cite{Baccetti:2012re}. There it was found that the effective stress-energy tensors that the bimetric interactions generate, i.e.~the negative of $V_{\mu\nu}^g$ and $V_{\mu\nu}^f$, never satisfy the null-energy condition simultaneously unless the two metrics are proportional. In more detail, Ref.~\cite{Baccetti:2012re} found that if $k^\mu$ is a null-vector with respect to $\gmn$, i.e.~$\gmn k^\mu k^\nu=0$, then one can always define $\bar{k}^\mu=(S^{-1})^\mu_{~\rho}k^\rho$ which turns out to be a null-vector with respect to $\fmn$, i.e.~$\fmn\bar{k}^\mu\bar{k}^\nu=\gmn k^\mu k^\nu=0$. This follows since $\fmn=g_{\mu\rho}S^\rho_{~\sigma}S^\sigma_{~\nu}=g_{\sigma\rho}S^\rho_{~\mu}S^\sigma_{~\nu}$ and can be used to demonstrate that if e.g.~$V_{\mu\nu}^g$ satisfy the null-energy condition $V_{\mu\nu}^gk^\mu k^\nu\leq0$ (where the direction of the inequality follows from our sign convention in defining $V^g_{\mu\nu}$) then $V_{\mu\nu}^f$ will satisfy an opposite inequality $V_{\mu\nu}^f\bar{k}^\mu\bar{k}^\nu\geq0$. Furthermore the inequalities only saturate for proportional solutions, when the interactions reduce to pure cosmological constant contributions (c.f.~section~\ref{sec:propbg}). Of course the null-energy condition is usually applied to the matter sector within the context of GR so it is not completely obvious what a violation actually means physically in this case. For example, if we consider the case where only $\gmn$ couples to a matter source $T_{\mu\nu}^g$ and take it that $V_{\mu\nu}^f$ does not (does) satisfy the null-energy condition. Then generically $V_{\mu\nu}^g$ will (will not) satisfy it which implies that the standard physical interpretation of the null-energy condition on the source term $T_{\mu\nu}^g$ may change. This interesting observation is something which deserves further study. One immediate consequence however was the prediction of worm-hole solutions in the theory, which were subsequently found and analysed in \cite{Sushkov:2015fma}.

\subsection{Black hole solutions}
Spherically symmetric solutions were first discussed in \cite{Comelli:2011wq, Volkov:2012wp} and have since then received considerable attention \cite{Enander:2013kza, Brito:2013yxa, Brito:2013xaa, Brito:2013wya, Kodama:2013rea, Babichev:2014fka, Babichev:2014tfa, Babichev:2014oua, Enander:2015kda, Ayon-Beato:2015qtt}. As already mentioned in the introduction of this section, there is no known analogue of Birkhoff's theorem in bimetric theory and therefore many of the uniqueness theorems of GR fail straight away. In fact, for spherically symmetric ans\"atze, it is fairly straightforward to make an initial separation into two classes of general solutions, the so-called bidiagonal and non-bidiagonal solutions. These labels are not that imaginative but they do keep to the point: The bidiagonal solutions are solutions for which both metrics can be brought to a simultaneously diagonal form while the non-bidiagonal solutions are those solutions which cannot. Another important class of solutions are the proportional solutions discussed in section \ref{sec:propbg}. Now, we can always use the isometries to choose coordinates such that at least one of the metrics is diagonal and therefore the proportional solutions fall into the broader class of bidiagonal solutions.

The most general spherically symmetric ans\"atze can be written in the form,
\begin{align}\label{spheransatz}
\gmn\td x^\mu\td x^\nu &=-U^2\td t^2+V^2\td r^2+r^2\td\Omega^2\,,\nn\\
\fmn\td x^\mu\td x^\nu &=-A^2\td t^2+B^2\td r^2+C^2\td t\td r+D^2\td\Omega^2\,,
\end{align}
where $U,V,A,B,C,D$ are all functions of the radial ($r$) and temporal ($t$) coordinates. We have made use of the spherical isometry to fix the angular and radial coordinates such that the angular measure $\td\Omega^2=\td\theta^2+\sin^2\theta\td\varphi^2$ is in the standard form and comes with the standard factor of $r^2$ for $\gmn$, which also puts $\gmn$ in diagonal form. This form has been heavily used in the literature and serves to simplify parts of our discussions, but it should be noted that in some situations, concerning in particular black hole solutions, it is preferable to work with coordinates that are regular at the horizon, such as the advanced Eddington-Finkelstein coordinates (see e.g.~\cite{Babichev:2014tfa, Babichev:2015xha}). Since $\gmn$ is diagonal it follows from the equations of motion,
\be
g^{\rho\mu}\mathcal{G}_{\mu\nu}+\frac{m^4}{m_g^2}\,g^{\rho\mu}V^g_{\mu\nu}=0\,,
\ee
that $g^{\rho\mu}V^g_{\mu\nu}$ must be diagonal on the solution. The only off-diagonal terms in $g^{\rho\mu}V^g_{\mu\nu}$ that do not vanish identically are,
\be
g^{t\mu}V^g_{\mu r}\sim g^{r \mu}V^g_{\mu t}
\sim C\left(\beta_1+2\beta_2\frac{D}{r}+\beta_3\frac{D^2}{r^2}\right)\,.
\ee
This can vanish either if $C=0$, which characterises the bidiagonal solutions, or if $D=\omega r$ with $\omega$ being a solution of,
\be\label{Domegasol}
\beta_1+2\beta_2\omega+\beta_3\omega^2=0\,.
\ee
Notice that the second option, which characterises the non-bidiagonal solutions, does not exist if two of the $\beta_n$ parameters with $n=1,2,3$ are zero. Furthermore, the condition forces $\omega$ to be a constant. We will now discuss the two classes of solutions in some more detail.

\subsubsection{Bidiagonal black hole solutions}\label{sec:bidibh}
We start by discussing the bidiagonal solutions, which are obtained from \eqref{spheransatz} with $C=0$. As a simple but interesting and illustrative warmup we recall the proportional background solutions of section \ref{sec:propbg}, with $\fmn=c^2\gmn$. If either metric is diagonal, then these are obviously bidiagonal. It should also be clear that any vacuum solutions of the standard GR equations will in fact also be solutions of the bimetric equations for the proportional backgrounds (with appropriate restrictions on the constant of proportionality). In particular we may now consider bidiagonal black hole metrics. Since these are proportional vacuum solutions the linearisation around these solutions follows the analysis of section \ref{sec:linspecBG} and the fluctuations obey the massless equations,
\be\label{FPeqsMeq0_2}
\delta E^{\mathrm{0}}_{\mu\nu}=
{\mathcal{E}}_{\mu\nu}^{\ph\mu\ph\nu\rho\sigma}\delta G_{\rho\sigma}
-\Lambda\left(\delta G_{\mu\nu}-\frac1{2}\gmn g^{\rho\sigma}\delta G_{\rho\sigma}\right)=0\,,
\ee
and the massive equations,
\be\label{FPeqsMneq0_2}
\delta E^{\mathrm{M}}_{\mu\nu}=
{\mathcal{E}}_{\mu\nu}^{\ph\mu\ph\nu\rho\sigma}\delta M_{\rho\sigma}
-\Lambda\left(\delta M_{\mu\nu}-\frac1{2}\gmn g^{\rho\sigma}\delta M_{\rho\sigma}\right)
+\frac{m_{\mathrm{FP}}^2}{2}\left(\delta M_{\mu\nu}
-\gmn g^{\rho\sigma}\delta M_{\rho\sigma}\right)=0\,,
\ee
where, in terms of the original fluctuations $\delta\gmn$ and $\delta \fmn$, the massless field is given by $\delta G_{\mu\nu}=\delta\gmn+\alpha^2\delta\fmn$, and the massive field is obtained from $2c\,\delta M_{\mu\nu}=\delta\fmn-c^2\delta\gmn$. We also recall that (with $\nabla_\mu$ associated to the background field $\gmn$),
\begin{align}\label{gEinstOp_2}
\mathcal{E}_{\mu\nu}^{\ph\mu\ph\nu\rho\sigma}\delta M_{\rho\sigma}
\equiv -\tfrac1{2}\Big[\delta^\rho_\mu\delta^\sigma_\nu\nabla^2+g^{\rho\sigma}\nabla_\mu\nabla_\nu 
&-\delta^\rho_\mu\nabla^\sigma\nabla_\nu-\delta^\rho_\nu\nabla^\sigma\nabla_\mu\nn\\
&-g_{\mu\nu} g^{\rho\sigma}\nabla^2 +g_{\mu\nu}\nabla^\rho\nabla^\sigma\Big]\delta M_{\rho\sigma}\,.
\end{align}
The massless equations \eqref{FPeqsMeq0_2} are precisely those of linearised GR and a standard treatment shows that they propagate the two ($\pm2$) helicity states of a massless spin-2 field. Let us therefore focus on the massive field equations \eqref{FPeqsMneq0_2}. Due to the Bianchi identities obeyed by the Einstein tensor (including the $\Lambda$ terms) a covariant divergence of these yields,
\be
\nabla^\mu\delta E^{\mathrm{M}}_{\mu\nu}=\frac{m_{\mathrm{FP}}^2}{2}\left(
\nabla^\mu\delta M_{\mu\nu}-g^{\rho\sigma}\nabla_\nu\delta M_{\rho\sigma}\right)\,,
\qquad 
\nabla^\mu\delta M_{\mu\nu}=g^{\rho\sigma}\nabla_\nu\delta M_{\rho\sigma}\,,
\ee
where the last equality follows on-shell and provides four non-dynamical constraint equations for $\delta M_{\mu\nu}$. A second covariant divergence of these equations yields,
\be\label{propdiv}
\nabla^\mu\nabla^\nu\delta E^{\mathrm{M}}_{\mu\nu}=\frac{m_{\mathrm{FP}}^2}{2}\left(
\nabla^\mu\nabla^\nu\delta M_{\mu\nu}-g^{\rho\sigma}\nabla^2\delta M_{\rho\sigma}\right)\,.
\ee
Tracing instead the field equations \eqref{FPeqsMneq0_2} we get,
\be\label{proptrace}
g^{\mu\nu}\delta E^{\mathrm{M}}_{\mu\nu}=
g^{\mu\nu}\nabla^2\delta M_{\mu\nu}-\nabla^\mu\nabla^\nu\delta M_{\mu\nu}
+\left(\Lambda-\frac{3m_{\mathrm{FP}}^2}{2}\right)g^{\mu\nu}\delta M_{\mu\nu}\,.
\ee
It then follows that the particular linear combination,
\be\label{FPcomb}
2\nabla^\mu\nabla^\nu\delta E^{\mathrm{M}}_{\mu\nu}+
m_{\mathrm{FP}}^2g^{\mu\nu}\delta E^{\mathrm{M}}_{\mu\nu}=
\frac{m_{\mathrm{FP}}^2}{2}\left(2\Lambda-3m_{\mathrm{FP}}^2\right)g^{\mu\nu}\delta M_{\mu\nu}\,,
\ee
constitutes another on-shell scalar constraint,\footnote{An exceptional situation occurs when the parameters satisfy $2\Lambda=3m_{\mathrm{FP}}^2$, saturating the so called Higuchi bound \cite{Higuchi:1986py}. In this case the left-hand side of (\ref{FPcomb}) vanishes identically, such that a new linear scalar gauge symmetry emerges and the theory propagates a partially massless spin-2 field with only four degrees of freedom \cite{Deser:1983mm,Deser:2001us}. We will get back to this in section \ref{sec:PM}.} $g^{\mu\nu}\delta M_{\mu\nu}=0$, which then also implies $\nabla^\mu\delta M_{\mu\nu}=0$. Together these equations correspond to five constraints that can be used to remove five degrees of freedom from the original 10 components of the symmetric tensor perturbations $\delta M_{\mu\nu}$. Enforcing these constraints, the equations of motion for the massive spin-2 fluctuations can therefore be reduced to the following system,
\be\label{FP_GlF}
\left(\nabla^2-m_{\mathrm{FP}}^2\right)\delta M_{\mu\nu}
+2R_{\mu\ph\rho\nu}^{\ph\mu\rho\ph\nu\sigma}\delta M_{\rho\sigma}=0\,,\qquad
\nabla^\mu\delta M_{\mu\nu}=0\,,\qquad g^{\mu\nu}\delta M_{\mu\nu}=0\,.
\ee
This brief discussion provides a generalisation of the constraint analysis of section \ref{sec:linmflatsp}, when augmented to the case of constant curvature backgrounds, and also complements the analysis of section \ref{sec:linspecBG}. Moreover, as first recognised in \cite{Babichev:2013una}, the dynamical equations for $\delta M_{\mu\nu}$ in the above form are actually identical to the equations of a $5D$ black string when projected down onto its $4D$ sub-manifold, namely,
\be\label{blackstr}
\left(\nabla^2-k^2\right)\delta M_{\mu\nu}
+2R_{\mu\ph\rho\nu}^{\ph\mu\rho\ph\nu\sigma}\delta M_{\rho\sigma}=0\,,
\ee
where $k^2$ denotes the wave number of the transverse Fourier mode. This equation is identical to \eqref{FP_GlF} with the replacement $m_{\mathrm{FP}}\rightarrow k$. It has immediate and interesting implications since it is well known that the solution of \eqref{blackstr} exhibits the so called Gregory-Laflamme instability~\cite{Gregory:1993vy} (for a broad review on this subject see~\cite{Harmark:2007md}). Namely, the solution is unstable in the parameter region $0< k<k_c$, where $k_c$ is of the order of the inverse Schwarzschild radius. In the case of the $5D$ black string this instability concerns the mode longitudinal to the $5${\it th} direction along the black string and the end point of the instability is known to result in a sort of ``pinching" of the black string into an open necklace with beads of $4D$ Schwarzschild black holes strung on it. In the case of the bidiagonal bimetric solution when the metrics are proportional Schwarzschild (or Schwarzschild-de Sitter) metrics it follows that it too will have a similar instability.\footnote{An interesting exception is the Schwarzschild-de Sitter case with parameters fixed to the partially massless model discussed in section \ref{sec:PM}. For that case the instability is absent~\cite{Brito:2013yxa}.} But since there is no $5${\it th} dimension, no such interpretation is available and the status of the end point is presently unknown (one possibility is mentioned below). It is clear however that the instability only manifests itself over a characteristic time-scale comparable to the Hubble time $H_{0}^{-1}$ for a small enough mass parameter $m_{\mathrm{FP}}\sim H_{0}$ and it would therefore be hard to detect any signature connected with this. A similar study of the bidiagonal Kerr solutions have revealed that these additionally suffer from a super-radiant instability~\cite{Brito:2013wya}. For more details and discussions on this we refer the interested reader to the reviews~\cite{Volkov:2013roa,Volkov:2014ooa,Babichev:2015xha} and references therein.

Another simple and illustrative example is to consider the metric ans\"atze \eqref{spheransatz} as a static perturbation of proportional Minkowski backgrounds, i.e.~$\bar{f}_{\mu\nu}=c^2\bar{g}_{\mu\nu}=c^2\eta_{\mu\nu}$.\footnote{Note that these backgrounds do not solve the bimetric equations in general, but the existence of such solutions requires fixing either $c$ or one of the $\beta_n$ parameters in terms of the others.} Thus we consider the functions in \eqref{spheransatz}, with $C=0$ and no temporal dependence, as given by
\begin{align}
U&=1+\delta U\,,\qquad V=1+\delta V\,,\nn\\
A&=c(1+\delta A)\,,\qquad B=c(1+\delta B)\,,\qquad D=c(r+\delta D)\,.
\end{align}
The equations of motion for the perturbations can then be solved perturbatively to give (see e.g.~\cite{Comelli:2011wq, Enander:2013kza}),
\begin{align}\label{sphspert}
\delta U&=-\frac{M_1}{r}+\frac{M_2}{r}\,e^{-r\,m_{\mathrm{FP}}}\,,\nn\\
\delta V&=\frac{M_1}{r}-\frac{M_2(1+r\,m_{\mathrm{FP}})}{2r}\,e^{-r\,m_{\mathrm{FP}}}\,,\nn\\
\delta A&=-\frac{M_1}{r}-\frac{M_2}{\alpha^2c^2r}\,e^{-r\,m_{\mathrm{FP}}}\,,\nn\\
\delta B&=\frac{M_1}{r}
-(1+r\,m_{\mathrm{FP}})[\alpha^2c^2r^2m_{\mathrm{FP}}^2+2(1+\alpha^2c^2)]\,
\frac{M_2}{2\alpha^2c^2m_{\mathrm{FP}}^2r^3}\,e^{-r\,m_{\mathrm{FP}}}\,,\nn\\
\delta D&=(1+\alpha^2c^2)[1+r\,m_{\mathrm{FP}}+r^2m_{\mathrm{FP}}^2]\,
\frac{M_2}{2\alpha^2c^2m_{\mathrm{FP}}^2r^2}\,e^{-r\,m_{\mathrm{FP}}}\,,
\end{align}
where $M_1$ and $M_2$ are integration constants and we recall that the Fierz-Pauli mass $m_{\mathrm{FP}}$ is given by,
\be
m_{\mathrm{FP}}^2=\frac{m^4}{m_g^2}\frac{1+\alpha^2c^2}{\alpha^2c^2}
(c\beta_1+2c^2\beta_2+c^3\beta_3)\,.
\ee
The above expressions in (\ref{sphspert}) clearly have the form of Yukawa suppressed GR solutions for spherically symmetric perturbations of flat space-time. They exhibit the vDVZ discontinuity since, in the small mass limit $m_{\mathrm{FP}}\rightarrow0$, the combination,
\be
\delta U+\delta V=\frac{M_2(1-r\,m_{\mathrm{FP}})}{2r}
e^{-r\,m_{\mathrm{FP}}}\rightarrow\frac{M_2}{2r}\,,
\ee
does not vanish as it would in GR. On the other hand, we also see that both $\delta B$ and $\delta D$ diverge for $m_{\mathrm{FP}}\rightarrow0$, so this limit is in fact not well-defined on the solution. These divergences in the small-mass limit are remedied via the Vainshtein mechanism when going to higher orders in perturbation theory. Although the full nonlinear forms of these solutions are not known analytically their perturbative and numerical existence provides some initial hope of such a completion. The perturbative form, going up to second order, has been used for initial studies of e.g.~strong lensing~\cite{Enander:2013kza} and in confirming the Vainshtein mechanism in bimetric theory~\cite{Volkov:2012wp, Babichev:2013pfa, Enander:2015kda}. Making a full perturbative ansatz has also allowed to find numerical solutions to the nonlinear equations which are asymptotically flat and have massive hair. It has been conjectured that these solutions are the end point of the linear bidiagonal Schwarzschild instability discussed above~\cite{Brito:2013xaa}. Matching these solutions however requires the black hole to be of cosmological size and they are therefore unlikely to be of astrophysical interest.

In more general setups, away from proportional backgrounds, all known analytical solutions correspond to both metrics being of standard GR form and do not have massive hair. On the other hand, it is known that the bidiagonal class allows also for numerical solutions which are of non-GR form and do contain massive hair, typically with anti-de Sitter asymptotics. So far these more exotic solutions have only been studied numerically and we refer the reader to \cite{Volkov:2014ooa,Babichev:2015xha} for more discussions of their explicit behaviour.

\subsubsection{Non-bidiagonal black hole solutions}\label{sec:nonbdBH}
The non-bidiagonal solutions have $C\neq0$ and $D=\omega r$ in \eqref{spheransatz}, with $\omega$ being a constant solution of \eqref{Domegasol}. Furthermore, the Bianchi constraint $\nabla^\mu V^g_{\mu\nu}=0$ implies that (see e.g.~\cite{Volkov:2014ooa}),
\be\label{Domegasol2}
\left(\beta_2+\beta_3\omega\right)\left[(\omega-S^t_{~t})(\omega-S^r_{~r})
+\left(\frac{U}{V}\,S^t_{~r}\right)^2\right]=0\,,
\ee
where $S^\rho_{~\nu}$ are components of the square-root matrix $S=\sqrt{g^{-1}f}$. This condition together with \eqref{Domegasol} suggests that the non-bidiagonal solutions are a very non-generic class in the sense that the metric coefficients are forced to have a quite peculiar dependence on the $\beta_n$ parameters of the action. If we assume that $\omega$ is a solution of \eqref{Domegasol} and that \eqref{Domegasol2} is satisfied, we find that the equations of motion decouple into two separate Einstein equations,
\be
\mathcal{G}_{\mu\nu}+\frac{m^4}{m_g^2}\,\lambda_g\gmn=0\,,\qquad
\tilde{\mathcal{G}}_{\mu\nu}+\frac{m^2}{\alpha^2}\lambda_f\fmn=0\,,
\ee
where the cosmological terms are given by,
\be
\lambda_g=\beta_0+2\beta_1\omega+\beta_2\omega^2\,,\qquad
\lambda_f=\frac{1}{\omega^2}\left(\beta_2+2\beta_3\omega+\beta_4\omega^2\right)\,.
\ee
Therefore, this peculiar way of fixing the metric functions in terms of the $\beta_n$ parameters is an alternative and more complicated way of tuning the interactions to be purely cosmological, as opposed to the more obvious proportional solutions. In this context, recall the results of section~\ref{sec:genprop}, showing that whenever one metric is a solution to Einstein's equations, the other one must be an Einstein solution as well.
That such solutions (both Einstein, but with different cosmological constants) exist could perhaps have been anticipated by the complicated matrix structure of the interaction terms. 
 All black hole solutions in the non-bidiagonal class are therefore of standard GR type. An initial treatment of perturbations around these solutions can be found in~\cite{Babichev:2014oua}. For radial modes the analysis simplifies drastically and it is possible to get analytical expressions for the perturbations. It turns out that, while these are regular at the horizon, they are not regular at infinity. This implies that unstable radial modes are not allowed on physical grounds and that the corresponding black holes may in fact be stable, at least in the perturbative sense. For a detailed discussion on the current status of perturbations of various known black hole solutions we refer the reader to~\cite{Babichev:2015xha}.

\subsection{Cosmological solutions}

On small scales GR makes very good predictions and, from a phenomenological point of view, there is no need to look for a modification of the theory. On cosmological scales, however, it seems that either quantum field theory or gravity (or both) have to be modified in order to explain the observed value of the cosmic acceleration in a satisfactory way. In order for this review to be self-contained and to set up some notation we have provided a condensed summary of standard GR cosmology in appendix~\ref{app:cosmo}.

Due to the conceptual problems of explaining the observed cosmic acceleration within GR, the implications of a consistent theory of modified gravity for cosmology are of immediate interest. 
Unfortunately, the original idea that a large vacuum energy contribution from the matter sector could be screened out simply by a non-zero graviton mass that weakens gravity at large scales turned out not to be realisable without fine tuning, as discussed in~\cite{Hassan:2011vm}. Nevertheless, one can take a less ambitious approach and assume that another, as of yet unknown, mechanism (such as supersymmetry, if it was realised at low energies) is at work and removes all vacuum energy. In this case, ordinary GR would predict a universe without cosmic acceleration, while in modified gravity theories, ``self-accelerating" solutions certainly exist in various forms. Massive gravity is a particularly interesting candidate for this purpose since its interaction potential breaks diffeomorphism invariance and therefore the spin-2 mass scale is expected to be protected from receiving large quantum corrections. The hope would therefore be that the interactions of the graviton could give rise to a small rate of cosmic acceleration, which, unlike the pure cosmological constant, can be regarded as being ``technically natural'' in the sense of ~'t\,Hooft~\cite{'tHooft:1979bh} (see e.g.~\cite{Burgess:2013ara} for a recent review on naturalness in the context of cosmology). Of course, it should be pointed out that although the guide of naturalness is philosophically appealing, in the end it may turn out that nature does not follow that principle. Nevertheless, like Occam's razor, it is a useful tool in discriminating between alternative hypotheses.

Shortly after the construction of the ghost-free potential, it was discovered that the dRGT model of massive gravity (with flat or even with general reference metric) does not possess stable homogeneous and isotropic solutions~\cite{D'Amico:2011jj}. 
For more work on cosmology in massive gravity with fixed $\fmn$, see~\cite{deRham:2010tw, Gumrukcuoglu:2011ew, Gumrukcuoglu:2011zh, Vakili:2012tm, DeFelice:2012mx, Fasiello:2012rw, DeFelice:2013awa}. 
On the other hand, in bimetric theory with dynamical reference metric the desired class of solutions does exist~\cite{Volkov:2011an, vonStrauss:2011mq, Comelli:2011zm}. 
The simplest examples are the proportional backgrounds considered in section~\ref{sec:propbg} which, however, only solve the equations with proportional sources. In this case, the effective cosmological constant in (\ref{lambdas}) receives contributions not only through vacuum energy from the matter sector (as captured by $\beta_0$) but also from all terms in the interaction potential. Even in the absence of vacuum energy (i.e., for $\beta_0=0$), cosmological backgrounds can be accelerating and the scale of acceleration is proportional to the technically natural mass scale $m$ (see e.g.~\cite{Akrami:2012vf}).

In the following we provide a brief summary of results in bimetric cosmology. For a more detailed discussion, we refer the reader to~\cite{Solomon:2015hja}.

\subsubsection{Homogeneous \& isotropic backgrounds in bimetric theory}\label{sec:homiso}

We will outline the derivation of cosmological solutions in bimetric theory, following mainly~\cite{vonStrauss:2011mq}. 
For simplicity, the source $T^f_{\mu\nu}$ of the $\fmn$ sector is set to zero, which allows us to interpret $\gmn$ as the physical metric in the usual way, provided that it has standard matter couplings. In other analyses, the second source term is kept nonzero~\cite{Akrami:2013ffa, Comelli:2014bqa, DeFelice:2014nja}, which in principle can serve to mimic a dark matter component~\cite{Aoki:2014cla, Bernard:2014psa, Blanchet:2015sra, Blanchet:2015bia}.

To keep the comparison of the bimetric solutions to those of GR as simple as possible, it is convenient to make the usual Friedmann-Lema\^itre-Robertson-Walker (FLRW) ansatz for the metric $\gmn$ which is coupled to matter,\footnote{More general non-bidiagonal ans\"atze for the metrics have been studied in~\cite{Volkov:2011an}. As we discussed in section \ref{sec:nonbdBH} these are identical to GR backgrounds.}
\beqn\label{FRWg}
\gmn \md x^\mu\md x^\nu=-\md t^2+a(t)^2\left(\frac{1}{1-kr^2}\md r^2+r^2\md \Omega\right)\,.
\eeqn 
It is now important to notice that, in order to arrive at this form, we have used time reparametrisations to remove one time dependent function in the most general homogeneous and isotropic ansatz. We have therefore fixed the entire diffeomorphism invariance and there is no gauge symmetry left to do the same to $\fmn$.
As a result, the most general homogeneous and isotropic, bi-diagonal ansatz for the second metric reads,
\beqn\label{FRWf}
\fmn\md x^\mu\md x^\nu=-X(t)^2\md t^2+Y(t)^2\left(\frac{1}{1-kr^2}\md r^2+r^2\md \Omega\right)\,,
\eeqn
involving two time dependent functions $X(t)$ and $Y(t)$.  Note that we have assumed the curvatures $k$ of the two metrics to be the same. This does not constitute a loss of generality because, starting with two different values for $k$, one can show that the Bianchi constraint forces them to be equal~\cite{Nersisyan:2015oha}.

Furthermore, as in GR, we use the perfect fluid form ${(T^g)^{\mu}}_\nu=\mathrm{diag}(-\rho, p, p, p)$ for the stress-energy tensor $T^g_{\mu\nu}$ of matter coupled to $\gmn$.
In GR any source is automatically covariantly conserved as a consequence of the Bianchi identity whereas in bimetric theory the sources for $\gmn$ and $\fmn$ are not necessarily conserved. We therefore need to make the additional assumption of diffeomorphism invariance of the matter coupling which gives $\nabla_g^\mu T^g_{\mu\nu}=0$. 
This in turn implies that the continuity equation for matter (see equation~(\ref{conteq})) is also valid in the model under consideration.

Next, we consider the bimetric equations of motion in (\ref{mateom2}), with $T^f_{\mu\nu}=0$ and with one index raised by the respective inverse metrics. 
In what follows we shall refer to their $00$-components simply as the $\gmn$ and~$\fmn$ equation. Plugging the ans\"atze of the previous section into (\ref{geomm2}) we can make use of known GR results to straightforwardly obtain the $\gmn$ equation,
\beqn\label{gmncos}
\left(\tfrac{\dot{a}}{a}\right)^2+\tfrac{k}{a^2}-\tfrac{\mu^2}{3}\left[\beta_0+3\beta_1\tfrac{Y}{a}+3\beta_2\left(\tfrac{Y}{a}\right)^2+\beta_3\left(\tfrac{Y}{a}\right)^3\right]=\tfrac{\rho}{3m_g^2}\,,
\eeqn
where $\mu^2=m^4/m_g^2$. The first two terms are the same as in the ordinary Friedmann equation of GR, c.f.~(\ref{Friedmann}).
In order to determine how the source $\rho$ influences the cosmological evolution of the scale factor $a(t)$, we need to determine the function $Y(t)$ from the other equations. We start with the Bianchi constraint~$\nabla^\mu V_{\mu\nu}^g=0$, which, when evaluated on the homogeneous and isotropic ansatz, reduces to, 
\beqn\label{bianchiev}
\tfrac{3\mu^2}{a}\left[\beta_1+2\beta_2\tfrac{Y}{a}+\beta_3\left(\tfrac{Y}{a}\right)^2\right]\left(\dot{Y}-\dot{a}X\right)=0\,.
\eeqn
In addition to these, we have to consider the $\fmn$ background equation whose form is slightly more complicated than~(\ref{gmncos}) due to the presence of the additional function $X(t)$ in the ansatz for $\fmn$. Instead of presenting its complete form here, we first replace $X(t)=\dot{Y}/\dot{a}$, which is the dynamical solution to the Bianchi constraint above.\footnote{The other algebraic solution, obtained from setting the first bracket in (\ref{bianchiev}) to zero, will be discussed below.} Then the $\fmn$ equation becomes,
\beqn
\left(\tfrac{\dot{a}}{Y}\right)^2+\tfrac{k}{Y^2}-\tfrac{\mu^2}{3\alpha^2}\left[\beta_1\left(\tfrac{a}{Y}\right)^3+3\beta_2\left(\tfrac{a}{Y}\right)^2+3\beta_3\tfrac{a}{Y}+\beta_4\right]=0\,,
\eeqn
where, as usual, $\alpha\equiv m_f/m_g$
Multiplying this equation by $Y^2/a^2$ and subtracting it from the $\gmn$ equation (\ref{gmncos}) yields an algebraic equation,
\begin{align}\label{polynom}
\frac{\beta_3}{3}\Upsilon^4+\left(\beta_2-\tfrac{\beta_4}{3\alpha^2}\right)\Upsilon^3&+\left(\beta_1-\tfrac{\beta_3}{\alpha^2}\right)\Upsilon^2\nn\\
&+\left(\rho_*+\tfrac{\beta_0}{3}-\tfrac{\beta_2}{\alpha^2}\right)\Upsilon-\tfrac{\beta_1}{3\alpha^2}=0\,,
\end{align}
which we have expressed in terms of the dimensionless quantities,
\beqn
\Upsilon=\frac{Y}{a}\,,\qquad \rho_*=\frac{\rho}{3m^4}\,.
\eeqn
From the quartic polynomial equation (\ref{polynom}) one can obtain a solution for $\Upsilon$ which can be inserted into the $\gmn$ equation~(\ref{gmncos}). After that, the latter will only contain $a(t)$, $\dot{a}(t)$ and $\rho(t)$, and thus becomes a modified Friedmann equation.

\subsubsection{Classification of solutions}\label{sec:clascosm}

In general, the cosmological evolution equations derived above allow for several branches of solutions which have been categorised and studied in detail~\cite{Volkov:2011an, vonStrauss:2011mq, Comelli:2011zm, Akrami:2012vf, Akrami:2013pna, Konnig:2013gxa, Konnig:2014xva, Cusin:2014psa, Konnig:2015lfa, Akrami:2015qga, Cusin:2015pya}. Here we summarise the main results.

To start with, there are two options to satisfy the Bianchi constraint equation~(\ref{bianchiev}); either the first or the second bracket must vanish.
\begin{itemize}

\item[\textbf{(I)}] {\bf The non-dynamical (algebraic) branch:} If at least two out of $\beta_1$, $\beta_2$ and $\beta_3$ are non-vanishing, then a constant $b$ can be chosen such that $\beta_1+2b\beta_2+b^2\beta_3=0$. In this case, an algebraic solution to the Bianchi constraint is $Y(t)=b \,a(t)$. Even though in this case the 00-components of the two metrics are not necessarily proportional, the contribution from the interaction potential in~(\ref{gmncos}) reduces to a cosmological constant term. The $\gmn$ equation hence simply becomes the ordinary Friedmann equation~(\ref{Friedmann}) with a cosmological constant $\Lambda=m^2(\beta_0+3b\beta_1+3b^2\beta_2+b^3\beta_3)$ and thus the background solutions for $\gmn$ are degenerate with those of GR. At the level of linear perturbations, several degrees of freedom appear without kinetic terms and are thus expected to be strongly coupled~\cite{Gumrukcuoglu:2011zh, Comelli:2012db, Comelli:2014bqa}. On top of that, this branch seems to give rise to a non-perturbative ghost~\cite{DeFelice:2012mx} and a late-time instability for the tensor modes~\cite{Cusin:2015tmf}.
Due to these pathologies, most of the literature focusses on branch II solutions. 
Note also that branch I does not exist if two out of $\beta_1$, $\beta_2$ and $\beta_3$ are equal to zero.

\item[\textbf{(II)}] {\bf The dynamical branch:} The alternative solution to the Bianchi constraint is $X(t)={\dot{Y}}/{\dot{a}}$. In contrast to branch~I, this is an evolution equation and allows for much more general solutions that can be very different from GR. In particular, $\Upsilon$ is not constant on this branch but a function of the matter density $\rho$, determined by~(\ref{polynom}).

\end{itemize}

The second branch further splits up into several  subbranches, corresponding to different solutions of the quartic polynomial equation~(\ref{polynom}). These can be classified according to the evolution of the ratio of scale factors, $\Upsilon=Y/a$~\cite{Konnig:2013gxa}.

\begin{itemize}

\item[\textbf{(IIa)}] {\bf Infinite branch:} This branch is characterised by an $\Upsilon$ that becomes infinitely large at early times and whose value decreases when moving forward in time.

\item[\textbf{(IIb)}] {\bf Finite branch:} The solution is such that the ratio of scale factors $\Upsilon$ evolves towards a finite asymptotic value in the infinite past. This branch cannot be obtained in models with $\beta_1=0$ (assuming $\beta_0=0$)~\cite{Konnig:2013gxa} .

\item[\textbf{(IIc)}] {\bf Exotic branches:} These are all other branches which generically lead to bouncing cosmologies or static universes in the asymptotic past or future~\cite{Konnig:2015lfa}. Even though it is not impossible to make such scenarios compatible with observations, these branches correspond to rather unconventional cosmologies and have not been studied in great detail so far.
\end{itemize}

Using phenomenologically inspired requirements, one can immediately rule out a set of bimetric models based on their cosmological background evolutions. For instance, out of the models with only one non-vanishing $\beta_n$, only the $\beta_0$- and the $\beta_1$-model can give rise to a viable cosmology at the background level. Allowing for two non-vanishing $\beta_n$, one can generically discard models with $\beta_0=\beta_1=0$.
More details on the viability conditions and exclusion of bimetric models based on background cosmology can be found in~\cite{Akrami:2012vf, Konnig:2013gxa}.

\subsubsection{Self-acceleration \& technical naturalness of the Hubble scale}

In view of the dark energy problem, it is of particular interest to investigate whether bimetric theory can give rise to an expansion history compatible with observations even when there is no vacuum energy arising from the matter sector. If a background solution can mimic the behaviour of vacuum energy, it is referred to as ``self-accelerating". Clearly, it would be even more interesting if it was possible to include a large vacuum energy contribution and still obtain a small acceleration rate. This can indeed be achieved in bimetric theory by (fine-)tuning the $\beta_n$ parameters to cancel terms from the interaction potential against the vacuum contribution. However, there is no mechanism (or symmetry) protecting the tiny difference of these two contributions and thus a configuration with small effective cosmological constant cannot be regarded as technically natural in this setup. At this stage bimetric theory is therefore in no better position than GR to solve the old cosmological constant problem. 

Nevertheless, it is still possible to view the spin-2 interaction potential as the source of dynamical dark energy, much in the same spirit as in quintessence models~\cite{Ratra:1987rm}. 
To this end one assumes that all vacuum contributions in the source $T^g_{\mu\nu}$ vanish and furthermore sets the bare cosmological constant term to zero by fixing $\beta_0=0$.
A thorough comparison to observational data of the whole bimetric parameter space but with vanishing $\beta_0$ was performed in~\cite{Akrami:2012vf}.
Self-accelerating solutions are found for many different combinations of nonzero $\beta_n$ parameters, and the fits to data are just as good as for the ordinary $\Lambda$CDM model of GR.
As could have been anticipated, the best fit value for the mass of the massive spin-2 field is on the order of the Hubble scale $H_0$. This tiny mass scale creates the hierarchy problem in GR with an ordinary cosmological constant which is not protected against large quantum corrections by any symmetry.
The situation is different in bimetric theory where in the zero-mass limit the full diffeomorphism symmetry that transforms the two metrics separately is restored. This symmetry protects the spin-2 mass scale against large corrections and renders the small Hubble scale technically natural. For explicit calculations confirming this last statement, we refer the reader to~\cite{deRham:2013qqa, Heisenberg:2014rka}.

\subsubsection{Perturbations}

In order to further study the viability of bimetric cosmology, one derives the linear equations for perturbations around the above cosmological backgrounds. In contrast to what we encountered for the proportional backgrounds in section~\ref{sec:propbg}, it is not possible to diagonalise the fluctuations into spin-2 mass eigenstates around more general solutions. For the homogeneous and isotropic backgrounds, the best one can do is to perform a decomposition into tensor, vector and scalar modes and try to decouple their dynamics as much as possible. Rather general analyses of the spectrum were already performed in~\cite{Comelli:2012db, Berg:2012kn} (and later in~\cite{Solomon:2014dua}), but since the resulting differential equations are not always easy to handle, developing a complete understanding of cosmological perturbation theory is still an ongoing process.  

As already mentioned in section~\ref{sec:clascosm}, branch I solutions are plagued by pathologies such as strong coupling behaviour and a non-perturbative ghost instability. Let us therefore focus on linear perturbations in the dynamical subbranches.

\paragraph{Infinite branch IIa:} 
While specific infinite branch solutions for models with $\beta_2=\beta_3=0$ are free of scalar instabilities~\cite{Konnig:2014xva}, in general the perturbations around this class of FLRW backgrounds suffer from ghosts in both the scalar and tensor sectors~\cite{Lagos:2014lca, Cusin:2014psa, Konnig:2015lfa}. Due to these pathologies, branch IIa solutions do not correspond to consistent backgrounds and must be discarded on theoretical grounds. Nevertheless, for an interesting study of the integrated Sachs-Wolfe-effect in this branch which appeared before its problems were established, see~\cite{Enander:2015vja}.

 \paragraph{Finite branch IIb:} Around this branch linear perturbations are generically ghost-free and well-behaved, except for a scalar gradient instability which occurs at early times~\cite{Comelli:2012db, Konnig:2014dna, Konnig:2014xva, Lagos:2014lca}. While, from a theoretical point of view, this growing scalar mode is not a consistency problem, it does invalidate the use of linear perturbation theory. For generic parameters, the instability sets in at recent times and, as a consequence, bimetric theory cannot be invoked to predict phenomena in the early universe, at least not with standard perturbative techniques. 
However, it was demonstrated in~\cite{Akrami:2015qga} that, in the GR limit (c.f.~section~\ref{sec:GRlim}), the scalar instability is pushed backwards to arbitrarily early times. Hence, for small enough $\alpha=m_f/m_g$, linear perturbation theory around the finite branches remains valid and the predictions for cosmology automatically resemble those made by GR. Another suggestion is that nonlinear effects related to the Vainshtein mechanism may render the instability irrelevant~\cite{Mortsell:2015exa}.

\paragraph{Exotic branches IIc:} As already mentioned, the background evolution on these branches is rather unusual and hence this type of solutions has not received much attention. The perturbations around these backgrounds generically seem to have pathologies~\cite{Konnig:2015lfa}.

To summarise, the only well-studied models that give rise to viable cosmological backgrounds and perturbations lie on the finite branch. In order to maintain the predictivity of cosmological perturbation theory, it is furthermore necessary to bring bimetric theory close to GR, either by
requiring a sufficiently small value of $\alpha=m_f/m_g$ or by invoking the Vainshtein mechanism. The advantage of the bimetric setup is the occurrence of a technically natural dark energy scale set by the spin-2 mass. However, eventually all of this will be useful only if one finds an explanation for the absence of the large vacuum energy contribution coming from the matter sector.

\newpage
\section{Partial Masslessness}\label{sec:PM}

A massive spin-2 field in de Sitter (dS) background exhibits several interesting features that are shared neither by lower-spin particles nor by spin-2 excitations in flat space. For a special value of the Fierz-Pauli mass in units of the background curvature, the linear theory possesses an additional gauge symmetry which removes one of the spin-2 helicity components. In this case, where the field loses one propagating degree of freedom, it is referred to as \textit{partially massless} (PM). The particular mass value for which this situation occurs is called \textit{Higuchi bound} and it divides the parameter region into unitary and non-unitary sub-sectors~\cite{Higuchi:1986py}. 
Below the Higuchi bound the helicity-zero mode of the spin-2 particle develops a ghost instability, whereas above the bound all helicity states are well-behaved.

The existence of a nonlinear theory that involves a massive spin-2 field enables us to address the question whether the concept of partial masslessness is restricted to the linear theory around de Sitter backgrounds or whether it may be extended to the nonlinear level. By definition, the demand on a nonlinear PM theory is the presence of an extra gauge symmetry, even away from de Sitter backgrounds.

\subsection{Partially massless spin-2 field on de Sitter background}\label{sec:PMdS}

Let us begin by reviewing the linear theory of a PM spin-2 particle in a de Sitter background, as first discussed in a sequence of papers by Deser~\textit{et.al.}~\cite{Deser:1983mm, Deser:2001pe, Deser:2001us, Deser:2001wx, Deser:2001xr, Deser:2004ji, Deser:2006zx}. 
Consider a de Sitter background metric $\bar{G}_{\mu\nu}$ whose curvature satisfies $R_{\mu\nu}(\bar{G})=\Lambda \bar{G}_{\mu\nu}$ with positive cosmological constant $\Lambda$. 
The linearised action for a massive spin-2 fluctuation propagating on this background is,
\beqn\label{linactds}
S_\mathrm{dS}=\tfrac{1}{2}\int \dd^4 x\,\Big[h_{\mu\nu}\bar{\mathcal{E}}^{\mu\nu\rho\sigma}h_{\rho\sigma}-\Lambda\left(h^{\mu\nu}h_{\mu\nu}-\tfrac{1}{2} h^2\right)
+\tfrac{m^2_\mathrm{FP}}{2}\Big(h^{\mu\nu}h_{\mu\nu}- h^2\Big)\Big]\,,
\eeqn
where the linearised Einstein operator ${\bar{\mathcal{E}}_{\mu\nu}}^{\rho\sigma}$ is the same as in (\ref{kinopds}).
The corresponding equations of motion for $h_{\mu\nu}$ reads,
\beqn\label{lineqds}
{\bar{\mathcal{E}}_{\mu\nu}}^{~~\rho\sigma}h_{\rho\sigma}-\Lambda\left(h_{\mu\nu}-\tfrac{1}{2}\bar{G}_{\mu\nu} h\right)
+\tfrac{m^2_\mathrm{FP}}{2}\Big(h_{\mu\nu}-\bar{G}_{\mu\nu} h\Big)=0\,.
\eeqn
For general values of the Fierz-Pauli mass, this equation possesses no gauge symmetries. As we showed in section~\ref{sec:bidibh}, it gives rise to five constraint equations that reduce the number of propagating degrees of freedom to five, as appropriate for a massive spin-2 particle.  

A remarkable feature of (\ref{lineqds}) without any analogue in flat space is that if the Fierz-Pauli mass saturates the Higuchi bound, $m^2_\mathrm{FP}=\frac{2}{3}\Lambda$, the equation is invariant under a new gauge symmetry. The corresponding linear gauge transformation reads,
\beqn\label{linPMsym}
\Delta (\hmn) = \bar{\nabla}_\mu\partial_\nu\xi(x)+\tfrac{\Lambda}{3} \bar{G}_{\mu\nu}\, \xi(x)\,,
\eeqn
in which $\xi(x)$ is a local gauge parameter.

Besides computing the variation of~(\ref{lineqds}) under the gauge transformation explicitly, there is another straightforward way to see the existence of a gauge symmetry in the equation for the massive spin-2 field with mass at the Higuchi bound. First take the double divergence of~(\ref{lineqds}) to arrive at,
\beqn\label{doublediv}
\tfrac{m_\mathrm{FP}^2}{2}\left(\bar{G}^{\mu\rho}\bar{G}^{\nu\sigma}-\bar{G}^{\mu\nu} \bar{G}^{\rho\sigma}\right)\bar{\nabla}_\nu\bar{\nabla}_\mu h_{\rho\sigma}=0\,.
\eeqn
The kinetic terms have dropped out after using the linearised Bianchi identity.
Furthermore, the trace of (\ref{lineqds}) is given by,
\beqn
\left(\bar{G}^{\mu\nu} \bar{G}^{\rho\sigma}-\bar{G}^{\mu\rho}\bar{G}^{\nu\sigma}\right)\bar{\nabla}_\nu\bar{\nabla}_\mu h_{\rho\sigma}+\left(\Lambda-\tfrac{3}{2}m^2_\mathrm{FP}\right)h=0\,.
\eeqn
If the mass is at the Higuchi bound, the terms without derivatives in this equation vanish identically, while the derivative terms are identical to those in~(\ref{doublediv}).
Hence we find that for $m^2_\mathrm{FP}=\frac{2}{3}\Lambda$ the traced equations of motion are identical to their double divergence or, in other words, the sum of the double divergence and the (correctly normalised) trace is identically zero. 
A gauge identity of this type even implies that the linearised action~(\ref{linactds}) is invariant under the corresponding gauge transformation. This can be seen by noticing that invariance of the action under~(\ref{linPMsym}) requires,
\beqn
\Delta S_\mathrm{dS}=\frac{\delta S_\mathrm{dS}}{\delta \hmn}\Delta (\hmn)=\frac{\delta S_\mathrm{dS}}{\delta \hmn}\Big(\bar{\nabla}_\mu\partial_\nu\xi(x)+\tfrac{\Lambda}{3} \bar{G}_{\mu\nu}\, \xi(x)\Big)=0\,.
\eeqn
Since $\tfrac{\delta S}{\delta \hmn}$ is nothing else than the equation of motion for $\hmn$, after integrating by parts, we find that $\Delta S$ indeed vanishes due to the gauge identity. 

The conclusion is that, owing to the gauge redundancy, the spin-2 field described by (\ref{lineqds}) with mass at the Higuchi bound has only four dynamical degrees of freedom, one less than is the case for an ordinary massive spin-2 mode. The existence of the additional gauge symmetry around de Sitter background is in agreement with the representation theory of the de Sitter group SO$(1,4)$, which allows for ``short" representations of higher-spin fields. These representations contain less degrees of freedom than the massive ones and 
have been dubbed partially massless (PM). 
For a discussion of this topic, consult e.g.~\cite{Garidi:2003ys}. 

An interesting feature of PM fields is that the enhanced symmetry could potentially improve the quantum behaviour of spin-2 theories. 
For instance, the gauge invariance protects the difference of cosmological constant and spin-2 mass against receiving large quantum corrections. Since the spin-2 mass scale itself is protected by diffeomorphism invariance of the massless theory, this could render a small value for the cosmological constant technically natural.
Unfortunately, matter couplings of $\hmn$ destroy the gauge symmetry already at the linear level, unless the matter source is conformally invariant. An idea how to avoid this problem is presented in~\cite{Gabadadze:2008uc}.
It is also worth mentioning that the linear PM theory possesses interesting properties similar to the electromagnetic duality in Maxwell's theory~\cite{Deser:2013xb, Hinterbichler:2014xga}.


\subsection{The search for a nonlinear PM theory}\label{sec:PMnonlin}

After the linear PM theory for spin-2 fields on de Sitter background had been discovered, it was soon attempted to construct higher-order interactions for the perturbation $\hmn$ that would leave the gauge symmetry intact. At cubic order in $\hmn$ around dS backgrounds such a construction turned out to be possible in $d=4$ space-time dimensions~\cite{Zinoviev:2006im, Joung:2012rv, Joung:2012hz}. On the other hand, in $d>4$, it was found that gauge invariant cubic vertices can only exist if one also includes higher-derivative terms into the theory \cite{Zinoviev:2006im}. The construction of a fully nonlinear action with a PM gauge symmetry remains an open task.
Many recent findings point towards the fact that such a theory cannot exist, in particular not as a theory involving nothing but a partially massless spin-2 field~\cite{Deser:2012qg, Deser:2013bs, Deser:2013uy, Deser:2013gpa, deRham:2013wv, Fasiello:2013woa, Joung:2014aba, Garcia-Saenz:2014cwa}. We will summarise these arguments in section~\ref{sec:counter}.

It is natural to expect the nonlinear PM theory (if it exists) to be found within the family of ghost-free nonlinear theories for massive spin-2 fields. A first investigation of this possibility was carried out in \cite{deRham:2012kf} where the authors aimed to identify a PM theory in nonlinear massive gravity with the reference metric taken to be a fixed de Sitter background. Their construction used St\"uckelberg fields in a generalised decoupling limit and showed that, for a certain choice of the $\beta_n$ parameters in the potential, a scalar degree of freedom is not propagating.
Here we will follow the slightly simpler strategy of~\cite{Hassan:2012gz} which derived the parameters for the PM candidate theory in bimetric theory.\footnote{Note that the parameters of the PM model in massive gravity obtained in~\cite{deRham:2012kf} are different from those first derived in~\cite{Hassan:2012gz} and later confirmed by~\cite{Deser:2013uy, deRham:2013wv}. The reason for this mismatch can be traced back to a wrong choice of background, and correcting for this one can use the method of~\cite{deRham:2012kf} to arrive at the parameters that are consistent with the other references.}
As it turns out, this particular bimetric model possesses several rather unexpected properties which could be of interest even if the no-go results against a nonlinear PM theory cannot be evaded. It is also possible that the nonlinear PM theory requires the input of additional degrees of freedom (e.g.~higher-spin fields) and the framework of bimetric theory allows us to study merely the remnants of the enhanced symmetry present in the unknown extended setup. 

All our considerations here will be in four space-time dimensions. It is possible to generalise some of the results to higher dimensions if the bimetric action is augmented by the Lanczos-Lovelock terms, which we shall come back to in section~\ref{sec:LL}.

\subsubsection{Identifying PM candidates}\label{sec:idPM}

Our first observation is that, in bimetric theory, the equation of motion (\ref{dMeq}) for the massive fluctuation $\delta M_{\mu\nu}$ around proportional backgrounds is of the same form as (\ref{lineqds}) with $\Lambda=\bar{\Lambda}_g$.  
As a consequence, if the cosmological constant is assumed to be positive and if the Fierz-Pauli mass in (\ref{FPmass}) is on the Higuchi bound, a gauge symmetry of the form (\ref{linPMsym}) is present in the linear theory around proportional backgrounds. In bimetric notation, the corresponding infinitesimal gauge transformation of the massive fluctuation reads
\beqn\label{linPMsym2}
\Delta (\delta M_{\mu\nu}) = \bar{\nabla}_\mu\partial_\nu\xi(x)+\tfrac{\bar{\Lambda}_g}{3} \bar{G}_{\mu\nu}\, \xi(x)\,.
\eeqn
On the other hand, the massless fluctuation transforms under the PM symmetry at most by a term that resembles a coordinate transformation, $\Delta(\delta G_{\mu\nu})\sim \bar{\nabla}_\mu\partial_\nu\xi(x)$.  
This follows from the fact that its equation of motion (\ref{dGeq}) does not have any extra gauge symmetry besides the usual linearised diffeomorphism invariance. Hence, we can always undo the PM transformation of $\delta G_{\mu\nu}$ by a coordinate transformation. Modulo GCTs, we can therefore demand $\Delta(\delta G_{\mu\nu})=0$. 

The question is now under what conditions the above symmetry transformations for the fluctuations around dS background have a chance to be extendable to the nonlinear level. 
Since the nonlinear theory is formulated in terms of the variables $\gmn$ and $\fmn$, we first need to translate the transformations of mass eigenstates at the linear level into transformations of the fluctuations of the metrics. 
Assuming that we simultaneously perform a GCT to achieve $\Delta(\delta G_{\mu\nu})=0$, this translation can be obtained uniquely using the expressions for the massive and massless fluctuations given in (\ref{dGanddM}). The result is
\beqn\label{fgtrans}
\Delta (\delta\gmn)=-\alpha^2\,\Delta (\delta\fmn)=\tfrac{-2(\alpha c)^{2}}{c(1+(\alpha c)^{2})}\Delta (\delta M_{\mu\nu})\,.
\eeqn
The crucial observation now is that, in a theory with a gauge symmetry at the nonlinear level, it must be possible to shift these transformations of the fluctuations $\delta\gmn$ and $\delta\fmn$ to the backgrounds $\bgmn$ and $\bfmn$. This follows simply from the fact that the split into background and fluctuations is not unique and one can always take out part of the fluctuation to redefine the background. Therefore we demand that the symmetry transformations (\ref{fgtrans}) should also leave the background equations invariant. But since we are dealing with the proportional solutions, on which the background equations are reduced to an Einstein equation for $\bar{g}_{\mu\nu}$ along with $\Lambda_g=\Lambda_f$, c.f.~(\ref{propbackeq}), we restrict ourselves to constant gauge transformations $\bar{\Delta}$ with $\xi(x)=\xi_0=$\,const., such that $\Delta (\delta M_{\mu\nu}) = \tfrac{\xi_0\bar{\Lambda}_g}{3} \bar{G}_{\mu\nu}$. In this way we ensure that the transformation does not take us away from the proportional backgrounds and avoid unnecessary complication. 
The restriction to constant gauge transformations is a strong simplification, but as we will see now, demanding the background equation to be invariant under these is constraining enough to identify the PM candidate theory. Namely, it should be evident from (\ref{fgtrans}) that shifting the constant transformations to the backgrounds $\bgmn$ and $\bfmn=c^2\bgmn$ results in a shift in $c^2$, 
\beqn
\bgmn\bar{\Delta}(c^2)= \bar{\Delta}(\bfmn)-c^2\bar{\Delta}(\bgmn)=\bar{g}_{\mu\nu}
\frac{2}{3}\,\Lambda_g\,c\,\xi_0\,.
\eeqn
In general this cannot lead to an integrable\footnote{Integrability of the symmetry transformation means that it can be performed more than once and still leave the equations invariant. This is a natural requirement on any gauge symmetry. Without the integrability condition it is sufficient to set the mass to the Higuchi bound in order to have an invariant background.} symmetry of the background equations because $c$ is determined by $\Lambda_g=\Lambda_f$ and can therefore not be shifted. 
If this is the case, then a nonlinear PM symmetry cannot exist. The only possibility to avoid this immediate no-go statement is to demand that $c$ is not fixed by the background equation. 
Then, the equation $\Lambda_g=\Lambda_f$ that is automatically satisfied in this case (i.e.~without fixing $c$) can be thought of as the gauge identity evaluated on the proportional backgrounds. Since the proportional ansatz for the two metrics partly fixes the gauge, only constant scalings are left as residual transformations.

\subsubsection{PM candidate theory}

In four space-time dimensions, the explicit expressions~(\ref{lambdas}) for the cosmological constants in terms of the $\beta_n$ parameters, the ratio of Planck masses $\alpha$ and the proportionality constant $c$ can be used to write the background condition $\Lambda_g=\Lambda_f$ as a polynomial equation for $c$,
\begin{align}
\beta_1+(3\beta_2-\alpha^2\beta_0)c&+3(\beta_3-\alpha^2\beta_1)c^2\nn\\
&+(\beta_4-3\alpha^2\beta_2)c^3-\alpha^2\beta_3c^4=0\,.
\end{align}
This equation clearly fixes $c$ unless all coefficients of different powers of $c$ vanish separately. A proportionality constant $c$ that is undetermined by the background equation therefore requires the following parameter choices in the bimetric interaction potential,
\beqn\label{PMpara}
\beta_1=\beta_3=0\,,\qquad \alpha^2\beta_0=3\beta_2=\alpha^{-2}\beta_4\,.
\eeqn
We will frequently refer to these values as the \textit{PM parameters}. Note that our requirement on the $\beta_n$ parameters fixes all but one of them uniquely in terms of the others and~$\alpha$. The one remaining parameter is of course degenerate with the scale~$m^4$ in the interaction potential and sets the scale for the Fierz-Pauli mass and the cosmological constant. Moreover, it is easy to check using the expressions (\ref{FPmass}) and (\ref{barlamb}) for $\bar{m}_\mathrm{FP}^2$ and $\bar{\Lambda}_g$, that the parameter choice (\ref{PMpara}) automatically puts the mass on the Higuchi bound. Therefore the theory of linear perturbations around the backgrounds at hand exhibits the usual PM gauge symmetry. It is worth emphasising that we did not demand this in any way; it followed from an independent requirement on the background equations.\footnote{ On the other hand, note that if we assume that there is a unique PM theory, then requiring $\bar{m}_\mathrm{FP}^2=\frac{2}{3}\bar{\Lambda}_g$ is sufficient to determine its parameters. Namely, in this case, the symmetry (\ref{esymprop}) of the interaction potential enforces $\alpha^{4-n}\beta_n=\alpha^{n}\beta_{4-n}$ because otherwise the theory obtained from replacing $\alpha^{4-n}\beta_n\rightarrow\alpha^{n}\beta_{4-n}$ would also be PM, contradicting the uniqueness requirement.
It is easy to see that this constraint on the $\beta_n$ parameters together with the Higuchi bound condition already implies (\ref{PMpara}).} Starting from the proportional backgrounds the finite form of the scaling symmetry is given, for any constant $a$, by~\cite{Hassan:2012gz},
\be
c\rightarrow c+a\,,\qquad\bar g_{\mu\nu}\rightarrow\frac{1+(\alpha c)^2}{1+\alpha^2(a+c)^2}\,\bar g_{\mu\nu}\,.
\ee

By restricting to constant gauge transformations, clearly, we are not dealing with the full PM symmetry at the nonlinear level. Note, however, that since the set of constant transformations is a subset of the full gauge group, the theory we obtain by requiring invariance under this subset must contain the theory with the full gauge group (if it exists). Moreover, since in four dimensions the restriction to constant gauge transformations is sufficient to uniquely determine the $\beta_n$ parameters, we conclude that the resulting theory is already the unique candidate for having the full PM symmetry.

The scale invariance of the equations of motion for proportional backgrounds is not the only interesting property of the theory specified by the PM parameters. The PM candidate exhibits additional astonishing features that further support the existence of a gauge symmetry at the nonlinear level. For instance, consider again the homogeneous and isotropic solutions which we presented in section~\ref{sec:homiso}. We saw that, after solving the Bianchi constraint, it was possible to arrive at (\ref{polynom}), an algebraic equation for the ratio of the two scale factors, $Y(t)/a(t)$. Inserting the PM parameters (\ref{PMpara}) into this equation in vacuum, we find that it becomes an identity. Clearly, it is the analogue of the equation $\Lambda_g=\Lambda_f$ for proportional backgrounds. But instead of a constant, the equations evaluated on the cosmological ansatz now leave a time-dependent function undetermined. This shows that the cosmological evolution equations of PM bimetric theory are invariant under a symmetry that is local in time. 

Since, in four dimensions, the proportional backgrounds correspond to maximally symmetric spacetimes with ten independent Killing vectors and the homogeneous and isotropic solutions still possess six isometries, one might speculate that the presence of a symmetry on these solutions could somehow be related to the amount of symmetry of the underlying geometry. To obtain more general results, it is therefore important to investigate the structure of our PM candidate theory beyond the proportional and cosmological backgrounds. 
We shall come back to this point in the next subsection where we show that the equations in the PM theory are Weyl invariant to lowest order in a derivative expansion.

\paragraph{Massive gravity limit.}
Let us comment briefly on the massive gravity limit. The PM parameters (\ref{PMpara}) in four space-time dimensions provide an example for a theory in which the coefficients in the interaction potential depend on the ratio $\alpha=m_f/m_g$ that is taken to infinity in the massive gravity limit which we discussed in section~\ref{sec:mglimit}. In this case, we have to be more careful when taking the limit because the $\beta_n$ parameters will scale as well. The conditions (\ref{PMpara}) fix the relative scale among them but their absolute scale may still be chosen freely. Suppose that $\beta_2$ does not scale when the limit $\alpha=m_f/m_g\rightarrow\infty$ is taken. Then $\beta_0\rightarrow 0$ and $\beta_4\rightarrow\infty$, whereas $\Lambda={\beta_4\,m^4}/{m_f^2}=\,$const.  In the massive gravity limit, the equations of motion for the PM theory thus reduce to
\beqn
\mathcal{G}_{\mu\nu}(g)
+\frac{\Lambda}{3}\,g_{\mu\lambda}{\big(Y^{(2)}\big)^\lambda}_{\nu}(S)=0\,, \qquad
\mathcal{G}_{\mu\nu}(f)
+\Lambda\fmn=0 \,.\label{fpm}
\eeqn
These are exactly the equations singled out in \cite{Deser:2013uy, deRham:2013wv} which investigated the possibility of realising nonlinear partial masslessness in massive gravity with fixed reference metric. The methods invoked in those references are completely different from ours and could provide independent support for the existence of a nonlinear PM symmetry. 
On the other hand, as we already mentioned previously, references \cite{Deser:2013uy, deRham:2013wv} also provided evidence for the absence of an additional gauge symmetry present in (\ref{fpm}). These results may extend to the bimetric case but such a generalisation is not obvious due to the singular nature of the massive gravity limit~\cite{Hassan:2014vja}.

\subsection{Connection to conformal gravity}

As shown in~\cite{Hassan:2013pca}, the PM candidate model is closely related to the well-known theory of conformal gravity whose action is invariant under Weyl transformations of the metric. In the following we will briefly review conformal gravity and then summarise the steps that establish its connection to the PM bimetric model.

\subsubsection{Review of conformal gravity}

In four space-time dimensions, there exists a particular higher-derivative action with an additional gauge symmetry~\cite{Stelle:1976gc}, 
\beqn\label{CGact}
    S_\mathrm{CG}=-\sigma\int\dd^4x\sqrt{g}\,\left(R^{\mu\nu}R_{\mu\nu}-\frac{1}{3}R^2\right)\,,
\eeqn
with dimensionless coefficient $\sigma$.
This action is invariant under Weyl transformations of the metric, 
\beqn\label{conftrafo}
\gmn\longmapsto \xi^2(x)\gmn\,,
\eeqn
where $\xi(x)$ is a local gauge parameter.
The cosmological constant and the Einstein-Hilbert term are not invariant under the transformation (\ref{conftrafo}); hence they do not appear in the above conformal gravity action.

The equations of motions that follow from variation of the conformal gravity action (\ref{CGact}) imply the vanishing of the Bach tensor for~$\gmn$~\cite{Bach},
\beqn\label{Bacheq}
   B_{\mu\nu}\equiv -\nabla^2 P_{\mu\nu}+\nabla^{\rho}\nabla_{(\mu}P_{\nu)\rho}+W_{\rho\mu\nu\sigma}P^{\rho\sigma}=0\,.
\eeqn 
Here, we have given the definition of $B_{\mu\nu}$ in terms of the Schouten tensor,
\beqn\label{schouten}
P_{\mu\nu}\equiv R_{\mu\nu}-\frac{1}{6}\gmn R\,,
\eeqn
as well as the Weyl tensor,\footnote{The conformal gravity action~(\ref{CGact}) can also be expressed, modulo the Euler invariant, as the square of the Weyl tensor, $\mathcal{L}_\mathrm{CG}\propto\sqrt{g}\,W_{\rho\mu\nu\sigma}W^{\rho\mu\nu\sigma}$.}
\beqn
W_{\rho\mu\nu\sigma}\equiv R_{\rho\mu\nu\sigma}+g_{\mu[\nu}R_{\sigma]\rho}-g_{\rho[\nu}R_{\sigma]\mu}+\frac{1}{3}g_{\rho[\nu}g_{\sigma]\mu}R\,.
\eeqn 
The Bach equation (\ref{Bacheq}) is also invariant under Weyl transformations in four space-time dimensions.

The conformal gravity action is closely related to the linear theory for a partially massless spin-2 field discussed in section~\ref{sec:PMdS}. In order to see this, one can introduce an auxiliary field $\varphi_{\mu\nu}$ and a parameter $\Lambda$ to write down an equivalent action~\cite{Kaku:1977pa},
\begin{align}\label{CGauxact}
    S_\mathrm{CG}^{(2)}=4\sigma\Lambda\int\dd^4x\sqrt{g}\,\Big[ \tfrac{1}{6}(R&-2\Lambda)+\varphi^{\mu\nu}\mathcal{G}_{\mu\nu} \nn\\
    &+ \Lambda\varphi+ \Lambda(\varphi^{\mu\nu}\varphi_{\mu\nu}-\varphi^2)\Big]\,.
\end{align}
Linearisation around a constant curvature background and diagonalisation into mass eigenstates shows that, for $\sigma>0$, the theory propagates a healthy massless and a ghost-like massive spin-2 particle. Moreover, the Fierz-Pauli mass of the massive fluctuation has the value corresponding to the Higuchi bound, $m_\mathrm{FP}^2=\frac{2}{3}\Lambda$, which implies the presence of the PM gauge symmetry (\ref{linPMsym}) for $\varphi_{\mu\nu}$. The massless fluctuation does not transform under this symmetry.
In fact, it is a simple exercise to check that, in addition to the obvious diffeomorphism invariance, the full nonlinear auxiliary action (\ref{CGauxact}) is invariant under the following linear gauge transformations~\cite{Deser:2012qg},
\beqn
\delta\gmn=2\xi(x)\gmn\,,\qquad \delta\varphi_{\mu\nu}=(\nabla_\mu\partial_\nu
+\tfrac{\Lambda}{3}\gmn)\xi(x)\,.
\eeqn
Note that the nonlinear field $\gmn$ transforms under the conformal part of the PM symmetry because its fluctuation does not correspond to the massless mode but is a linear superposition of mass eigenstates. Its transformation is of course nothing but the linear version of the Weyl transformation (\ref{conftrafo}).

Conformal gravity, or its equivalent form (\ref{CGauxact}), therefore describes only six propagating degrees of freedom instead of seven that would correspond to a massless and a massive spin-2 field. For more detailed discussions of its spectrum we refer the reader to \cite{Fradkin:1981iu, Lee:1982cp, Riegert:1984hf, Maldacena:2011mk}. 
Like any other Weyl invariant theory, $S_\mathrm{CG}$ does not contain any dimensionful couplings and therefore avoids the non-renormalisibility problem of GR. Being a renormalisable field theory, conformal gravity has been suggested as a quantum theory for gravity. It has also been shown to allow for viable cosmological solutions and even fit galaxy rotation curves, providing a possible solution for at least part of the dark matter problem~\cite{Mannheim:2011ds}. 

Unfortunately, all of its features remain irrelevant unless a cure for the ghost problem in conformal gravity is found. Suggestions for altering the theory in order to make it healthy include, for instance, a modification of quantum mechanics~\cite{Bender:2007wu} and the addition of specific boundary conditions~\cite{Maldacena:2011mk}, but none of these have been sufficiently convincing for the theory to be accepted as a consistent alternative to GR. 
In order to avoid the inconsistencies that plague theories with a finite number of higher derivatives, another possible solution is to complete the equations of motion~(\ref{Bacheq}) of conformal gravity with an infinite number of derivatives for which Ostrodgradski's theorem does not hold. Higher-derivative terms will, of course, break the conformal symmetry because they necessarily enter with a suppressing mass scale. On the other hand, one could imagine that the symmetry transformation needs to be completed with an infinite number of higher-derivative terms as well, in order to be a gauge symmetry of the full theory. In that case, both the equations of motion and the symmetry transformation could be thought of as a perturbative expansion in derivatives. Order by order, the equations would be invariant under the gauge symmetry, starting with the Bach tensor and its Weyl invariance at lowest order. Although this idea sounds promising, without any further input it is difficult to guess the form of the higher-derivative corrections to the Bach equation that could give rise to such a gauge symmetry. It would therefore be helpful to have a guideline telling us how to construct these terms. This is where the PM candidate of bimetric theory comes into play.

\subsubsection{Perturbative solution to the $\gmn$ equation}

We review here the results of~\cite{Hassan:2013pca}.
Our aim is to combine the bimetric equations of motion to eliminate one of the metrics and derive an effective equation involving only the other. To this end, we will solve the $\gmn$ equation algebraically for the square-root matrix ${S^\mu}_\nu$ as a perturbation series in curvatures of~$\gmn$. From this we deduce a perturbative solution for $\fmn$ that can be used to eliminate $\fmn$ from its own equation of motion, resulting in a perturbative equation for $\gmn$. We will derive the lowest orders of this equation for general $\beta_n$ parameters and then see that it exhibits remarkable features when we restrict to the PM parameters~(\ref{PMpara}) in the subsequent subsection.
Of course, we could switch the roles of the metrics and in a similar manner derive an effective equation for $\fmn$.

It will prove convenient to first rewrite the Einstein tensor in terms of the Schouten tensor defined in (\ref{schouten}) and raise one index with $g^{\mu\nu}$. This gives,\footnote{Mainly for the sake of notational simplicity, we have set the source for $\gmn$ to zero. This simplification constitutes no loss of generality because in the final results of our computation it can always be reinstated by making the replacement,
$
{P^\mu}_\nu~\rightarrow~ {P^\mu}_\nu-m_g^{ -2}\left( {T^{g\mu}}_\nu -\tfrac{1}{3}\Tr\,T^g\,\delta^\mu_\nu\right)\,.
$\label{foot:PtoPpT}}
\beqn\label{eomrewr}
\Tr\,P\,\delta^\mu_\nu-{P^\mu}_\nu=\mu^2\sum_{n=0}^{3}(-1)^n\beta_n {(Y^{(n)})^\mu}_\nu(S)\,,\qquad \mu^2\equiv\tfrac{m^4}{m_g^2}\,.
\eeqn
Then we make the following perturbative ansatz for the square-root matrix $S=\sqrt{g^{-1}f}$,
\begin{align}\label{ansatzs}
{S^\mu}_\nu=a\delta^\mu_\nu&+\tfrac{1}{\mu^2}\Big(b_1{P^\mu}_\nu+b_2\Tr\,P\,\delta^\mu_\nu\Big) \nn\\
&+ \tfrac{1}{\mu^4}\Big( c_1{(P^2)^\mu}_\nu+c_2\Tr\,P\,{P^\mu}_\nu  
+c_3\Tr\,P^2\delta^\mu_\nu+c_4(\Tr\,P)^2\delta^\mu_\nu \Big) 
+\mathcal{O}\left(\tfrac{P^3}{m^6}\right)\,,
\end{align}
with arbitrary complex coefficients $a, b_i, c_i,\hdots$ Allowing the parameters to assume complex values is reasonable as long as the coefficients that will finally appear in the effective equation for $\gmn$ remain real. 

In the model where only $\beta_0$, $\beta_1$ and $\beta_4$ are non-vanishing, the solution takes the very simple closed form,
\beqn\label{ssolbeta1}
{S^\mu}_\nu=-\frac{\beta_0}{3\beta_1}\delta^\mu_\nu+\frac{1}{\beta_1\mu^2}{P^\mu}_\nu(g)\,,\qquad \text{if}~~\beta_2=\beta_3=0\,.
\eeqn
For more general parameters, the expansion does not terminate and we can only determine the coefficients in the ansatz order by order in curvatures. As a side-remark: The existence of this exact solution in the $\beta_1$ model is rather remarkable. For example, linearising the equations of motion in this model we can use the above relation to fully remove any occurrence of $\fmn$ in the linearised equations. In particular, in the massive gravity limit this means that it is possible to obtain equations for a massive spin-2 field propagating on any background without any reference to a second metric~\cite{Bernard:2014bfa, Bernard:2015mkk}.

In order to simplify the expressions in the following, we introduce a new set of linear combinations of the $\beta_n$ parameters,
\beqn\label{snalphan}
s_n\equiv \sum_{k=n}^{3}{3-n\choose k-n}a^k\beta_k\,,
\eeqn
Note that on proportional backgrounds, $\fmn=c^2\gmn$ and $s_0$ is proportional to the cosmological constant $\Lambda_g$ defined in~(\ref{lambdas}) if in that expression one replaces $c$ by $a$. 

The lowest orders in the solution for $S$ are determined to be of the following form,
\begin{align}\label{sols}
{S^\mu}_\nu=a\delta^\mu_\nu+\tfrac{a}{s_1\mu^2}{P^\mu}_\nu +\tfrac{as_2}{s_1^3\mu^4}\Big[{(P^2)^\mu}_\nu-\Tr\,P\,{P^\mu}_\nu+\tfrac{1}{3}e_2( P)\delta^\mu_\nu\Big] 
+\mathcal{O}\left(\tfrac{P^3}{\mu^6}\right)\,.
\end{align}
We can then use $\fmn=g_{\mu\rho}{(S^2)^\rho}_\nu$ to arrive at,
\begin{align}\label{fsolP}
\fmn=a^2\gmn+\tfrac{2a^2}{s_1\mu^2}P_{\mu\nu}+\tfrac{a^2(s_1+2s_2)}{s_1^3\mu^4}{(P^2)^\mu}_\nu
+\tfrac{2a^2s_2}{s_1^3\mu^4}\Big[\tfrac{1}{3}e_2( P)\delta^\mu_\nu&-\Tr\,P\,{P^\mu}_\nu\Big]\nn\\
&+\mathcal{O}\left(\tfrac{P^3}{\mu^6}\right)\,.
\end{align}
This is the most general covariant solution for $\fmn$ obtained from the $\gmn$ equation.
An immediate consequence of equation (\ref{fsolP}) is that if the solution for $\gmn$ has constant curvature, $P_{\mu\nu}\propto\gmn$, then $\fmn\propto\gmn$, i.e.~the two metrics are proportional to each other. But, as we already mentioned in section~\ref{sec:genprop}, there exist solutions for which both metrics have constant curvature while not being proportional to each other. This class of (non-covariant) solutions is therefore not captured by our ansatz for $S$ in~(\ref{ansatzs}).

Next we use the solution~(\ref{fsolP}) for $\fmn$ to eliminate it in its own equation of motion. This will lead to a set of effective equations for $\gmn$ containing an infinite number of derivatives, due to the presence of the inverse $f^{\mu\nu}$ in these equations.
Inserting (\ref{fsolP}) into the Einstein tensor for $\fmn$ we find,  
\begin{align}
{\mathcal{G}}_{\mu\nu}(f)={\mathcal{G}}_{\mu\nu}(g)-\tfrac{1}{s_1\mu^2}
\Big( \nabla^2P_{\mu\nu}&+\nabla_\mu\nabla_\nu 
P-\nabla^\rho\nabla_\mu P_{\rho\nu}-\nabla^\rho\nabla_\nu P_{\rho\mu} \nn\\
&+3PP_{\mu\nu}-\gmn\left[P^{\alpha\beta}P_{\alpha\beta}+\tfrac{1}{2}P^2
  \right] \Big)+\mathcal{O}\left(\tfrac{P^3}{\mu^4}\right)\,.
\end{align}
The contributions from the interaction potential to the $\fmn$ equation evaluated on~(\ref{fsolP}) become, 
\begin{align}
\tfrac{\mu^2}{\alpha^2}\tilde{V}_{\mu\nu}&=\tfrac{\mu^2\Omega}{a^2\alpha^2} 
\gmn+\tfrac{1}{a^2\alpha^2}\mathcal{G}_{\mu\nu}+
\tfrac{2\Omega}{a^2\alpha^2s_1}P_{\mu\nu}\nn\\
&+\tfrac{1}{a^2\alpha^2s_1^3\mu^2}\left[c_1P^\rho_\mu P_{\rho\nu}+
c_2 PP_{\mu\nu}+\tfrac{1}{6}\gmn(c_3P^{\alpha\beta}P_{\alpha\beta}-c_2P^2)
\right]+\mathcal{O}\left(\tfrac{P^3}{\mu^4}\right)\,,
\end{align}
in which the expansion coefficients are given by,
\begin{align}
	c_1=2s_1^2+\Omega(s_1+2s_2)\,,\qquad
	c_2=-3s_1^2-2s_2\Omega\,,\qquad
	c_3=3s_1^2-2s_2\Omega\,,
\end{align}
where we have defined,
\begin{align}
	\Omega=a\beta_1+3a^2\beta_2+3a^3\beta_3+a^4\beta_4\,.
\end{align}
Note that this would be proportional to $\Lambda_f$ in \eqref{lambdas} if $\fmn=a^2\gmn$.
Combining the kinetic and potential terms, we can write the entire $\fmn$ equation of motion
as a higher-derivative equation for $\gmn$, 
\begin{align}\label{fEq2nd}
&\tfrac{\Omega}{a^2\alpha^2}\,\gmn+\tfrac{1}{\mu^2}
\left[1+\tfrac1{a^2\alpha^2}\right]\mathcal{G}_{\mu\nu}
+\tfrac{2\Omega}{a^2\alpha^2s_1\mu^2}\,P_{\mu\nu} 
\nn\\
&+\tfrac{1}{\mu^4 s_1}B_{\mu\nu}
+\tfrac{\Omega}{a^2\alpha^2s_1^3\mu^4}\Big[(s_1+2s_2)P_\mu^{~\rho}
P_{\rho\nu} -2s_2PP_{\mu\nu}-\tfrac{s_2}{3}\gmn\left(P_{\rho\sigma}
P^{\rho\sigma}-P^2\right)\Big]
\nn\\
&-\tfrac{1}{s_1\mu^4}(1+\frac{1}{\alpha^2 a^2})\left[3P P_{\mu\nu}
-2P_\mu^{~\rho} P_{\rho\nu} -\frac{1}{2}g_{\mu\nu}(P^2-P^{\alpha\beta}
P_{\alpha\beta})\right]+\mathcal{O}\left(\tfrac{P^3}{\mu^6}\right) 
=0\,,
\end{align}
where we have collected some of the terms with four derivatives
into the Bach tensor defined in~(\ref{Bacheq}). We have thus arrived at an effective equation involving only $\gmn$ in which the terms with more than four derivatives can be computed order by order following the same procedure as outlined above. The terms in the first line of (\ref{fEq2nd}) are the cosmological constant, the Einstein tensor and a correction proportional to the Schouten tensor. The equation hence reduces to GR in the small curvature limit only if $(\Omega/a^2\alpha^2)(1/s_1\mu^2)\rightarrow0$. Note that the first of these brackets directly sets the size of the observed cosmological constant. The phenomenological relevance of this equation for weak gravitational fields has not been studied so far, but doing so would require adding a source in accordance with our remark in footnote~\ref{foot:PtoPpT}.

Before coming to the partially massless case, let us make one more remark: Plugging the perturbative solution (\ref{fsolP}) back into the bimetric action results in an effective action of the form,
\begin{align}\label{HDactobt}
    S_\mathrm{HD}=m_g^{2}\int\dd^4x\sqrt{g}\,\Big[ c_\Lambda+c_RR(g)
    -\frac{c_{RR}}{m^2}\Big(R^{\mu\nu}R_{\mu\nu}-\frac{1}{3}R^2\Big) \Big]~+~\mathcal{O}\left(\tfrac{P^3}{m^6}\right)\,,
\end{align}
where $c_\Lambda$, $c_R$ and $c_{RR}$ are functions of the bimetric parameters. We recognise the conformal gravity combination at fourth order in derivatives and notice that the above action, at quadratic order in curvatures, represents the generalisation of three-dimensional New Massive Gravity~\cite{Bergshoeff:2009hq} to four dimensions which is also known to propagate seven degrees of freedom~\cite{Stelle:1976gc, Stelle:1977ry}. However, the action~(\ref{HDactobt}) is not equivalent to the original bimetric action since we have used the equation for $\gmn$ instead of the one for $\fmn$ to obtain a solution for $\fmn$. The equations derived from the above action differ from the bimetric equations by the product of a differential operator. Only by restricting to solutions where the zero modes of this operator are absent will the equations give the same solutions. The additional terms that would arise using the correct procedure, i.e.~integrating out $\fmn$ by its own equations of motion, are nonlocal because a derivative operator needs to be inverted in order to solve the $\fmn$ equation for $\fmn$. This is discussed in detail in~\cite{Hassan:2013pca, Cusin:2014zoa}. In the following, we will not work with any effective action but restrict ourselves to the equations of motion where this ambiguity does not arise.

\subsubsection{Partially massless case}\label{secPMCG}

Let us determine the set of parameter values for which the higher-derivative equation~(\ref{fEq2nd}) obtained from bimetric theory reduces to the Bach equation~(\ref{Bacheq}) at lowest order in derivatives. This means that we require,
\beqn\label{condCG}
\Omega=0\,,\qquad a^2=-\alpha^{-2}\,.
\eeqn
In addition to this, satisfying the $\gmn$ equation at lowest order in curvatures requires that the combination $s_0$ defined in~(\ref{snalphan}) vanishes. Using $a^2=-\alpha^{-2}$ in the expressions for $\Omega$ and $s_0$, it is easy to see that these requirements combine into two complex equations,
\beqn
\beta_0+\frac{3i}{\alpha}\beta_1-\frac{3}{\alpha^2}\beta_2-\frac{i}{\alpha^3}\beta_3&=&0\,,\nn\\
\beta_0+\frac{4i}{\alpha}\beta_1-\frac{6}{\alpha^2}\beta_2-\frac{4i}{\alpha^3}\beta_3+\frac{1}{\alpha^4}\beta_4&=&0\,,
\eeqn
which need to be solved for the $\beta_n$ parameters.
The real and imaginary parts must vanish separately and the unique solution is,
\beqn\label{PMparz}
\beta_1=\beta_3=0\,,\qquad \alpha^2\beta_0=3\beta_2=\alpha^{-2}\beta_4\,.
\eeqn
Remarkably, these values precisely corresponds to the PM parameter choice in (\ref{PMpara}). We conclude that the PM candidate is the unique member of the family of bimetric theories with the feature that to lowest order in a derivative expansion the effective equation for $\gmn$ is identical to the Weyl invariant equation of conformal gravity,
\beqn\label{bachpm}
B_{\mu\nu}+\mathcal{O}\left(\tfrac{P^3}{m^6}\right)=0\,.
\eeqn
Moreover, it is straightforward to show that the method that we used here to arrive at an effective equation for $\gmn$ can yield a similar result for $\fmn$:
The effective equation for $\fmn$, obtained from solving the $\fmn$ equation for $\gmn$ and plugging the solution into the $\gmn$ equation, also sets the Bach tensor for $\fmn$ to zero at lowest order in derivatives. This can directly be deduced from the symmetry property~(\ref{sympotcontr}) of the equations of motion and the invariance of the PM parameter choice under $\alpha^{4-n}\beta_n\rightarrow \alpha^{n}\beta_{4-n}$.

In this way, the nonlinear PM bimetric theory proposes a ghost-free completion of conformal gravity. The presence of a gauge symmetry to lowest order in a derivative expansion can be regarded as further support of the existence of an additional gauge symmetry in the theory. In particular, note that the above analysis is not based on de Sitter (nor FLRW) background.

If existent, the PM symmetry could be viewed as the generalisation of the gauge invariance of conformal gravity to higher orders in derivatives. In principle, it is possible to explicitly demonstrate the invariance of the equations under symmetry transformations that are extended to higher orders in derivatives. Using the perturbative solution to the $\gmn$ equation together with the analogous expression obtained from the $\fmn$ equation, the gauge invariance has been shown to exist up to sixth order in derivatives~\cite{Hassan:2015tba}.

\subsection{Arguments against a nonlinear PM theory}\label{sec:counter}

As mentioned before, the literature contains a large variety of no-go theorems that forbid the existence of a nonlinear theory for partially massless spin-2 fields. Most of these counter arguments refer to an action involving no other degrees of freedom besides the PM field but some of them may eventually also apply to bimetric theory. Let us summarise them briefly:

\begin{itemize}

\item In the massive gravity limit of the PM bimetric model, first obstructions were already encountered by the authors of~\cite{deRham:2013wv} and it was shown in~\cite{Deser:2013bs, Deser:2013uy} that the equations of motion do not satisfy the nonlinear Bianchi identity which is expected in a PM theory. More explicitly, the authors showed that the nonlinear covariant constraint which removes the Boulware-Deser ghost does not identically vanish. This result was argued to extend to bimetric theory~\cite{Deser:2013gpa} where, however, the situation is less clear since the nonlinear version of the constraint is not known. In fact, a covariant constraint does not even seem to exist for general backgrounds~\cite{ourpaper}.

\item  The linear spectrum of conformal gravity always contains a ghost~\cite{Deser:2012qg}. Although the full nonlinear bimetric theory is ghost-free, it is possible that the appearance of the Weyl invariant Bach tensor~(\ref{bachpm}) in our perturbative approach is intimately related to the fact that we are expanding in small curvatures, i.e.~around flat solutions which suffer from the same pathology as conformal gravity. Flat backgrounds in the PM theory require choosing $c^2=-1$, which renders the kinetic term of the massive perturbation ghost-like~\cite{Hassan:2015tba}. Hence, the Weyl invariance at lowest order in the equations of motion~(\ref{bachpm}) seems to be obtainable only at the cost of giving up unitarity.

\item The authors of~\cite{Fasiello:2013woa} argued that, in order to give rise to a nonlinear PM symmetry, the scalar mode of the massive field needs to disappear entirely from a ``decoupling limit" of bimetric theory. 
It was then demonstrated that, even though the pure scalar interactions are indeed absent, the mode reappears in interactions with the vectors.\footnote{From our point of view, it is not obvious if this analysis is actually able to rule out the gauge symmetry because, firstly, the backgrounds that~\cite{Fasiello:2013woa} assumed for the metrics do not solve the equations of motion and, secondly, the constraint that removes the ghost is not imposed. As a consequence of the latter, the vector-scalar interactions involve higher-derivative terms of the scalar mode that are known to disappear after accounting for the constraint.} The fact that the vectors vanish on maximally symmetric backgrounds explains why one sees a PM gauge symmetry to linear order around dS space but not beyond.

\item 
Imposing a closure condition on the PM transformations, the authors of~\cite{Garcia-Saenz:2014cwa} showed that no nonlinear Lagrangian involving at most two derivatives on the fields can realise the required symmetry for a  single spin-2 field in a gravitational background. The setup is more general than ghost-free massive gravity but it still does not include the bimetric case.

\item
More group theoretical evidence against a nonlinear PM symmetry for spin-2 fields coupled to gravity was provided in~\cite{Joung:2014aba} where it was shown that no unitary theory exists and the unique non-unitary example is conformal gravity. However, for technical reasons which are discussed in~\cite{Hassan:2015tba}, these arguments do not directly carry over to bimetric theory.

\item
In \cite{Garcia-Saenz:2015mqi} the authors provide arguments against the existence of a partially massless theory with a non-abelian Yang-Mills like extension. Again, these arguments cannot directly be applied to the bimetric setup.

\end{itemize}

If the above results eventually turn out to also be extendable to bimetric theory, this would imply that the PM symmetry can indeed not be realised as a nonlinear theory involving only spin-2 fields. But even in this case, the bimetric candidate model seems to possess interesting properties that could pave the way towards understanding partial masslessness from a background-independent point of view. A promising future direction could be the combination of bimetric theory with higher-spin degrees of freedom. Finally, let us point out that finding an additional scalar symmetry is not merely an interesting exercise but could indeed prove to be very useful from several perspectives. First of all, it would guarantee that the helicity-zero mode is absent even nonlinearly. This will have an enormous effect on the phenomenology since the scalar mode is usually responsible for various instability issues and causes the most tension with observations. Secondly it would provide an argument for why a small cosmological constant is technically natural since its value is tied to the mass of the spin-2 field via a symmetry, whilst a small mass is itself technically natural since its vanishing restores full diffeomorphism invariance. Thirdly, the absence of the scalar mode may lead to improved quantum behaviour of the theory. Indeed, since the symmetry if it exists is given by Weyl scalings to lowest order in a derivative expansion, the theory may have much better renormalisation properties than GR.

\newpage
\section{Extensions of Bimetric Theory}\label{sec:extbim}

Extensions of massive gravity have been proposed mainly by introducing new scalar degrees of freedom and in the context of cosmology. For instance, a scenario in which the spin-2 mass is obtained through a scalar condensate has been suggested in~\cite{D'Amico:2011jj} and a quasi-dilaton extension of the massive gravity action was constructed in~\cite{D'Amico:2012zv, DeFelice:2013dua}. Similarly $f(R)$ extensions have been studied in e.g.~\cite{Nojiri:2012zu, Nojiri:2012re, Kluson:2013yaa, Bamba:2013fha, Nojiri:2015qyc} and other scalar fields with non-minimal coupling terms have also been added to the bimetric action, as in~\cite{Darabi:2015ssa}.

In the following we shall focus on more direct generalisations which preserves the structure of the bimetric theory itself without adding additional degrees of freedom, except the most natural generalisation to include interactions of multiple massive spin-2 fields.\footnote{Interactions between multiple massless spin-2 fields are forbidden on quite general grounds~\cite{Boulanger:2000rq}.} First, we discuss its generalisation to higher-dimensions, where new kinetic terms satisfy the requirement of classical consistency. Then we shall review interactions of multiple spin-2 fields and, in the same context, also present the vierbein formulation of bimetric theory.

\subsection{Higher derivative Lanczos-Lovelock extension}\label{sec:LL}

According to the constructive consistency proof in section~\ref{sec:construction}, the interaction potential of the Hassan-Rosen bimetric theory includes all  ghost-free non-derivative terms. Therefore, the only option to obtain an extended version of the theory is to add more derivative terms to the action. The literature contains several no-go theorems on generalising the kinetic structure of the spin-2 fields in four dimensions to anything beyond the Einstein-Hilbert term~\cite{deRham:2013tfa, deRham:2014tga, deRham:2015rxa}.\footnote{See, however,~\cite{Folkerts:2011ev, Hinterbichler:2013eza, Li:2015izu}. Moreover, note that the structure of the kinetic terms is of course only determined up to field redefinitions.} On the other hand, in dimensions greater than four, the Lanczos-Lovelock (LL) invariants, which either vanish or are topological in $d=4$, are expected not to reintroduce the Boulware-Deser ghost~\cite{Lanczos:1938sf, Lovelock:1971yv, Lovelock:1972vz, Deser:2011zk, Sisman:2012rc}.
These are defined as totally antisymmetric contractions of Riemann tensors $R^{\alpha\beta}_{\phantom{\alpha\beta}\mu\nu}$,
\beqn\label{LLdef}
\mathcal{L}_{(n)}=\frac{1}{2^nn!}\delta^{\mu_1}_{[\alpha_1}\delta^{\nu_1}_{\beta_1}\hdots\delta^{\mu_n}_{\alpha_n}\delta^{\nu_n}_{\beta_n]}
\prod_{k=1}^nR^{\alpha_k\beta_k}_{\phantom{\alpha_k\beta_k}\mu_k\nu_k}\,.
\eeqn
The definition of the antisymmetric product and relations to analogue expressions in terms of Levi-Civita symbols is given in appendix~\ref{app:technical}. In spite of being higher-derivative operators, the LL invariants are believed to avoid inconsistencies that are usually introduced by such terms. This is due to the antisymmetric structure in (\ref{LLdef}) which ensures the absence of more than two time derivatives acting on one field in the action. It is interesting that the same structure appears in the interaction potential of ghost-free bimetric theory, but of course this is not a complete coincidence: In the decoupling limit of nonlinear massive gravity it can be seen that the antisymmetric structure removes higher time derivatives from the longitudinal modes of the St\"uckelberg fields~\cite{deRham:2010ik, deRham:2010kj}.  

\subsubsection{Extended action and equations of motion}

In order to extend ghost-free bimetric theory to $d$ dimensions, we first write the Hassan-Rosen bimetric action in the more general form,
\begin{align}\label{SbmHRd}
	S_\mathrm{HR} =\, m_g^{d-2}\int\td^dx\sqrt{g}\,R(g)&+m_f^{d-2}\int\td^dx\sqrt{f}\,R(f)\nn\\
	&-2m^d\int\td^dx\sqrt{g}\sum_{n=0}^d\beta_ne_n\left(\sqrt{g^{-1}f}\right)\,,
\end{align}
where the elementary symmetric polynomials are still defined through the same recursion formula~(\ref{recursen}). The consistency proof of~\cite{Hassan:2011zd}, as outlined in sections~\ref{sec:construction} and~\ref{sec:aogbt}, generalises straightforwardly to the higher-dimensional case. It has also been shown that the mass spectrum is very similar to the four-dimensional case and contains a massless and massive excitation~\cite{Hassan:2012rq}.

Next, guided by the above considerations, we extend the Hassan-Rosen action by the following terms, as done in~\cite{Paulos:2012xe, Hassan:2012rq},
\begin{align}\label{LLaction}
S_\mathrm{LL}~&=~m_g^{d-2}\int\dd^dx\sum_{n=2}^{\lfloor d/2 \rfloor}\sqrt{-g}~l_n^g\,\mathcal{L}_{(n)}(g)\nn\\
&+~m_f^{d-2}\int\dd^dx\sum_{n=2}^{\lfloor d/2 \rfloor}\sqrt{-f}~l_n^f\,\mathcal{L}_{(n)}(f)\,,
\end{align}
where we have introduced two sets of couplings, $l_n^g$ and $l_n^f$, of mass dimension $2(1-n)$.
It is well known that the LL invariant $\mathcal{L}_{(n)}$ is topological in $d=2n$ and vanishes for $2n>d$ which is why the sums in (\ref{LLaction}) terminate at integer part of $d/2$.

The equations of motion for $\gmn$ and $\fmn$ that follow from the LL extended bimetric action are
\beqn\label{LLeom}
\mathcal{G}_{\mu\nu}(g)+\sum_{n=2}^{\lfloor d/2 \rfloor}l_n^g\mathcal{G}^{(n)}_{\mu\nu}(g)+\frac{m^d}{m_g^{d-2}}V^g_{\mu\nu}&=&0\,, \nn\\
\mathcal{G}_{\mu\nu}(f)+\sum_{n=2}^{\lfloor d/2 \rfloor}l_n^f\mathcal{G}^{(n)}_{\mu\nu}(f)+\frac{m^d}{m_g^{d-2}}V^f_{\mu\nu}&=&0\,.
\eeqn
Here, $\mathcal{G}^{(n)}_{\mu\nu}$ are the Lovelock tensors that follow from variation of (\ref{LLaction}) and $V^g_{\mu\nu}$ and $V^f_{\mu\nu}$ are the contributions from the interaction potential which were defined in (\ref{potconbim}) and in which the summations now run from $0$ to $d$. We will not need the explicit expressions for the Lovelock tensors, but let us note the important property,
\beqn\label{usefid}
g^{\mu\nu}\mathcal{G}^{(n)}_{\mu\nu}(g)=\frac{2n-d}{2}\mathcal{L}_{(n)}(g)\,,
\eeqn
which will be useful later on.

\subsubsection{Proportional backgrounds}

As in pure bimetric theory in $d=4$ we are interested in finding the proportional background solutions to (\ref{LLeom}) and study perturbations around those. Again, these backgrounds correspond to maximally symmetric spacetimes, for which the curvatures satisfy
\beqn\label{bgcurv}
R_{\mu\nu\rho\sigma}(\bar{g})&=&\frac{2\lambda}{(d-1)(d-2)}(\bar{g}_{\mu\rho}\bar{g}_{\nu\sigma}-\bar{g}_{\nu\rho}\bar{g}_{\mu\sigma})\,,\nn\\
R_{\mu\nu}(\bar{g})&=&\frac{2\lambda}{d-2}\bgmn\,,\qquad R(\bar{g})~=~\frac{2d\,\lambda}{d-2}\,,
\eeqn
with cosmological constant $\lambda$. On such backgrounds with constant curvature, the $n$-th order LL invariant is proportional to $\lambda^n$,
\beqn\label{LLbg}
\mathcal{L}_{(n)}=N_n(d)\,\lambda^n\,,\quad\text{with}~~N_n(d)\equiv\frac{2^n d!}{(d-1)^n(d-2)^n(d-2n)!}\,.
\eeqn
Next, we trace the equations of motion (\ref{LLeom}) in order to obtain two scalar equations and insert (\ref{bgcurv}) for the curvatures.
Furthermore, we make use of the identity (\ref{usefid}) and plug in the proportional ansatz, $\bfmn=c^2\bgmn$. 
Finally, in the equation for $\fmn$, we use $\mathcal{L}_{(n)}(c^2 \bar{g})=c^{-2n}\mathcal{L}_{(n)}( \bar{g})$. The resulting two equations read
\beqn\label{LLbgeq}
\lambda + \sum_{n=2}^{\lfloor d/2 \rfloor}l^g_n\frac{d-2n}{2d}N_n(d)\lambda^n - \Lambda_g&=&0\,,\nn\\
\lambda + \sum_{n=2}^{\lfloor d/2 \rfloor}c^{2-2n}l^f_n\frac{d-2n}{2d}N_n(d)\lambda^n - \Lambda_f&=&0\,.
\eeqn
Here, $\Lambda_g$ and $\Lambda_f$ are the contributions from the potential evaluated on the proportional backgrounds which were defined earlier in (\ref{lambdas}). Note, however, that these no longer correspond to the physical cosmological constants and are not necessarily equal. 

For general parameters $\beta_n$, $l_n^g$, and $l_n^f$, the two algebraic equations in (\ref{LLbgeq}) serve to determine the background curvature $\lambda$ as well as the proportionality constant $c^2$. Note also that in the absence of the LL contributions, i.e.~for $l_n^g=l_n^f=0$, we re-arrive at the bimetric background equations, $R(\bar{g})=\tfrac{2d}{d-2}\Lambda_g$ and $\Lambda_g=\Lambda_f$.

\subsubsection{Mass spectrum}

Computing the linearised equations of motion around general background solutions to (\ref{LLeom}) is quite involved, even when the potential contributions vanish. However, as has been observed in \cite{Sisman:2012rc}, around constant curvature backgrounds with cosmological constant $\lambda$, the fluctuation equations of GR augmented by the LL terms with couplings $l_n^g$ assume the same form as the ones obtained from Einstein's equations. The only difference is that, in the linearised equations of the LL extended theory, the Planck mass $m_g$ needs to be replaced by an effective mass parameter $\tilde{m}_g$ defined through
\beqn\label{modmg}
\frac{\tilde{m}_g^{d-2}}{m_g^{d-2}}=1+(d-3)!\sum_{n=2}^{\lfloor d/2 \rfloor}\frac{n(d-2n)l_n^g}{(d-2n)!}\left(\frac{2\lambda}{(d-1)(d-2)}\right)^{n-1}.
\eeqn
It is a remarkable feature of the Lovelock invariants that, in spite of being higher-derivative operators in the nonlinear theory, they give rise to linearised equations that are only second order in derivatives.

For the bimetric case the above result implies that the linear equations for the theory with LL terms are the same as in pure bimetric theory if one replaces $m_g$ by $\tilde{m}_g$ as above and $m_f$ by $\tilde{m}_f$ given by
\beqn\label{modmf}
\frac{\tilde{m}_f^{d-2}}{m_f^{d-2}}=1+(d-3)!\sum_{n=2}^{\lfloor d/2 \rfloor}\frac{n(d-2n)c^{2-2n}l_n^f}{(d-2n)!}\left(\frac{2\lambda}{(d-1)(d-2)}\right)^{n-1}.
\eeqn
Hence, using the results of~\cite{Sisman:2012rc}, without any further computation we can conclude that the linear equations in a maximally symmetric background with cosmological constant $\lambda$ will again diagonalise into a massless and massive equation, where in the latter the Fierz-Pauli mass (\ref{FPmass}) is now given in terms of $\tilde{m}_g$ and $\tilde{m}_f$.

\subsubsection{Partial masslessness in $d > 4$}

To conclude the discussion of the Lanczos-Lovelock extension of bimetric theory, let us comment on the possibility to extend the linear symmetry of partially massless (PM) fields to the nonlinear level within this generalised setup. Since the proportional backgrounds and their perturbations possess the same structure as in pure bimetric theory in $d=4$, the linear equations are again invariant under a PM symmetry if the mass is on the Higuchi bound which in $d$ dimensions corresponds to the value $m_{\mathrm{FP}}^2=\frac{2}{d-1}\lambda$. 
The arguments given in section~\ref{sec:idPM} leading to a PM candidate theory can straightforwardly be applied to the higher-dimensional case. The criterion of leaving the proportionality constant $c$ undetermined by the background equations~(\ref{LLbgeq}) singles out unique PM candidate theories which have been identified in~\cite{Hassan:2012rq} up to $d=8$. All the interaction parameters are fixed with respect to each other in these models and, in particular, the Lanczos-Lovelock coefficients $l^g_n$ and $l^f_n$ are nonzero. This result confirms earlier findings~\cite{Zinoviev:2006im} which revealed that the linear PM symmetry cannot be extended in $d>4$ unless higher-derivative terms are included into the action.

\subsection{Multiple interacting spin-2 fields}\label{sec:vierbein}

Ghost-free bimetric theory contains the correct number of degrees of freedom for a massless and a massive spin-2 mode. While no-go theorems forbid consistent interactions among more than one massless spin-2 field~\cite{Boulanger:2000rq}, it is possible to extend bimetric theory by additional massive spin-2 degrees of freedom. Hinterbichler and Rosen first wrote down these ghost-free interactions using vierbeins instead of metrics~\cite{Hinterbichler:2012cn}.\footnote{The formulation of the full ghost-free bimetric potential in terms of vierbeins first appeared in~\cite{Volkov:2011an}.} In the following we will derive the consistent multi-spin-2 interactions in the metric formulation and then review their formulation in terms of vierbeins.
More work on the vierbein formulation of massive gravity, bimetric and multimetric theory can be found in~\cite{Hassan:2012wt, Volkov:2012cf, Zinoviev:2012yb, Deffayet:2012nr, Kluson:2013aca, Bergshoeff:2013xma, Tamanini:2013xia, deRham:2013awa, Alexandrov:2013rxa, Noller:2013yja, Deser:2014hga, Noller:2014sta, Soloviev:2014eea, Scargill:2014wya, Noller:2015eda, Hinterbichler:2015yaa, deRham:2015rxa}.

\subsubsection{Multiple bimetric couplings}  

When constructing a theory for more than two spin-2 fields, of course, the main requirement on its structure remains the absence of the Boulware-Deser ghosts. The interaction potential for $\mathcal{N}$ fields therefore has to be chosen such that the action in Hamiltonian formulation gives rise to $(4+\mathcal{N}-1)$ constraints. Out of these, four should be first class constraints, generating the group of general coordinate transformations, while the remaining $(\mathcal{N}-1)$ are expected to be second class and serve to remove all Boulware-Deser ghosts from the spectrum.

An obvious way to satisfy these requirements is to simply add up several copies of the Hassan-Rosen action~(\ref{SbmHR}) for fields $\fmn$ and $\gmn (I)$ with $I=1,\hdots, \mathcal{N}-1$, to obtain a theory in which one of the metrics is coupled to all the others,
\beqn\label{centercoup}
S_\mathrm{centre}\big[f,\{g(I)\}\big]=\sum_{I=1}^{\mathcal{N}-1}S_{\mathrm{HR}}\big[g(I),f\big]\,.
\eeqn
This will be consistent because in every interaction potential one can redefine the ADM variables for $\gmn(I)$ to make it linear in the lapse and shift of~$\fmn$ as well as the lapse of $\gmn(I)$, such that one obtains the desired $(4+\mathcal{N}-1)$ constraints. 

The left panel of Figure~1 shows how such a coupling can be visualised in a graph~\cite{Hinterbichler:2012cn}. Each solid dot represents a different spin-2 field and the lines stand for standard Hassan-Rosen bimetric couplings between them. Fields corresponding to dots that are not connected by a line do not directly interact. Due to the corresponding picture with one metric in the centre, we shall refer to interactions of the form~(\ref{centercoup}) as ``centre couplings".

Another option to obtain consistent multiple bimetric interactions is to build a ``chain" of $\mathcal{N}$ coupled spin-2 fields $\gmn(I)$ where $I=1,\hdots,\mathcal{N}$. That is to say, we take an action of the form,
\beqn
S_\mathrm{chain}\big[\{g(I)\}\big]=\sum_{I=1}^{\mathcal{N}-1}S_{\mathrm{HR}}\big[g(I),g(I+1)\big]\,,
\eeqn
for which the corresponding graph is depicted in the right panel of Figure~1.
This construction works because in the first term one can redefine the ADM variables for $g(1)$ such that it becomes linear in the lapse and shift of $g(2)$ as well as the lapse of $g(1)$. In the second term one performs a similar redefinition for the shift of $g(2)$ which, since the redefinition is linear in the lapses, does not destroy the linearity in the lapse of $g(2)$ in the first term. Note however that the second term will no longer be linear in the shift of $g(2)$. Continuing with this procedure in all the terms results in an action that is linear in the lapse and shift of $g(\mathcal{N})$ and in all the other lapses.

    \begin{figure}[h]
    \begin{center}
    \includegraphics[width=210pt]{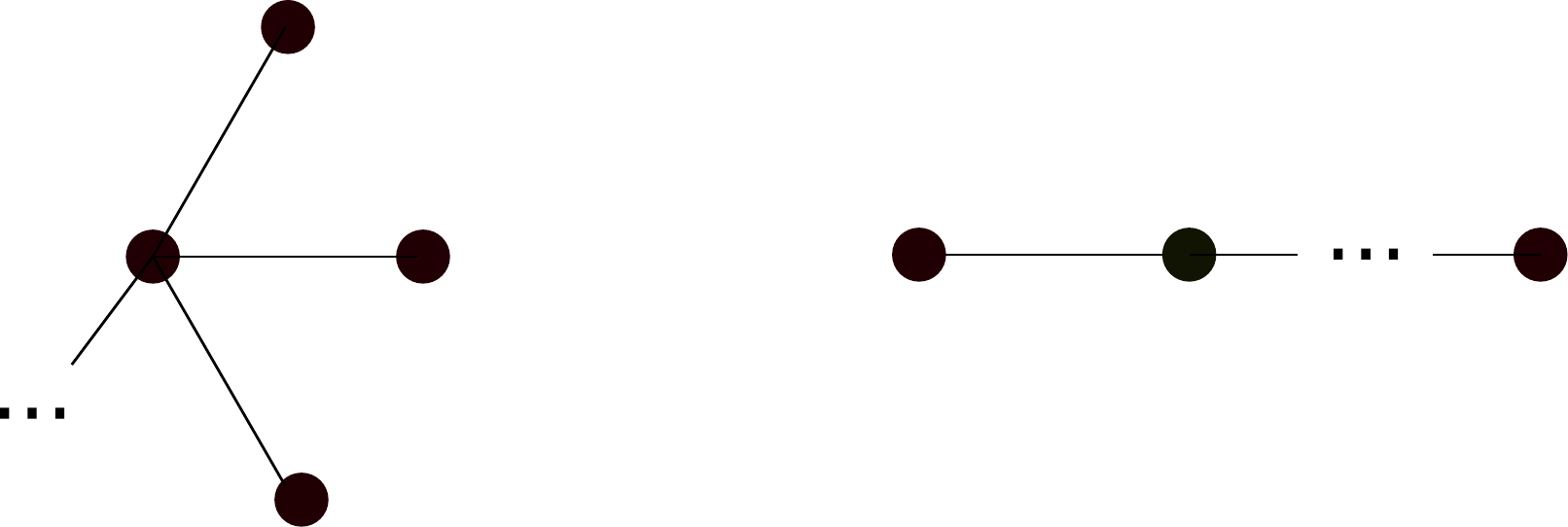}
    \caption{The centre couplings \textit{(left)} and the chain couplings \textit{(right)}.}
    \end{center}\label{figcentchaint}
    \end{figure}

In fact, it is also possible to combine the two above constructions in the following manner: First introduce a chain coupling for fields $f(K)$, $K=1,\hdots,\mathcal{K}$, and then to each $f(K)$ attach $\mathcal{I}_K$ fields $g_{(K)}(I_K)$, $I_K=1,\hdots\mathcal{I}_K$, in a centre coupling. This gives rise to the most general ghost-free multimetric action,
\beqn\label{gencv}
S_\mathrm{multi}~=~S_\mathrm{chain}\big[\{f(K)\}\big]+\sum_{K=1}^\mathcal{K}S_\mathrm{centre}\big[f(K),\{g_{(K)}(I_K)\}\big]\,.
\eeqn
In this theory the total number of fields is $\mathcal{N}=\mathcal{K}+\sum_{K=1}^\mathcal{K}\mathcal{I}_k$. An example for a graph is shown in the left panel of Figure~2.
  \begin{figure}[h]
    \begin{center}$
    \begin{array}{cc}
    \includegraphics[width=140pt]{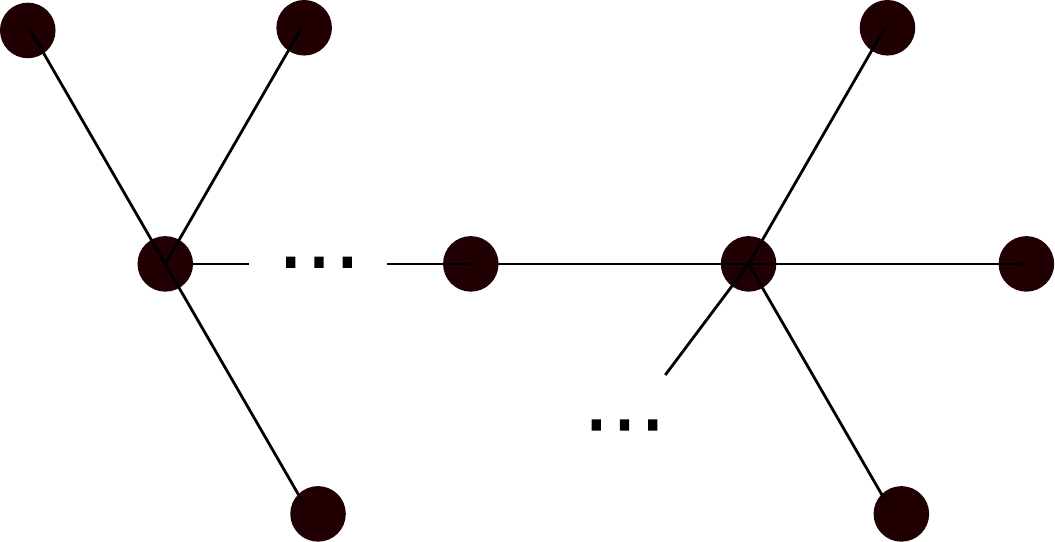}~~~~~~~~ & ~~~~~~~~
    \includegraphics[width=45pt]{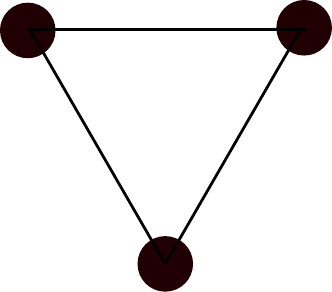}
    \end{array}$
    \caption{The combined chain and centre couplings \textit{(left)} and a forbidden loop coupling \textit{(right)}.}
    \end{center}
    \end{figure}
Note that the action~$S_\mathrm{multi}$ still consists purely of multiple copies of the Hassan-Rosen bimetric action; we have not introduced any type of new interaction term. 
Of course, in the above construction, it is also possible to add further chain or centre couplings to each of the $g_{(K)}(I_K)$.
However, it is important to realise that not all possible bimetric couplings are allowed. In particular, one may not couple the metrics in a ``loop", which means adding terms such as $S_{\mathrm{HR}}\big[f(K),f(K+2)\big]$ or $S_{\mathrm{HR}}\big[f(K+1),g_{(K)}(I_K)\big]$ to (\ref{gencv}). To see why this fails, consider a simple loop coupling among three metrics,
\beqn\label{loopact}
S_\mathrm{loop}(g,f,h)=S_{\mathrm{HR}}(g,f)+ S_{\mathrm{HR}}(f,h)+S_{\mathrm{HR}}(h,g)\,,
\eeqn
whose graph is depicted in the right panel of Figure~2.
Now imagine that in the first term one redefines the shift of $g$ to render it linear in the lapse and shift of $f$ and the lapse of $g$. Then, in the second term redefine the shift of $f$ such that the first two terms afterwards are linear in the lapse and shift of $h$ and the lapses of $g$ and $f$. In the third term, one would now have to redefine the shift of $h$, but from the particular form~(\ref{shiftred}) of the redefinition it is clear that afterwards, the action will again contain the original shift variable for $g$ which cannot be expressed in terms of the redefined variables only. Therefore there is at least no obvious consistent redefinition of ADM variables that renders (\ref{loopact}) linear in all the lapses and we conclude that the loop couplings most likely give rise to Boulware-Deser ghosts. This is in agreement with results of a similar analysis performed in~\cite{Nomura:2012xr}.

\subsubsection{Vierbein formulation}  \label{sec:vbform}

The multimetric action~(\ref{gencv}) consists of several copies of the Hassan-Rosen action~(\ref{SbmHR}). It is possible to rewrite the latter in terms of different variables which avoid the appearance of the square-root matrix.\footnote{For suggestions to remove the square-root matrix by introducing auxiliary fields, see~\cite{Golovnev:2011aa, Schmidt-May:2014tpa}.}
To this end, let us introduce the tetrads or vierbeins defined for each metric,
\beqn\label{defvielbein}
g_{\mu\nu}(I)={E^a}_\mu(I)\eta_{ab}{E^b}_\nu(I)\,,
\eeqn  
or in matrix notation, $g(I)= E^\mathrm{T}(I)\eta E(I)$. The expression for the metric in terms of vierbeins is invariant under local Lorentz transformations $\Lambda^a_{~b}(I)$ that act on the upper index of ${E^b}_\mu(I)$ according to $E(I)\rightarrow \Lambda(I)E(I)$ with $\Lambda^\mathrm{T}\eta \Lambda=\eta$. Associated to each vierbein, there are thus six gauge invariances which serve to reduce the number of physically relevant components from 16 to 10.
Any metric theory that is reformulated in terms of vierbeins is automatically invariant under these transformations and, in particular, the Einstein-Hilbert terms $S_\mathrm{EH}(I)$, in which the metric $g(I)$ is replaced by the vierbein, are Lorentz invariant.

In terms of the respective vierbeins, the square-root matrix $\sqrt{g(1)^{-1}g(2)}$ built out of two metrics $g(1)$ and $g(2)$ becomes,
\beqn
S=\sqrt{g(1)^{-1}g(2)}=\sqrt{E(1)^{-1}\eta^{-1}  (E(1)^\mathrm{T})^{-1} E(2)^\mathrm{T}\eta E(2)}\,.
\eeqn
It is now convenient to fix the Lorentz gauge for one of the vierbeins such that $ E(2) E(1)^{-1}\eta^{-1}$ is a symmetric matrix. We thus impose the so-called ``symmetric" gauge (or Deser-van Nieuwenhuizen gauge~\cite{Deser:1974cy}),
\beqn\label{vbsymcond}
\eta^{-1}(E(1)^\mathrm{T})^{-1} E(2)^\mathrm{T}=E(2) E(1)^{-1}\eta^{-1}\,.
\eeqn
In this particular Lorentz frame, we can easily evaluate the square-root as,
\beqn\label{sqrtev}
S=\sqrt{E(1)^{-1}E(2) E(1)^{-1}  E(2)}=E(1)^{-1}  E(2)\,.
\eeqn
Using the identity,
\beqn
{\epsilon}_{c_1\hdots c_d}=\big(\det E\big)\,{\epsilon}_{\nu_1\nu_2\nu_3 \nu_4}{(E^{-1})^{\nu_1}}_{c_1}
{(E^{-1})^{\nu_2}}_{c_2}{(E^{-1})^{\nu_3}}_{c_3}{(E^{-1})^{\nu_4}}_{c_4}\,,
\eeqn
we can then write the bimetric action in the form
\begin{align}\label{vielbeinpot1}
S_\mathrm{HR}=S_\mathrm{EH}(1)+S_\mathrm{EH}(2)-S_\mathrm{int} \,,\qquad
S_\mathrm{int}\equiv 2m^4\int\dd^4x~\sum_{n=0}^4V_n\,,
\end{align}
where $S_\mathrm{EH}(I)$ are the standard Einstein-Hilbert actions for $E(I)$ and
\begin{align}\label{vbpotdef}
V_n&=b_n{\epsilon}^{\mu_1\hdots\mu_n\mu_{n+1}\hdots \mu_4}{\epsilon}_{a_1\hdots a_n a_{n+1}\hdots a_4}{E^{a_1}}_{\mu_1}(1)\hdots {E^{a_n}}_{\mu_n}(1)
{E^{a_{n+1}}}_{\mu_{n+1}}(2){E^{a_4}}_{\mu_4}(2)\,,
\end{align}
where $b_n=\frac{2\beta_n}{n!(4-n)!}$ are the same parameters as in~(\ref{completep}). Since we have only fixed one of the two Lorentz gauges, this interaction potential still has one overall local Lorentz invariance under which both vierbeins transform in the same way. 

It is important to note that the vierbeins in~(\ref{vbpotdef}) must satisfy the symmetrisation condition~(\ref{vbsymcond}), otherwise the vierbein formulation is not equivalent to bimetric theory for general values of $b_n$. More precisely, if we started from~(\ref{vbpotdef}) without imposing the condition by hand, then the dynamics of the vierbein theory would allow for configurations that do not give back the ghost-free bimetric formulation. As discussed in~\cite{Hinterbichler:2012cn}, for each vielbein there exist six combinations of its equations of motion that do not contain any derivatives. These read,
\beqn\label{constvierb}
\frac{\delta S_\mathrm{int}}{\delta {E^a}_\mu(I)}{E^c}_\mu(I)\eta_{cb}-\frac{\delta S_\mathrm{int}}{\delta {E^b}_\mu(I)}{E^c}_\mu(I)\eta_{ca}=0\,.
\eeqn
For models with $b_2=b_3=0$ or $b_1=b_2=0$, these equations imply the symmetrisation condition~(\ref{vbsymcond}) and the equivalence to bimetric theory is dynamically guaranteed. For more general parameters, the symmetrised vierbeins still solve the equations but other solutions also exist~\cite{Deffayet:2012zc, Banados:2013fda}. These disconnected branches give rise to different theories which contain the Boulware-Deser ghost instability~\cite{deRham:2015cha}.

Another important aspect, first pointed out in~\cite{Deffayet:2012zc}, is that one cannot always arrive at the symmetrised form~(\ref{vbsymcond}) by acting with a Lorentz transformation on general vierbeins. For configurations that do not allow this, the square-root cannot be evaluated as in~(\ref{sqrtev}). In fact, the condition on the combination of vierbeins to be symmetrisable by a Lorentz transformation is exactly the same as the one we derived in section~\ref{sec:exofred} on the metric components by requiring that the square-root exists~\cite{Hassan:2014gta}:
Ensuring that the Lorentz boost velocity $v_a$ satisfies $v_a \delta^{ab}v_b<1$ is equivalent to requiring the existence of a real solution to the scalar square-root $\sqrt{x}$ that appears in the ADM decomposition of $\sqrt{g^{-1}f}$ in (\ref{exprab}). This condition in turn ensures that one can always impose the ``symmetric" gauge by a Lorentz transformation.

\subsubsection{More general interactions?}  

So far we have used the earlier results of the bimetric ADM analysis in order to couple several spin-2 fields to each other using only bimetric interactions. The next question is whether there exist more general couplings than the bimetric vertex. We introduce graphs for two examples of such $n$-point vertices in Figure~3. 
\begin{figure}[h]
    \begin{center}
    \includegraphics[width=150pt]{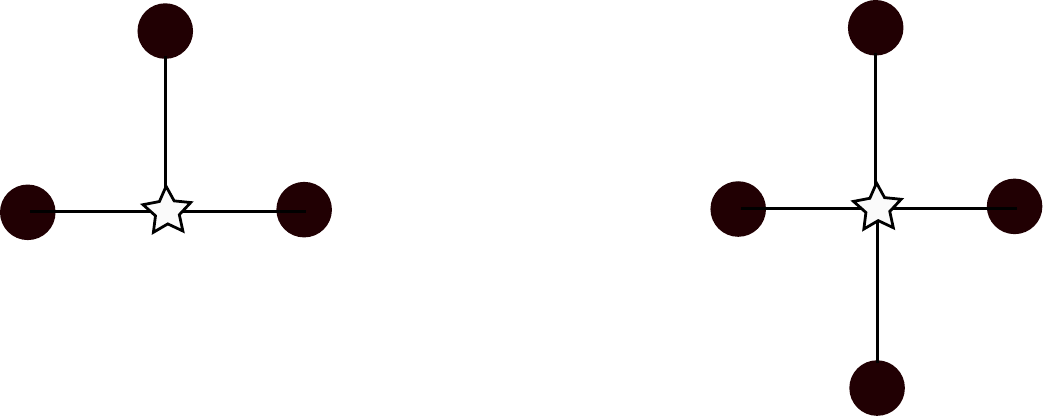}
       \caption{A three- and a four-point vertex.}
    \end{center}
    \end{figure}
A priori, such couplings could be very different from the bimetric ones and the ADM analysis has to be redone from the start.

In~\cite{Hinterbichler:2012cn} it was proposed that the interactions of vierbeins could be generalised to,
\begin{align}\label{vielbeinpot}
S_\mathrm{int}&=\frac{m^4}{4}\sum_{\{I\}}\int\dd^4x~T^{I_1 I_2 I_3 I_4}U_{I_1 I_2 I_3 I_4}\,,\nn\\
\text{with}~~~~U_{I_1\hdots I_4}&={\epsilon}^{\mu_1\mu_2\mu_3\mu_4}{\epsilon}_{a_1a_2 a_3 a_4}{E^{a_1}}_{\mu_1}(I_1){E^{a_2}}_{\mu_2}(I_2){E^{a_3}}_{\mu_3}(I_3){E^{a_4}}_{\mu_4}(I_4)\,,
\end{align}
where $T^{I_1 I_2 I_3 I_4}$ are coupling constants symmetric in all the indices $I_n\in \{I\}$ which run from 1 to $\mathcal{N}$. The total multivielbein action would then be given by
\beqn\label{multac}
S_\mathrm{multi}=\sum_{I=1}^\mathcal{N}S_\mathrm{EH}(I)+S_\mathrm{int}\,.
\eeqn
Note that the antisymmetric structure of (\ref{vielbeinpot}) ensures that there cannot be more than four vierbeins (or in $d$ dimensions, not more than~$d$ vielbeins) interacting in one vertex. 

However, the results obtained in~\cite{deRham:2015cha} severely constrain the structure of the ``tensor" of coupling constants $T^{I_1\hdots I_4}$ and only diagrams of the type displayed in Figure~1 are allowed.
The more general couplings displayed in Figure~3 which would arise from an unconstrained $T^{I_1\hdots I_4}$ and give rise to diagrams with at most $4$ lines ending in one vertex are excluded. The same holds for the loop couplings in the right panel of Figure~2.
Although these forbidden terms have been shown to possess a ghost-free decoupling limit, they do spoil the consistency of the full theory~\cite{deRham:2015cha, deRham:2015rxa} and the reason can be traced back to the fact that the corresponding equations of motion are incompatible with the symmetrisation condition~(\ref{vbsymcond}). 
In summary, the only consistent vierbein couplings are those that give rise to a metric formulation of the type~(\ref{gencv}).


\newpage
\section{Discussion and Outlook}\label{sec:discuss}

In this article we summarised the recent developments of nonlinear theories involving massive spin-2 fields. It was shown how to construct the unique ghost-free interaction potential of nonlinear massive gravity with fixed reference metric, which we afterwards generalised to the fully dynamical bimetric theory. The latter describes the interactions of a massless spin-2 field with a massive one and, when coupled to matter, captures the behaviour of gravity in the presence of an additional tensor field. We have seen that there exists a limit in the parameter space of bimetric theory which takes the theory arbitrarily close to general relativity. In this limit, viable cosmological solutions can be found where, in the absence of vacuum energy, the dark energy scale is set by the spin-2 mass which is protected against receiving large quantum corrections. We have discussed the idea of nonlinear theories for partially massless spin-2 fields and in the process established a connection between bimetric theory and conformal gravity. Finally, we have reviewed the extension of bimetric theory to dimensions greater than four and to couplings between more than two spin-2 fields.

On the phenomenological side, one of the most interesting open questions is what possible implications the existence of a massive spin-2 field in nature could have on the dark matter problem. Several scenarios where dark matter components are attributed to the matter sector of the second metric $\fmn$ have already been considered in~\cite{Aoki:2014cla,Bernard:2014psa, Blanchet:2015sra, Blanchet:2015bia}. From our point of view, however, the more interesting option is to regard the second metric as the dark matter field. Since the spin of the dark matter particle is unknown, such a scenario is not immediately excluded and, as we already discussed in section~\ref{sec:GRlim}, the largeness of the physical Planck mass automatically weakens the interactions of the massive spin-2 field with all Standard Model particles.

On the theoretical side, it would be of great importance to understand the origin of the mass term for the spin-2 field at a fundamental level. For spin-1 fields, it is well-known that the consistent picture at the quantum level is to generate their mass via spontaneous breaking of the gauge symmetry in the massless theory. To this end, it is necessary to invoke additional (Higgs) fields whose vacuum expectation values break gauge invariance and set the spin-1 mass scale. Since aspects of field theories generically do not become simpler with increasing spin, it is natural to expect an underlying mechanism that generates the mass term also for spin-2 fields. Such a ''spin-2 Higgs mechanism" should break the two independent copies of diffeomorphism invariance in the $\gmn$ and $\fmn$ sectors down to their diagonal subgroup, which is the residual gauge symmetry of bimetric theory. The precise nature of the fields that are needed to trigger this symmetry breaking spontaneously is still unknown; a realisation on anti-de-Sitter space has been suggested in~\cite{Porrati:2001db,Porrati:2003sa}. For a more detailed discussion of the topic, we refer the reader to the recent article~\cite{Goon:2014paa}.
\\
Related to the origin of the mass term is the behaviour of bimetric theory at the quantum level. Quantum corrections at one-loop order in the effective field theory picture have been computed in massive gravity~\cite{deRham:2013qqa} and bimetric theory~\cite{Heisenberg:2014rka} and the results confirmed the stability of the spin-2 mass scale. On the other hand, without knowing the underlying mechanism that generates the interaction potential for the two metrics, it is difficult to tell to what extend these results can be trusted.

Theories with extended symmetries are known to generically exhibit an improved quantum behaviour. An example for the spin-2 case is the action for conformal gravity with its Weyl symmetry. Unfortunately, this theory suffers from a ghost but we have seen that bimetric theory contains a ghost-free model which seems to be closely related to the Weyl invariant action. Understanding its properties further and investigating possibilities to realise the gauge symmetry for partially massless at the nonlinear level (most likely by invoking additional degrees of freedom) could give us important new insights on the nature of quantum gravity. 

Since consistent theories involving spin-2 fields are so rare, we do not expect to encounter a large variety of possibilities to extend them further. As it stands now, it does not seem to be possible to extend the kinetic sector in four dimensions~~\cite{deRham:2013tfa, deRham:2014tga, deRham:2015rxa} and the interaction potential is the only consistent one by construction. Up to field redefinitions, the structure of the ghost-free bimetric action is therefore unique. On the other hand, it is of course worth studying couplings to fields with different and, in particular, higher spins. The spectrum of presently known higher-spin theories contains only massless spin-2 fields (see, for instance,~\cite{Bekaert:2005vh, Taronna:2012gb, Didenko:2014dwa}). It would be interesting to see whether bimetric theory can give hints on the form of more general interactions including massive spin-2 degrees of freedom.
At this stage, we can only hope that one day we will be able to write down the most general interactions for massless and massive spin-2 fields.


\bigskip
{\bf Acknowledgements}: We would like to thank Yashar Akrami, Eugeny Babichev, Laura Bernard, C\'edric Deffayet, Jonas Enander, Dario Francia, Aleksander Garus, Fawad Hassan, Frank K\"onnig, Alessandro Sfondrini and Adam Solomon for fruitful discussions and comments on the draft. We are particularly grateful to Niklas Beisert who motivated us to write this review.\\
The research of MvS leading to these results has
received funding from the European Research Council under the European
Community's Seventh Framework Programme (FP7/2007-2013 Grant Agreement
no. 307934).
ASM is supported by ERC grant no.~615203 under the FP7 and the Swiss National Science Foundation through the NCCR SwissMAP.

\newpage

\appendix

\section{\label{app:technical} Technical details of interaction potential}
In the text, for notational simplicity, we frequently write the interactions in terms of elementary symmetric polynomials $e_n(S)$. These can be defined in various equivalent ways, for example via the following commonly used equalities which hold for any square $d\times d$ matrix S,
\begin{align}
e_n(S)&=S^{\mu_1}_{~[\mu_1}\cdots S^{\mu_{n}}_{~\mu_n]}
=\frac{1}{n!}\delta^{\mu_1\cdots\mu_n}_{\nu_1\cdots\nu_n}S^{\nu_1}_{~\mu_1}\cdots S^{\nu_n}_{~\mu_n}\nn\\
&=\frac{1}{n!(d-n)!}\,\epsilon^{\mu_1\mu_2\hdots\mu_n\lambda_{n+1}\hdots\lambda_d}\epsilon_{\nu_1\nu_2\hdots\nu_n\lambda_{n+1}\hdots\lambda_d}~S^{\nu_1}_{~\mu_1}\cdots S^{\nu_n}_{~\mu_n}\,.
\end{align}
Here we have expressed them in terms of antisymmetrisation, the generalised Kronecker delta and the Levi-Civita symbol respectively, all normalised by unit weight. Obviously the last equality only makes sense when $d$ is the dimension spanned by the indices.
They can also be defined through a dimension-independent recursion formula,
\beqn\label{recursen}
e_n(S)=\frac{(-1)^{n+1}}{n}\sum_{k=0}^{n-1}(-1)^{k}\tr (S^{n-k})e_k(S)\,,\quad \text{with}~~e_0(S)=1\,.
\eeqn
In $d$ dimensions, they satisfy $e_d(S)=\det(S)$ and $e_n(S)=0$ for $n>d$. Their explicit expressions for $n\leq 4$ read,
\begin{align}
e_0(S)&=1\,,\qquad 
e_1(S)= [S]\,,\qquad
e_2(S)=\tfrac{1}{2} \Big([S]^2-[S^2]\Big)\,,\nn\\
e_3(S)&=\tfrac{1}{6} \Big([S]^3-3[S][S^2]+2[S^3]\Big)\nn\\
e_4(S)&=\tfrac{1}{24} \Big([S]^4-6[S]^2[S^2]+3[S^2]^2+8[S][S^3]-6[S^4]\Big)\,,
\end{align}
where square brackets denote a matrix trace. Moreover, the $e_n$ obey the following useful identity for any matrix $S$ and parameter $\lambda$,
\be\label{app:enrel}
e_n(\mathbb{1}+\lambda S)=\sum_{k=0}^{n}{d-k\choose n-k}\lambda^ke_k(S)\,.
\ee
A special case ($n=d$) of this is,
\be\label{app:detrel}
\det(\mathbb 1+\lambda S)=\sum_{n=0}^{d}\lambda^n\,e_n(S)\,.
\ee
This means that the consistent interaction potential can be regarded as a \textit{deformed determinant}~\cite{Hassan:2011vm}. The variation of the $e_n(S)$ can be computed from e.g.~\eqref{recursen} and is given by,\footnote{See e.g.~\cite{Bernard:2015mkk} for an explicit derivation.}
\be\label{d_e_n}
\delta e_n(S)=-\sum_{k=1}^n(-1)^k\Tr[S^{k-1}\delta S]\,e_{n-k}(S)\,,\qquad n\geq1\,,
\ee
with $\delta e_0(S)=0$ (since $e_0(S)=1$). Together with $2\Tr[S^{k-1}\delta S]=\Tr[S^{k-2}\delta S^2]$ and $\delta S^2=-g^{-1}\delta g S^2 + g^{-1}\delta f$ (which follows from $S^2=g^{-1}f$) it is then straightforward to derive the field equations of the theory. Explicitly, the contributions from the interaction potential to the equations of motion in (\ref{potconbim}) are,
\beqn
(Y_{(n)})^\mu_{~\nu}(S)\equiv \sum_{k=0}^{n} (-1)^k (S^{n-k})^\mu_{~\nu}\,e_k(S)\,.
\eeqn
In $d$ dimensions, they satisfy $Y_{(n)}=0$ for $n\geq d$. In particular, $Y_{(d)}=0$ is simply the statement of the Cayley-Hamilton theorem, that any square matrix satisfies its own characteristic (or secular) equation.

Finally we note that all of the above expressions can easily be rewritten in terms of the inverse $S^{-1}$ by using the identity,
\be\label{enSSI}
e_n(S^{-1})=\frac{e_{d-n}(S)}{e_d(S)}\,.
\ee

\section{\label{App:redefinition} Derivation of redefined shift vector}

Here we derive the expressions for the redefinition~(\ref{shiftred}) of the ADM shift vector $N^i=\gamma^{ij}N_j$ and the matrices $\mA$ and $\mB$ in (\ref{exprab}), following~\cite{Hassan:2011tf}. We start from the most general ansatz for the redefinition, 
\beqn\label{redefans}
N^i=c^i+N d^i\,,
\eeqn
where $c^i$ and $d^i$ are functions of $\gamma_{ij}$, the new shift vector $n^i$ and the ADM components of $\fmn$. They will be determined in what follows. Recall that the redefinition has to be linear in the lapse $N$ in order not to introduce nonlinearities into the Einstein-Hilbert term~(\ref{gradm}). Next we turn to equation~(\ref{sadm}) and take the square of both sides to arrive at,
\beqn\label{sqeqs}
g^{-1}f=\frac{1}{N^2}\mA^2+\frac{1}{N}(\mA\mB+\mB\mA)+\mB^2\,.
\eeqn
The left-hand side of this equation can be expressed in terms of ADM variables using (\ref{invadm}) and (\ref{admf}). We get,
\beqn\label{ginvf}
&~&g^{\mu\rho}f_{\rho\nu} \nonumber\\
&=&\frac{1}{N^2}\begin{pmatrix}
L^2-L_l \phi^{lk}L_k  +N_l\gamma^{lk}L_k  ~&~    -L_j +N_l\gamma^{lk}\phi_{kj}\\
N^2\gamma^{ik}L_k-N^i(L^2-L_l \phi^{lk}L_k  +N_l\gamma^{lk}L_k ) ~&~N^2\gamma^{ik} \phi_{kj}+N^i(N^k\phi_{kj}-L_j)
\end{pmatrix}\,.\nonumber\\
\eeqn
In this expression we replace $N^i$ in terms of $n^i$ using our ansatz~(\ref{redefans}). Next, we collect the pieces according to their order in $1/N$, which gives,
\beqn
g^{-1}f=\frac{1}{N^2}\mathbb{E}_0+\frac{1}{N}\mathbb{E}_1+\mathbb{E}_2\,,
\eeqn
where, in terms of $a_0\equiv  L^2-L_l \phi^{lk}L_k + c^lL_l$ and $a_i\equiv-L_i+c^l\phi_{li}$, we have defined the matrices,
\beqn
\mathbb{E}_0=\begin{pmatrix}
a_0 ~&~ a_j\\ -a_0 c^i ~&~ -c^i a_j
\end{pmatrix}\,,
\quad
&~&\mathbb{E}_1=\begin{pmatrix}
d^l L_l ~&~ d^l\phi_{lj}\\ -(d^lL_lc^i+a_0d^i) ~&~ -(c^i d^l \phi_{lj}+d^i a_j)
\end{pmatrix}\nn\\
&~&\mathbb{E}_2=\begin{pmatrix}
0~&~ 0\\ (\gamma^{il}-d^id^l)L_l ~&~ (\gamma^{il}-d^id^l)\phi_{lj}
\end{pmatrix}\,.
\eeqn
Comparing this to the right-hand side of (\ref{sqeqs}) and equating the coefficients in front of equal powers of $1/N$, we see that,
\beqn\label{abitoe}
\mA^2=\mathbb{E}_0\,,\qquad
\mB^2=\mathbb{E}_2\,,\qquad
\mA\mB+\mB\mA = \mathbb{E}_1\,.
\eeqn
Since $\mathbb{E}_0$ is a projector on a one-dimensional subspace, its square-root can easily be evaluated, resulting in,
\beqn\label{maitoas}
\mA=\sqrt{\mathbb{E}_0}=\frac{1}{L\sqrt{x}}
\begin{pmatrix}
a_0~&~a_j\\
-a_0c^i~&~-c^ia_j
\end{pmatrix}\,,\qquad x\equiv \frac{1}{L^2}(a_0-c^la_l)\,.
\eeqn
The special structure of $\mathbb{E}_2$ also allows us to evaluate its square-root which gives,
\beqn\label{solfb}
\mB=\sqrt{\mathbb{E}_2}=\sqrt{x}
\begin{pmatrix}
0~&~0\\
{D^i}_kL^k~&~{D^i}_j
\end{pmatrix} \,,
\qquad D^i_{~j}\equiv \frac{1}{\sqrt{x}}\sqrt{(\gamma^{il}-d^id^l)\phi_{lj}}\,.
\eeqn
Using these expressions for $\mA$ and $\mB$ in the third equation of~(\ref{abitoe}), we obtain,
\beqn
\mA\mB+\mB\mA = \mathbb{E}_1 \qquad \Rightarrow \qquad d^i=\frac{1}{L}D^i_{~k}\big(c^k-\phi^{kl}L_l\big)\,,
\eeqn
where we have also made use of the symmetry property $\phi_{ik}D^k_{~j}=\phi_{jk}D^k_{~i}$ which follows directly from the definition of~$D$ in~(\ref{solfb}). This equation determines a particular combination of the unknown functions $d^i$ and $c^i$ in the redefinition. The redefinition is thus not uniquely determined. One simple option to fix the ambiguity is to choose the new shift vectors as $n^k=\frac{1}{L}(c^k-\phi^{kl}L_l)$. Then the above condition reduces to $d^i=D^i_{~k}n^k$ and the redefinition becomes,
\beqn
N^i=c^i+Nd^i=L^i+Ln^i+ND^i_{~k}n^k\,,
\eeqn
where $L^i\equiv \phi^{ij}L_j$. The definition of the matrix $D$ in~(\ref{solfb}) depends on $d^i$ and therefore gives rise to the following matrix equation that needs to be solved for $D$,
\beqn
\sqrt{x}\,D= \sqrt{(\gamma^{-1}-Dn(Dn)^\mathrm{T})\phi}\,,\qquad x\equiv 1-n^l\phi_{lk}n^k\,. 
\eeqn
As shown in~\cite{Hassan:2011tf}, the solution to this equation is given by~(\ref{defD}).

\section{Short summary of standard GR cosmology}\label{app:cosmo}

Here we very briefly recapitulate the derivation of the cosmological evolution equations in GR. Consider Einstein's equations of motions for $\gmn$,
\beqn\label{einsteq}
R_{\mu\nu}-\frac{1}{2}\gmn R=\frac{1}{M_\mathrm{Pl}^2}\left(T_{\mu\nu}-\rho_\Lambda\,\gmn\right)\,,
\eeqn
where $T_{\mu\nu}$ is the stress-energy tensor obtained from variation of the matter Lagrangian and we have defined the constant energy density $\rho_\Lambda\equiv\Lambda M_\mathrm{Pl}^2$. 
When looking for homogeneous and isotropic solutions in General Relativity, one makes a Friedmann-Robertson-Walker ansatz for the metric,
\beqn\label{FRWg1}
\gmn \md x^\mu\md x^\nu=-\md t^2+a(t)^2\left(\frac{1}{1-kr^2}\md r^2+r^2\md \Omega\right)\,,
\eeqn
in which $k=0,-1,+1$ parameterises the curvature of the universe (flat, open, closed), $a(t)$ is the scale factor, and $\md \Omega=\md \theta^2+\sin^2 \theta\,\md \varphi^2$. 
In accordance with homogeneity and isotropy, the stress-energy tensor is taken to be that of a perfect fluid, 
\beqn\label{setans}
{(T^g)^{\mu}}_\nu=\mathrm{diag}(-\rho, p, p, p)\,,
\eeqn
where $\rho(t)$ and $p(t)$ are the time dependent energy density and pressure of the fluid, respectively. 
As a consequence of the Bianchi identity, the source is automatically covariantly conserved which implies the continuity equation,
\beqn\label{conteq}
\dot{\rho}+3\frac{\dot{a}}{a}\big(\rho+p\big)=0\,,
\eeqn
where dots denote time derivatives. 
Plugging the diagonal ans\"atze~(\ref{FRWg1}) and~(\ref{setans}) into Einstein's equations~(\ref{einsteq}) gives a set of four equations. The 00-component is Friedmann's equation,
\beqn\label{Friedmann}
\left(\frac{\dot{a}}{a}\right)^2+\frac{k}{a^2}=\frac{\rho+\rho_\Lambda}{3M_\mathrm{Pl}^2}\,,
\eeqn 
which is often expressed in terms of the Hubble function $H\equiv \dot{a}/a$.
Due to the isotropic ansatz, the $ii$-components of the equations are all equivalent and therefore provide only one additional independent equation. The most common way to present it is by taking the traced Einstein equations and using~(\ref{Friedmann}) to arrive at the acceleration equation,
\beqn\label{acceq}
\frac{\ddot{a}}{a}=-\frac{1}{6M_\mathrm{Pl}^2}\left(\rho+\rho_\Lambda+3p\right)\,.
\eeqn
In order to characterise different components of the cosmological fluid, one introduces the equation of state parameter $w$ defined as the ratio of pressure and energy density, $p=w\rho$. One can then rewrite the continuity equation~(\ref{conteq}) as 
\beqn
\frac{\dd \rho}{\dd a}+3(1+w)\frac{\rho}{a}=0\,,
\eeqn
which is solved by
\beqn
\rho=\rho_0\, a^{-3(1+w)}\,.
\eeqn
Here, $\rho_0$ is a constant to be determined by boundary conditions. Inserting this solution into the Friedmann equation~(\ref{Friedmann}) with $k=0$ and $\rho_\Lambda=0$ gives the following evolution of the scale factor for $w\neq -1$,
\beqn
a~\propto~ t^{\frac{2}{3(1+w)}}\,.
\eeqn
The behaviour for $w=-1$ is
\beqn
a~\propto~ e^{Ht}\,,\qquad H\equiv \frac{\dot{a}}{a}=\mathrm{const}.
\eeqn
Moreover, from the acceleration equation~(\ref{acceq}) we infer that, in the absence of a bare cosmological constant, the expansion of the universe accelerates for $w<-1/3$ and decelerates for $w>-1/3$. Three cases are of particular interest: $w=0$ characterises non-relativistic matter, $w=1/3$ corresponds to relativistic matter (``radiation"),  and $w=-1$ describes a cosmological constant. The energy density for non-relativistic matter therefore scales as $a^{-3}$, i.e.~inversely proportional to the volume. Non-relativistic matter experiences an additional redshift and its energy density goes as $a^{-4}$. None of these can explain an accelerated expansion of the universe. In contrast, a cosmological constant which of course has constant energy density leads to an exponentially accelerated expansion. 

It is common to measure the contributions of different fluid components in terms of their density parameters,
\beqn
\Omega_i=\frac{\rho_i}{3M_\mathrm{PL}^2 H_0^2},
\eeqn
where $i$ stands for either radiation, non-relativistic matter or the cosmological constant and $H_0$ denotes the Hubble parameter at the present time. One also defines a curvature contribution, $\Omega_k=-\tfrac{k}{a^2H_0^2}$, and in terms of these the Friedmann equation becomes
\beqn
\Omega_\mathrm{rad}+\Omega_\mathrm{mat}+\Omega_\Lambda+\Omega_k=\frac{H^2}{H^2_0}\,.
\eeqn 
In particular, at the present time the density parameters add up to one. Latest observational data \cite{Adam:2015rua} suggest the following (approximate) values for the cosmological parameters at the present time,
\beqn
&~&H_0\sim 70\, \mathrm{km/s/Mpc}\,,\qquad \Omega_\mathrm{rad}\sim 10^{-5}\,,\qquad \Omega_\mathrm{mat}\sim 0.3 \,,\nn\\
&~&\Omega_\Lambda\sim 0.7\,,\qquad\qquad\qquad~~~\,   \Omega_k< 10^{-3}\,.
\eeqn
These values imply that our universe is flat and dominated by a cosmological constant component. As already discussed above, the energy density corresponding to this cosmological constant is extremely small compared to energy scales of the Standard Model of Particle Physics. Since so far we lack an explanation for this small value, the curious energy component $\rho_\Lambda$ is often referred to as \textit{dark energy}. On top of that, observations also show that the matter component $\Omega_\mathrm{m}$ is not dominated by familiar baryonic matter, but rather mainly consists of an unknown \textit{dark matter} component. 

Consequently, under the assumption that GR is indeed the theory of gravity, we must accept that 95$\%$ of the universe's energy content is not at all understood. Nevertheless, when this obscurity is ignored, the so-called $\Lambda$CDM model (GR with a cosmological constant and cold dark matter) fits the observational data very well.

\bibliographystyle{utphys}
\addcontentsline{toc}{section}{References}
\providecommand{\href}[2]{#2}\begingroup\raggedright\endgroup

\end{document}